\documentclass{ws-ijmpa}
\usepackage[super,compress]{cite}
\usepackage{graphicx}
\usepackage{hyperref}
\hypersetup{colorlinks,urlcolor=black,citecolor=black,linkcolor=black,filecolor=black}

\usepackage[todo,colour,check,front]{optional}
\usepackage{slashed}
\usepackage{multirow}
\usepackage{url}
\usepackage{setspace}
\usepackage{amsmath}
\usepackage{amssymb}
\usepackage{cite}
\usepackage{epsfig}
\usepackage{subfigure}
\usepackage{bbold}

\begin{document}

%
\catchline{}{}{}{}{}
%

\title{CENTRAL EXCLUSIVE PRODUCTION WITHIN THE DURHAM MODEL: A REVIEW}

\author{L. A. HARLAND-LANG}

\address{Department of Physics and Institute for Particle Physics Phenomenology, University of Durham, DH1 3LE, UK}

\author{V. A. KHOZE}

\address{Department of Physics and Institute for Particle Physics Phenomenology, University of Durham, DH1 3LE, UK}
\address{Petersburg Nuclear Physics Institute, NRC Kurchatov Institute, Gatchina, St. Petersburg, 188300, Russia}

\author{M. G. RYSKIN}

\address{Petersburg Nuclear Physics Institute, NRC Kurchatov Institute, Gatchina, St. Petersburg, 188300, Russia}

\author{W. J. STIRLING}

\address{Imperial College London, Exhibition Road, London, SW7 2AZ}

\maketitle


\begin{abstract}
We review recent results within the Durham model of central exclusive production. We discuss the theoretical aspects of this approach and consider the phenomenological implications in a variety of processes, comparing to existing collider data and addressing the possibilities for the future.

\end{abstract}

\ccode{PACS numbers: 12.38.Aw, 12.38.Bx, 12.38.Qk.}

\tableofcontents

\section{Introduction}\label{intro}

Over the last decade there has been a steady rise of theoretical and experimental interest in studies of central exclusive production (CEP) in high-energy hadronic collisions~\cite{Khoze:2001xm,FP420,Martin:2009ku,Albrow:2010yb,Albrow:2013yka,Royon:2013ala,HarlandLang:2013jf,Ronan}. The CEP of an object $X$ may be written in the form 
\begin{equation}\nonumber
pp({\bar p}) \to p+X+p({\bar p})\;,
\end{equation}
where `$+$' signs are used to denote the presence of large rapidity gaps, separating the system $X$ from the intact outgoing protons. On the theoretical side, the study of CEP requires the development of a framework which is quite different from that used to describe the inclusive processes more commonly considered at hadron colliders. The approach, often referred to as the `Durham model', represents a novel application of perturbative QCD, as well as requiring an account of soft diffractive physics. For such processes it is found that a dynamical selection rule operates, where $J_z^{PC}=0^{++}$ quantum number states (here $J_z$ is the projection of the object angular momentum on the beam axis) are expected to be dominantly produced. As we will show, this simple fact leads to many interesting and non--trivial implications for CEP processes, which are not seen in the inclusive case. Experimentally, CEP represents a very clean signal, with just the object $X$ and no other hadronic activity seen in the central detector (as least in the absence of pile--up). In addition, the outgoing hadrons can be measured by installing special `tagging' detectors, situated down the beam line from the central detector, which can provide information about the mass and quantum numbers of the centrally produced state. 

An important advantage of these reactions is that they provide an especially clean environment in which to investigate in details the properties of centrally produced resonance states (in particular to probe their nature and quantum numbers), from `old' Standard Model (SM) mesons to beyond the Standard Model (BSM) Higgs bosons~\cite{Kaidalov03,Heinemeyer:2007tu,HarlandLang:2010ep}. The CEP of, for instance, dijets, $\gamma\gamma$, heavy $(c,b)$ quarkonia, new charmonium--like states, and meson pairs offer a very promising framework within which to study various aspects of QCD, and can serve as `standard candle' processes with which we can benchmark predictions for new CEP physics  at the LHC. These processes have been the focus of recent detailed studies by the authors, including a Monte Carlo (MC) implementation in the \texttt{SuperCHIC} MC~\cite{SuperCHIC}. In this article we will review these results, considering both the phenomenological implications and interesting theoretical features in each case. We will also consider the case of SM and BSM Higgs boson production, considering the implications of the current LHC data on future measurement possibilities.

Currently a wealth of measurements of high--energy CEP have been made, both at the Tevatron~\cite{Mike}  and in Run I of the LHC, with events selected by vetoing on additional hadronic activity over particular rapidity intervals. At the LHC, new CEP data have come from LHCb\cite{Ronan}, CMS\cite {daSilveira:2013aaa}  and ALICE\cite{Zamora:2013dna,Reidt:2013kz}, with very encouraging prospects for the future. CEP measurements are also being made at RHIC, where forward proton taggers are already installed\cite{Leszek}.  The current experimental data are in reasonably good agreement with the Durham expectations\cite{Ronan,Mike}, but a much more detailed comparison with the theory will come after the whole accumulated statistics have been analyzed and as new results from the LHC Run II become available. In this review we will compare the Durham predictions for the processes discussed above to such existing data, as well as considering the prospects for future measurements.

Finally, we note that at the LHC there is a very promising CEP program with tagged protons, using the installed and proposed forward proton spectrometers. The possibility for such measurements with the ATLAS+ALFA detectors is currently under discussion\cite{Staszewski:2011bg}, while the first results of a combined TOTEM+CMS measurement\cite{Oljemarkeds,Oljemark:2013wsa,CMSeds,Antchev:2013hya}, based on a common run with integrated data taking in 2012, are expected to be available soon.  A wide program of CEP studies is also currently under discussion in the framework of the PPS\cite{Albrow:2013yka} and the AFP\cite{Royon:2013en,AFP} upgrade projects, which would allow an investigation of the region of centrally produced masses around 200--800 GeV, using proton detectors stationed  at roughly 210m  and 240m from the interaction points of ATLAS and CMS, respectively. In addition, during low pile--up LHC runs, forward shower counters (FSC) can help extend the rapidity coverage in the forward region and reduce the role of events with proton diffractive dissociation in CEP measurements. These instruments detect showers produced by very forward hadrons hitting the beam pipe and surrounding materials, and have been installed\cite{Ponzo}  in the CMS detector, and successfully used throughout 2012, in particular during the common CMS-TOTEM run. FSCs at LHCb (in the so-called HERSHEL Project \cite{Ronan,Paula2}) are also currently being installed. There is therefore a diverse and promising experimental CEP program at the LHC, with various possibilities for detector upgrades and future measurements, providing an additional strong motivation for studying such processes.

In this review we will present an overview of the recent quantitative theoretical studies by the authors of various CEP processes, within the framework of the Durham model, comparing to data where they exist and presenting predictions and considering the possibilities for future measurements. This review is organized as follows. In Section 2 we summarise the main aspects of the Durham model of CEP, emphasising the importance of the Sudakov factor, considering how soft survival effects may be included, and discussing the so--called `$J_z^{PC}=0^{++}$ selection rule'. In Section 3 we discuss heavy $c$ and $b$--quarkonium production, both of established and `exotic' states. In Section 4 we discuss $\gamma\gamma$ CEP. In Section 5 we discuss the CEP of meson pairs at sufficiently high meson transverse momentum $k_\perp$ that a perturbative approach may be taken. In Section 5 we discuss exclusive 2 and 3--jet production. In Section 6 we discuss the CEP of SM and BSM Higgs bosons, emphasising the implications of recent LHC data for future measurements. Finally in Section 7 we conclude.


\section{The Durham model}\label{dur}

\subsection{Theory}\label{durt}
 
 \subsubsection{Hard process}\label{durth}

\begin{figure}[t]
\begin{center}
\includegraphics[scale=1]{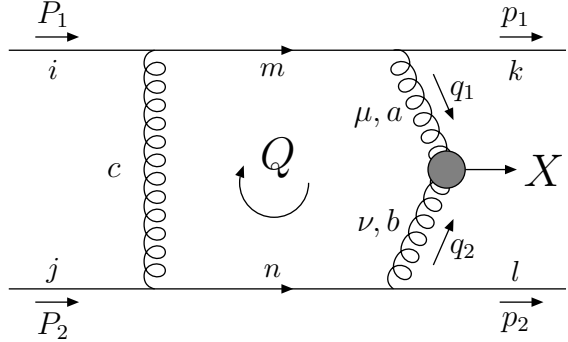}
\caption{Feynman diagram for $qq \to q\,+\,X\,+\,q$ process in perturbative QCD}
\label{qqX}
\end{center}
\end{figure}

The formalism used to calculate the perturbative CEP cross section is explained in detail elsewhere~\cite{Khoze97,Khoze00,Kaidalov03,HarlandLang:2010ep,Khoze:2001xm,Albrow:2010yb,Harland-Lang:2013xba} and we present a summary and motivation here. The lowest order QCD contribution to the CEP process (Fig.~\ref{qqX}) is due to the exchange of two $t$--channel gluons, with the second `screening' gluon, which is assumed not to couple to the system $X$, present to ensure that no colour is transferred between the incoming and outgoing quarks. In the limit of high c.m.s. energy squared $s$, when the object $X$ is produced centrally and the outgoing quarks travel in approximately the forward direction, the colour singlet nature of the exchange tells us that this amplitude will be predominantly imaginary. The calculation can therefore be simplified via the Cutkosky rules, which allow us to calculate the imaginary part of this amplitude in a relatively straightforward way. Applying these, and after colour averaging, attaching the fusing gluons to the quark lines with eikonal vertices and making use of the gauge invariance of the $gg \to X$ subprocess, we then arrive at an expression for the parton level amplitude $A$
\begin{equation}\label{qqH}
\frac{iA}{s}=\frac{8}{N_C^2-1}\alpha_s^2 C_F^2\int \frac{{\rm d}^2 Q_\perp}{Q_\perp^2 q_1^2 q_2^2}\,\overline{\mathcal{M}}\;.
\end{equation}
where the momenta are defined in Fig.~\ref{qqX}, and $\overline{\mathcal{M}}$ is the colour-averaged, normalised sub-amplitude for the $gg \to X$ process:
\begin{equation}\label{Vnorm}
\overline{\mathcal{M}}\equiv \frac{2}{M_X^2}\frac{1}{N_C^2-1}\sum_{a,b}\delta^{ab}q_{1_\perp}^\mu q_{2_\perp}^\nu V_{\mu\nu}^{ab} \; .
\end{equation}
Here the $q_{i\perp}$ are the transverse momenta of the fusing gluons, given by
\begin{align}\label{q1pdef}
q_{1\perp}&=Q_\perp-p_{1_{\perp}}\;, \\ \label{q2pdef}
q_{2\perp}&=-Q_\perp-p_{2_{\perp}}\;.
\end{align}
where $p_{1_{\perp}},p_{2_{\perp}}$ are the transverse momenta of the outgoing protons.

While it is relatively straightforward to write down this expression (\ref{qqH}) for the LO, parton--level, CEP amplitude, there are further corrections which must be included. Firstly, we can see that the integral over the loop momentum $Q_\perp$ is divergent in the infra--red. This issue is resolved by a more careful treatment of higher--order effects: in particular, as the CEP process involves the disparate scales of the object mass $M_X$ and the gluon transverse momentum $Q_\perp$, we will expect  large logarithms $\sim \ln (M_X^2/Q_\perp^2) $ to be present when we consider higher--order virtual corrections to the LO process. These can be resummed systematically in a Sudakov factor, given by
\begin{equation}\label{tsp}
T_g(Q_\perp^2,\mu^2)={\rm exp} \bigg(-\int_{Q_\perp^2}^{\mu^2} \frac{{\rm d} k_\perp^2}{k_\perp^2}\frac{\alpha_s(k_\perp^2)}{2\pi} \int_{0}^{1-k_\perp/M_X} \bigg[ z P_{gg}(z) + n_F P_{qg}(z) \bigg]{\rm d}z \bigg)\;.
\end{equation}
This resums these virtual logarithms in $M_X^2/Q_\perp^2$ which occur when the loop momenta in virtual diagrams become soft and/or collinear to the external particle directions, to next--to--leading logarithmic accuracy. That is, with these choices of lower and upper cutoffs on the $k_\perp$ and $z$ integrals, it takes into account all terms of order $\alpha_s^n\ln^m(M_X^2/Q_\perp^2)$, where $m=2n,2n-1$. More physically, it corresponds to the (Poissonian) probability of no extra parton emission from a fusing gluon, that is the probability that the gluon evolves from a scale $Q_\perp$ to the hard scale $\mu$ without additional real emission. This interpretation follows from the fact that these large higher--order logarithms are  being generated by a mis--match between real and virtual corrections which occurs due to the exclusivity requirement that no extra emissions be present.

Clearly, it is crucial to correctly account for these large corrections before giving a reliable cross section prediction. In particular, if we include only the double logarithmically enhanced contribution to (\ref{tsp}) and assume a fixed coupling $\alpha_s$ for simplicity, then we have
\begin{equation}\label{tspapprox}
T_g(Q_\perp^2,\mu^2=M_X^2)= \exp\left(-\frac{\alpha_sN_c}{4\pi}\ln^2\left(\frac{Q_\perp^2}{M_X^2}\right)\right)\;,
\end{equation}
which vanishes faster than any power of $Q_\perp^2$ that may come from the parton--level amplitude as $Q_\perp^2\to 0$, and so has the effect of making the loop integration in (\ref{qqH}) infrared finite; this result remains true when the full form (\ref{tsp}) is taken. It is therefore this requirement that there is no emission off the two gluons which ensures we are considering an infrared stable observable. Specifically, for length scales $\lambda_Q \gtrsim 1/Q_\perp$ the two--gluon system acts as a colour singlet, and so there will no additional emission with $k_\perp \lesssim Q_\perp$, as for these wavelengths ($\gtrsim 1/Q_\perp$) the individual gluon colour is not resolved by the radiation. However, as $Q_\perp \to 0$, the transverse size of the system increases, with the effect that this additional radiation can no longer be suppressed, and thus the amplitude for exclusive production vanishes in this infrared region.

\begin{figure}
\begin{center}
\includegraphics[scale=0.55]{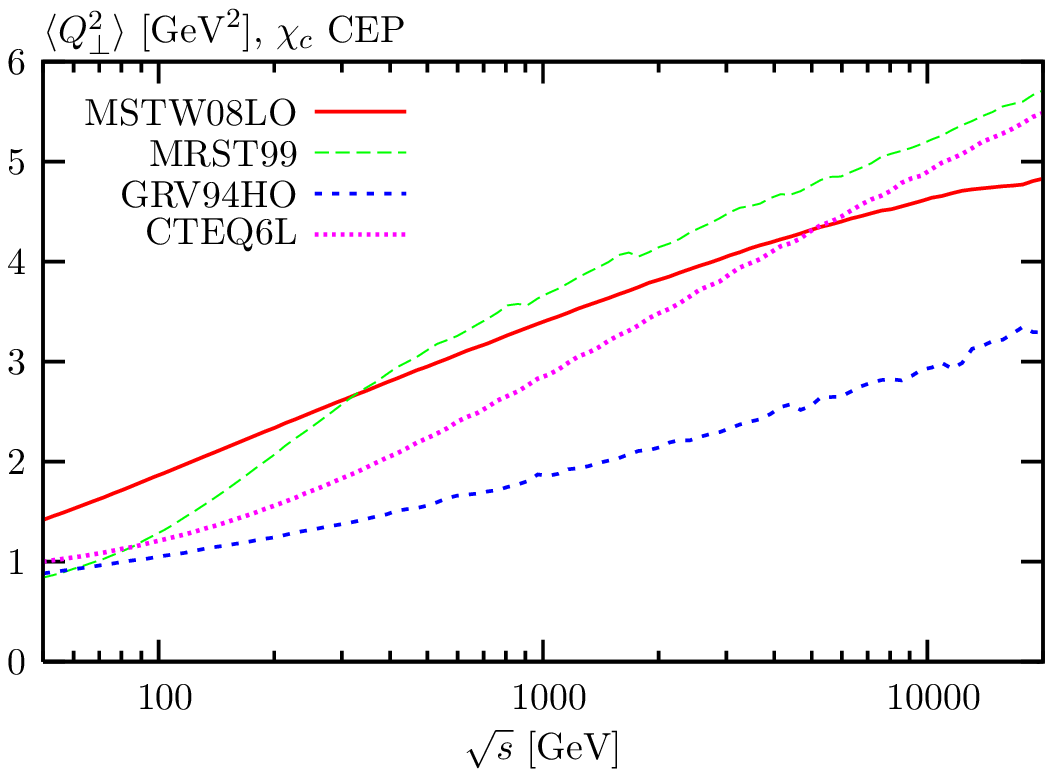}\qquad
\includegraphics[scale=0.55]{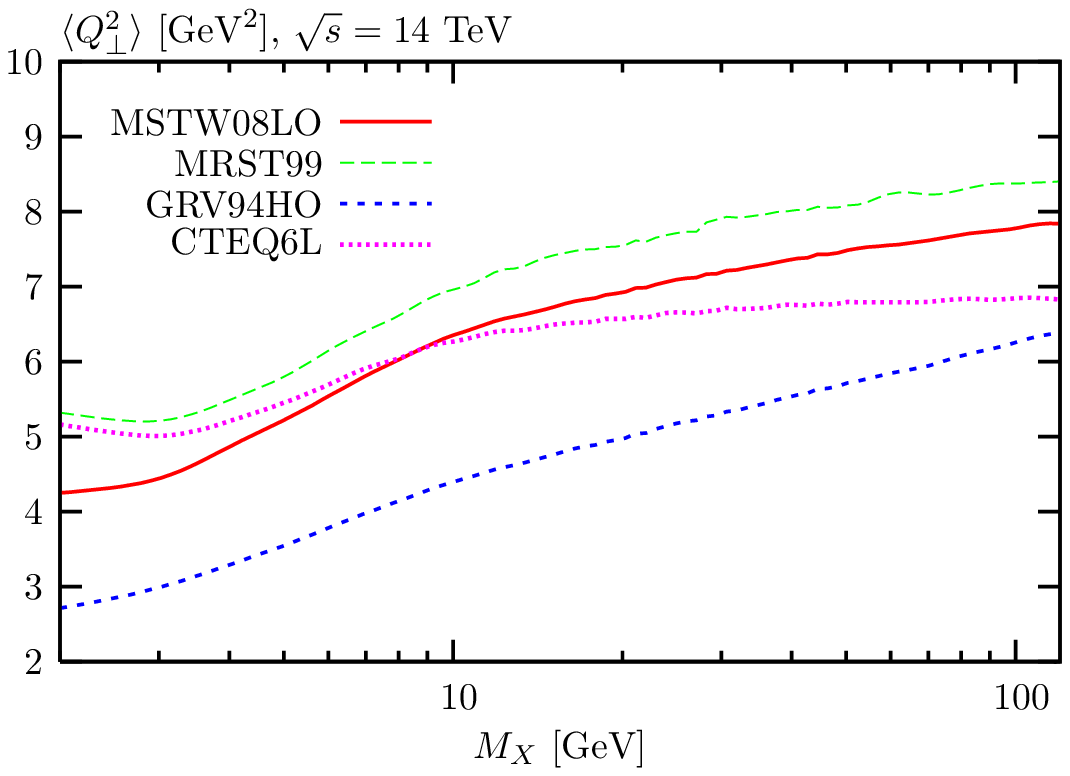}
\caption{The average gluon squared--transverse momentum $\langle Q_\perp^2\rangle$ in the integrand of (\ref{bt}) for the production of a $0^+$ scalar particle in the forward proton ($p_\perp=0$) limit, as a function of the c.m.s. energy $\sqrt{s}$, and of the object mass $M_X$, for different choices of the gluon PDF (GRV94HO~\cite{Gluck94}, MSTW08LO~\cite{Martin:2009iq}, CTEQ6L~\cite{Pumplin:2002vw} and MRST99~\cite{Martin:1999ww}).}\label{fgcomp1}
\end{center}
\end{figure}

As well as ensuring such an IR finite result, the Sudakov factor also ensures that the CEP cross section is {\it perturbative}, that is the average gluon transverse momentum $Q_\perp$ is safely in the perturbative regime. This is shown explicitly in Fig.~\ref{fgcomp1}, where the dependence of the expectation value $\langle Q_\perp^2\rangle$ of the loop integral for the final CEP amplitude (\ref{bt}) is plotted, for different $\sqrt{s}$ values and choices of PDF. We can see that the average $\langle Q_\perp^2\rangle$ increases both with the object mass $M_X$ and c.m.s. energy $\sqrt{s}$, independent of the PDF set used (although the value of $\langle Q_\perp^2\rangle$ does have some non--negligible PDF dependence), and that in all cases we have $\langle Q_\perp^2\rangle = O({\rm GeV}^2)$, safely in the perturbative regime.

In addition to this, we must also convert the parton--level amplitude (\ref{qqH}) to the hadron level. This is achieved by the introduction of the so--called `skewed' unintegrated PDFs $f_g$, by making the replacement in (\ref{qqH})~\cite{Ryskin:1992ui,Khoze97} 
\begin{equation}
\frac{\alpha_S C_F}{\pi} \to f_g(x,x',Q_\perp^2,\mu^2) \; ,
\end{equation}
where $\mu\sim M_X$ is the factorisation scale, and $x'$ ($x$) are the momentum fractions carried by the screening (fusing) gluon. As we are in fact interested  in the distribution of gluons in $Q_\perp$, which are evolved in energy up to the hard scale $M_X$, such that they are accompanied by no additional radiation, these objects involve both the gluon PDFs and the Sudakov factor (\ref{tsp}) in a non--trivial combination. In particular it can be shown that, in the $x' \ll x$ regime relevant to CEP, the $f_g$'s can be written as
\begin{equation}\label{fgskew}
f_g(x,x',Q_\perp^2,\mu^2) =\; \frac{\partial}{\partial \ln(Q_\perp^2)} \left[ H_g\left(\frac{x}{2},\frac{x}{2};Q_\perp^2\right) \sqrt{T(Q_\perp,\mu^2)} \right]\;,
\end{equation}
where $H_g$ is the generalised gluon PDF~\cite{Belitsky:2005qn}. For CEP kinematics this can be related to the conventional PDFs using the `Shuavev transform'~\cite{Shuvaev:1999ce}. More precisely, it has been shown recently~\cite{Harland-Lang:2013xba} that this can be written in a very simple form\footnote{We note that some previous expressions for the skewed PDFs are written in terms of a factor $R_g$~\cite{Harland-Lang:2013xba}. However frequently the correct $Q_\perp$ dependence of this factor is ignored when such an approach is used, and so we now prefer to use the more precise form given here.}
\begin{equation}
H_g\left(\frac{x}{2},\frac{x}{2},Q^2\right)\; = \frac{4x}{\pi} \int_{x/4}^1 \;{\rm d}y \; y^{1/2}(1-y)^{1/2}\,g\left(\frac{x}{4y},Q^2\right)\;.
\end{equation}
A careful treatment~\cite{Coughlin:2009tr} shows that (\ref{fgskew}) is the correct form for the skewed PDFs, with in particular the limits on the Sudakov factor (\ref{tsp}) determined by the requirement that all next--to--leading logarithms be resummed correctly.

Following from this discussion, we may therefore write down a final expression for the CEP amplitude
\begin{equation}\label{bt}
T\equiv \frac{iA}{s}=\pi^2 \int \frac{d^2 {\bf Q}_\perp\, \overline{\mathcal{M}}}{{\bf Q}_\perp^2 ({\bf Q}_\perp-{\bf p}_{1_\perp})^2({\bf Q}_\perp+{\bf p}_{2_\perp})^2}\,f_g(x_1,x_1', Q_1^2,\mu^2;t_1)f_g(x_2,x_2',Q_2^2,\mu^2;t_2) \; ,
\end{equation}
where $\mathcal{M}$ is given by (\ref{Vnorm}) and we have now introduced the $t$--dependence of the skewed PDFs (omitted above for simplicity): typically this is assumed to factorize out as a proton form factor, which we take to have the form $F_N(t)={\rm exp}(bt/2)$, with $b=4\,{\rm GeV}^{-2}$. Here $\mu$, as described above, is the hard scale of the process, and in what follows we take $\mu=M_X/2$ for concreteness.

\subsubsection{Soft corrections}\label{secsurv}

The expression (\ref{bt}) corresponds to the amplitude for the exclusive production of an object $X$ in a short--distance interaction, that is, with no perturbative emission. However, as we are requiring that there are no other particles accompanying this final state we must also include the probability that these are not produced in additional soft proton--proton interactions (or `rescatterings'), independent of the hard process, i.e. as a result of underlying event activity. This probability is encoded in the so--called `eikonal survival factor', $S^2_{\rm elk}$~\cite{Bjorken:1992er,Khoze:2006uj,Ryskin:2009tk,Ostapchenko:2010gt,Gotsman:2012rq,Khoze:2013dha,Khoze:2014aca,Gotsman:2014pwa}.

The survival factor is conventionally written in terms of the proton opacity $\Omega(s,b_t)$. The proton opacity is related via the usual elastic unitarity equations to such hadronic observables as the elastic and total cross sections as well as, combined with some additional physical assumption about the composition of the proton, the single and double diffractive cross sections. Thus, while the survival factor is a soft quantity which cannot be calculated using pQCD, it may be extracted from soft hadronic data~\cite{Khoze:2013dha,Khoze:2013jsa}. Although there is some uncertainty in the precise level of suppression (in particular in its dependence on the c.m.s. energy $\sqrt{s}$), this is found to be a sizeable effect, reducing the CEP cross section by about two orders of magnitude. 

The survival factor is not a simple multiplicative constant~\cite{HarlandLang:2010ep}, but rather depends on the distribution in impact parameter space of the colliding protons. In particular, in the simplest `one--channel' model, which ignores any internal structure of the proton, we can write the average suppression factor as
\begin{equation}\label{S2}
\langle S^2_{\rm eik} \rangle=\frac{\int {\rm d}^2 {\bf b}_{1t}\,{\rm d}^2 {\bf b}_{2t}\, |T(s,{\bf b}_{1t},{\bf b}_{2t})|^2\,{\rm exp}(-\Omega(s,b_t))}{\int {\rm d}^2\, {\bf b}_{1t}{\rm d}^2 {\bf b}
_{2t}\, |T(s,{\bf b}_{1t},{\bf b}_{2t})|^2}\;,
\end{equation}
where ${\bf b}_{it}$ is the impact parameter vector of proton $i$, so that ${\bf b}_t={\bf b}_{1t}+{\bf b}_{2t}$ corresponds to the transverse separation between the colliding protons, with $b_t = |{\bf b}_t|$.  $T(s,{\bf b}_{1t},{\bf b}_{2t})$ is the CEP amplitude (\ref{bt}) in impact parameter space, and $\Omega(s,b_t)$ is the proton opacity discussed above; physically, $\exp(-\Omega(s,b_t))$ represents the probability that no inelastic scattering occurs at impact parameter $b_t$. 

While the rescattering probability only depends on the magnitude of the proton transverse separation $b_t$, the hard matrix element may have a more general dependence. More specifically, $T(s,{\bf b}_{1t},{\bf b}_{2t})$ is the Fourier conjugate of the CEP amplitude (\ref{bt}), i.e. we have
\begin{equation}\label{Mfor}
T(s,{\bf p}_{1_\perp},{\bf p}_{2_\perp})=\int {\rm d}^2{\bf b}_{1t}\,{\rm d}^2{\bf b}_{2t}\,e^{i{\bf p}_{1_\perp}\cdot {\bf b}_{1t}}e^{-i{\bf p}_{2_\perp}\cdot {\bf b}_{2t}}T(s,{\bf b}_{1t},{\bf b}_{2t})\;,
\end{equation}
where the minus sign in the ${\bf p}_{2_\perp}\cdot {\bf b}_{2t}$ exponent is due to the fact that the impact parameter ${\bf b}_t$ is the Fourier conjugate to the momentum transfer ${\bf q}={\bf p}_{1_\perp}-{\bf p}_{2_\perp}$. We can therefore see that (\ref{S2}) is dependent on the distribution in the transverse momenta ${\bf p}_{i_\perp}$ of the scattered protons, being the Fourier conjugates of the proton impact parameters, ${\bf b}_{it}$. This connection can be made clearer by working instead in transverse momentum space, where we should calculate the CEP amplitude including rescattering effects, $T^{\rm res}$, by integrating over the transverse momentum ${\bf k}_\perp$ carried round the Pomeron loop (represented by the grey oval labeled `$S_{\rm eik}^2$' in Fig.~\ref{npip}). The amplitude including rescattering corrections is given by
\begin{equation}\label{skt}
T^{\rm res}(s,\mathbf{p}_{1_\perp},\mathbf{p}_{2_\perp}) = \frac{i}{s} \int\frac{{\rm d}^2 \mathbf {k}_\perp}{8\pi^2} \;T_{\rm el}(s,{\bf k}_\perp^2) \;T(s,\mathbf{p'}_{1_\perp},\mathbf{p'}_{2_\perp})\;,
\end{equation}
where $\mathbf{p'}_{1_\perp}=({\bf p}_{1_\perp}-{\bf k}_\perp)$ and $\mathbf{p'}_{2_\perp}=({\bf p}_{2_\perp}+{\bf k}_\perp)$, while $T^{\rm el}(s,{\bf k}_\perp^2)$ is the elastic $pp$ scattering amplitude in transverse momentum space, which is related to the proton opacity via
\begin{equation}\label{sTel}
T_{\rm el}(s,t)=2s \int {\rm d}^2 {\bf b}_t \,e^{i{\bf q} \cdot {\bf b}_t} \,T_{\rm el}(s,b_t)=2is \int {\rm d}^2 {\bf b}_t \,e^{i{\bf q} \cdot {\bf b}_t} \,\left(1-e^{-\Omega(s,b_t)/2}\right)\;,
\end{equation}
where $t=-{\bf k}_\perp^2$. We must add (\ref{skt}) to the `bare' amplitude excluding rescattering effects to give the full amplitude, which we can square to give the CEP cross section including eikonal survival effects
\begin{equation}\label{Tphys}
\frac{{\rm d}\sigma}{{\rm d}^2\mathbf{p}_{1_\perp} {\rm d}^2\mathbf{p}_{2_\perp}} \propto |T(s,\mathbf{p}_{1_\perp},\mathbf{p}_{2_\perp})+T^{\rm res}(s,\mathbf{p}_{1_\perp},\mathbf{p}_{2_\perp})|^2 \;,
\end{equation}
where here (and above) we have omitted the dependence of the cross section on all other kinematic variables for simplicity. In this way the expected soft suppression is given by 
\begin{equation}\label{seikav1}
\langle S_{\rm eik}^2\rangle= \frac{\int {\rm d}^2{\bf p}_{1_\perp}\,{\rm d}^2{\bf p}_{2_\perp}\,|T(s,\mathbf{p}_{1_\perp},\mathbf{p}_{2_\perp})+T^{\rm res}(s,\mathbf{p}_{1_\perp},\mathbf{p}_{2_\perp})|^2}{\int {\rm d}^2{\bf p}_{1_\perp}\,{\rm d}^2{\bf p}_{2_\perp}\,|T(s,\mathbf{p}_{1_\perp},\mathbf{p}_{2_\perp})|^2}\;.
\end{equation}
It can readily be shown that (\ref{S2}) and (\ref{seikav1}) are equivalent. As we expect, the soft suppression factor depends on the proton transverse momenta, and so may have an important effect on the distribution of the outgoing proton ${\bf p}_{\perp i}$, via (\ref{Tphys}). A simplified approach, where the soft survival suppression is simply included in the CEP cross section as an overall constant factor will completely omit this effect. We also note that as the survival factor depends on the $p_\perp$ structure of the hard process, the average suppression will depend, as we will see later,  on the object spin and parity.

\begin{figure}[b]
\begin{center}
\includegraphics[scale=1.15]{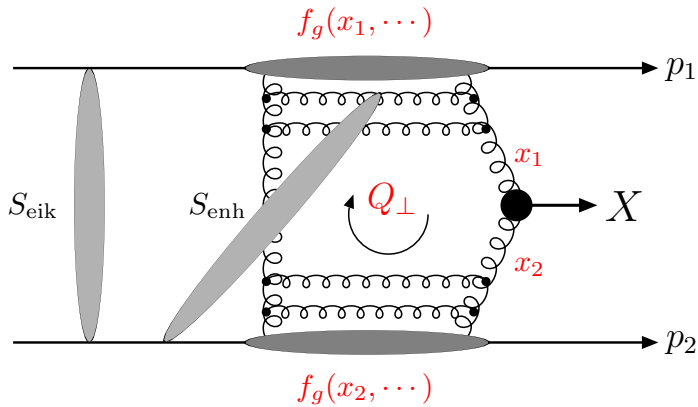}
\caption{The perturbative mechanism for the exclusive process $pp \to p\,+\, X \, +\, p$, with the eikonal and enhanced survival factors 
shown symbolically.}
\label{fig:pCp}
\end{center}
\end{figure}

Besides the effect of eikonal screening $S_{\rm eik}$, there is some suppression caused by the rescatterings of the protons with the intermediate partons~\cite{Ryskin:2009tk,Martin:2009ku,Ryskin:2011qe} (inside the unintegrated gluon distribution $f_g$). This effect is described by the so-called enhanced Reggeon diagrams and usually denoted as $S^2_{\rm enh}$, see Fig.~\ref{fig:pCp}. The value of $S^2_{\rm enh}$ depends mainly on the transverse momentum of the corresponding partons, that is on the argument $Q^2_i$ of $f_g(x,x',Q^2_i,\mu^2)$ in (\ref{bt}), and depends only weakly on the $p_\perp$ of the outgoing protons~\cite{Ryskin:2011qe}. The precise size of this effect is uncertain, but due to the relatively large transverse momentum (and so smaller absorptive cross section $\sigma^{\rm abs}$) of the intermediate patrons, it is only expected to reduce the corresponding CEP cross section by a factor of at most a `few', that is a much weaker suppression than in the case of the eikonal survival factor. The value of $S_{\rm enh}$ is also expected to depend crucially on the size of the available rapidity interval for rescattering $\propto \ln(s/M_X^2)$.

Combining these two effects, we may write down a final expression for the CEP cross section at $X$ rapidity $y_X$
\begin{equation}\label{ampnew}
\frac{{\rm d}\sigma}{{\rm d} y_X}=\langle S^2_{\rm enh}\rangle\int{\rm d}^2\mathbf{p}_{1_\perp} {\rm d}^2\mathbf{p}_{2_\perp} \frac{|T(\mathbf{p}_{1_\perp},\mathbf{p}_{2_\perp})|^2}{16^2 \pi^5} S_{\rm eik}^2(\mathbf{p}_{1_\perp},\mathbf{p}_{2_\perp})\; ,
\end{equation}
where $T$ is given by (\ref{bt}) and the factor $\langle S^2_{\rm enh}\rangle$ corresponds to the enhanced survival factor averaged (i.e. integrated) over the gluon $Q_\perp$, while $S_{\rm elk}^2$ is simply given by
\begin{equation}
S_{\rm eik}^2(\mathbf{p}_{1_\perp},\mathbf{p}_{2_\perp})=\frac{|T(s,\mathbf{p}_{1_\perp},\mathbf{p}_{2_\perp})+T^{\rm res}(s,\mathbf{p}_{1_\perp},\mathbf{p}_{2_\perp})|^2}{|T(s,\mathbf{p}_{1_\perp},\mathbf{p}_{2_\perp})|^2}\;,
\end{equation}
as can be seen by comparing (\ref{Tphys}) and (\ref{ampnew}).

Finally, we note that the formalism described above for the eikonal survival factor is only valid within the `one--channel' framework, which considers the pure elastic case, where the proton state is the correct degree of freedom for hadron--hadron scattering. More realistically, in particular to account for the possibility of (low mass) diffractive dissociation $p \to N^*$, a more sophisticated `multi--channel' framework is required, in which the incoming proton is considered to be in a coherent superposition of so--called diffractive eigenstates, which can each be described  by the above one--channel framework, that is with $S_{\rm elk}$ calculated for each pair $(i,k)$ of eigenstates. The above formalism therefore still corresponds to the basic physics input into the model of soft diffraction that we use, and the extension to the multi--channel case can be achieved in a quite straightforward manner~\cite{Ryskin:2009tj,Ryskin:2011qe}. Nonetheless it should be emphasised that the overall gap survival probability depends sensitively on the structure of the diffractive eigenstate decomposition, with different choices giving equally good fits to soft diffractive data, while predicting quite different survival factors~\cite{Ryskin:2012az,Khoze:2013dha}.

It is this expression (\ref{ampnew}) which will be used in the discussion that we present in this review, and which corresponds to the pQCD--based Durham model of CEP; this is shown schematically in Fig.~\ref{fig:pCp}. Before going on to consider some important phenomenological applications of this model, we will discuss the so--called `$J_z^{PC}=0^{++}$ selection' rule which is of great importance in such exclusive processes.

\subsection{$J^{PC}_z=0^{++}$ selection rule}\label{select}

When we consider the CEP process in the limit that the outgoing protons scatter at zero angle (corresponding to the proton $p_{i\perp}=0$), then this in fact obeys certain important selection rules which determine the quantum numbers of the centrally produced state $X$~\cite{Khoze:2004rc,Khoze:2000mw,Khoze00}.

First, as the fusing gluons in Fig.~\ref{qqX} must be in a colour singlet $C$--even state, the object $X$ must have positive $C$--parity. Second, as the initial-- and final--state protons both travel in the $z$--direction (where we define the $z$--axis as the beam direction) they both have $L_z=0$, with no angular momentum transfer between them, and therefore by conservation of angular momentum the object $X$ must also have $J_z=0$, where $J_z$ is the projection of the total object angular momentum on the $z$--axis. Finally, as we will discuss below, the object $X$ must also have even parity. We therefore have that in the forward proton limit the centrally produced state $X$ obeys a $J_z^{PC}=0^{++}$ selection rule. As the outgoing protons will in general have some non--zero transverse momentum, $p_{i\perp}\neq 0$, then there will be some violation of the $J_z^P$ part of this rule, as there is now some small angular momentum transfer between the initial and final--state protons. However as this 
$p_\perp$ must be small for such an elastic reaction, being limited by the proton form factor, we would na\"{i}vely expect such corrections to be correspondingly small, and for states with $J_z^{PC}=0^{++}$ to be dominantly produced, although this should be checked explicitly by inspecting the form of perturbative CEP amplitude.

To see how this selection rule arises in the case of the pQCD--based Durham model, we note that in the subprocess amplitude (\ref{Vnorm}) the gluon transverse momenta play the role of the gluon polarization vectors $\epsilon_i\sim q_{i\perp}$ which would couple to the $ggX$ vertex $V_{\mu\nu}$ in the usual on--shell $gg \to X$ process. In the forward proton limit we have $q_{1_{\perp}}=-q_{2_{\perp}}=Q_\perp$ 
and therefore the gluon polarizations satisfy $\epsilon_1=-\epsilon_2$, which exactly corresponds to the centrally produced object being in a $J_z=0$ state. More generally, we can decompose (\ref{Vnorm}) in terms of the incoming gluon polarization vectors, given by
\begin{align}\nonumber
\epsilon^{_{+(-)}}_{_{1(2)}}&=-\frac{1}{\sqrt{2}}(\hat{x}+i\hat{y})\;,\\ \label{pol}
\epsilon^{_{-(+)}}_{_{1(2)}}&=\frac{1}{\sqrt{2}}(\hat{x}-i\hat{y})\;,
\end{align}
where the $x-y$ plane is perpendicular to the direction of motion of the gluons in the $gg$ rest frame; in the on--shell approximation (valid up to small corrections of order $\sim q_\perp^2/M_X^2$), this $x-y$ plane coincides with the transverse plane in the lab frame. We can then invert (\ref{pol}) to change the incoming momenta vectors $q_\perp$ to the helicity basis, giving
\begin{align}
q_{1_\perp}^i q_{2_\perp}^j \mathcal{M}_{ij} =\begin{cases} &-\frac{1}{2} ({\bf q}_{1_\perp}\cdot {\bf q}_{2_\perp})(\mathcal{M}_{++}+\mathcal{M}_{--})\;\;(J^P_z=0^+)\\ 
&-\frac{i}{2} |({\bf q}_{1_\perp}\times {\bf q}_{2_\perp})|(\mathcal{M}_{++}-\mathcal{M}_{--})\;\;(J^P_z=0^-)\\ 
&+\frac{1}{2}((q_{1_\perp}^x q_{2_\perp}^x-q_{1_\perp}^y q_{2_\perp}^y)+i(q_{1_\perp}^x q_{2_\perp}^y+q_{1_\perp}^y q_{2_\perp}^x))\mathcal{M}_{-+}\;\;(J^P_z=+2^+)\\ 
&+\frac{1}{2}((q_{1_\perp}^x q_{2_\perp}^x-q_{1_\perp}^y q_{2_\perp}^y)-i(q_{1_\perp}^x q_{2_\perp}^y+q_{1_\perp}^y q_{2_\perp}^x))\mathcal{M}_{+-}\;\;(J^P_z=-2^+)
\end{cases}\label{Agen}
\end{align}
where $\mathcal{M}_{\lambda_1\lambda_2}$ are the $g(\lambda_1)g(\lambda_2)\to X$ helicity amplitudes\footnote{As both gluon momenta are defined as incoming, a $(++/--)$ helicity state corresponds to $J_z=0$ along the $gg$ axis.}. The quantum number assignments follow straightforwardly from the contributing helicity amplitudes. In the odd--parity case, we recall that the gluon polarizations must be in an antisymmetric state for the $gg$ state to be odd under a parity inversion. For the $J_z=\pm 2$ piece, we note that this can be written in the manifestly covariant form
\begin{equation}
\epsilon_{\mu\nu}^{(+2)} q_{1_\perp}^\mu q_{2_\perp}^\nu\epsilon^{_-}_1\epsilon^{_+}_2 \mathcal{M}_{-+}+\epsilon_{\mu\nu}^{(-2)} q_{1_\perp}^\mu q_{2_\perp}^\nu\epsilon^{_+}_1\epsilon^{_-}_2 \mathcal{M}_{+-}\;,
\end{equation}
where the $\epsilon_{\mu\nu}^{(\pm 2)}$ are the usual $|J_z|=2$ polarization tensors\cite{HarlandLang:2009qe} (evaluated in the rest frame of the $gg$ system).

In the $p_\perp\to 0$ limit the only non-vanishing term after the $Q_\perp$ integration in (\ref{bt}) is the first one, with
\begin{equation}
q_{1_\perp}^i q_{2_\perp}^j \mathcal{M}_{ij} \to \frac{1}{2}Q_\perp^2(\mathcal{M}_{++}+ \mathcal{M}_{--})\sim\sum_{\lambda_1,\lambda_2}\delta^{\lambda_1\lambda_2}\mathcal{M}_{\lambda_1\lambda_2}\;,
\end{equation}
as we expect from the $J_z^{PC}=0^{++}$ selection rule. For $p_{i\perp} \neq 0$, the non--$J_z=0$ terms in (\ref{Agen}) do not vanish upon the $Q_\perp$ integration, and can therefore contribute. After performing the $Q_\perp$ integral and squaring, we find that the $|J_z|=2$ amplitude is approximately suppressed by a factor~\cite{HarlandLang:2009qe,HarlandLang:2010ep}
\begin{equation}\label{simjz2}
\frac{|T(|J_z|=2)|^2}{|T(J_z=0)|^2}\sim \frac{\langle p_\perp^2 \rangle^2}{\langle Q_\perp^2\rangle^2}\;,
\end{equation}
where $\langle Q_\perp^2\rangle$ ($\sim 2-5$ ${\rm GeV}^2$, see Fig.~\ref{fgcomp1}) corresponds to the average $Q_\perp^2$ in the integrand of (\ref{bt}), and $\langle p_\perp^2 \rangle = 1/b\sim0.25 \,{\rm GeV}^2$ is the average proton transverse momentum. We therefore have
\begin{equation}
\frac{|T(|J_z|=2)|^2}{|T(J_z=0)|^2}\sim 1\%\;,
\end{equation}
justifying the statement that non--$J_z=0$ quantum numbers are expected to be strongly suppressed.

Considering now the odd--parity part of the selection rule, we have
\begin{equation}
V(gg \to 0^-) \sim \frac{i}{2} |({\bf q}_{1_\perp}\times {\bf q}_{2_\perp})|(\mathcal{M}_{++}-\mathcal{M}_{--})\;,
\end{equation}
which clearly vanishes in the forward ($q_{1\perp}=-q_{2\perp}=Q_\perp$) limit. The reason for this parity selection rule is not just due to the fact that the fusing gluons are in a $J_z=0$ state in this limit (the pseudoscalar decay $h(0^-) \to gg$, for example, can after all occur), but is rather due to the fact that the $gg \to X$ helicity \emph{amplitudes} are summed coherently. While for the individual $g_1(+)g_2(+)$ and $g_1(-)g_2(-)$ gluon helicity states in the $gg \to X$ process the gluons can be in either odd or even parity states, it is only the even parity combination `$g_1(+)g_2(+)+g_1(-)g_2(-)$' which contributes to the gluon `polarization tensor' $\delta^{ij}$ in (\ref{Agen}). More concretely, we have seen that in the forward limit the correlation between the fusing gluons forces their linear polarizations to be parallel, with (\ref{q1pdef}) and (\ref{q2pdef}) reducing to $q_{1\perp}=-q_{2\perp}=Q_\perp$. In this case it readily follows that the gluons must be in an even parity state: for example, a $gg$ 
state with gluon 1 travelling along the $z$--axis and linear polarization in the $+x$ direction, while gluon 2 travels along the negative $z$--axis and has linear polarization in the $-x$ direction is even under a ($x\to -x$ and $z\to -z$) parity transformation, which is equivalent to swapping the identical boson gluons. 

\section{CEP of heavy quarkonium}\label{chicintro}

Among the potential standard candle processes, the CEP of heavy quarkonium ($\chi_{(c,b)}$ and $\eta_{(c,b)}$) states plays a special role~\cite{HarlandLang:2009qe,HarlandLang:2010ep,HarlandLang:2010ys,Khoze:2004yb,Pasechnik:2009bq,Pasechnik:2009qc,Pumplin:1993xk,Yuan01,Petrov:2004nx,Petrov:2004hh,Bzdak:2005rp,Rangel:2006mm}. First, heavy quarkonium production provides  a valuable tool to test the ideas and methods of the QCD physics of bound states, such as  effective field theories, lattice QCD, NRQCD, etc\cite{Bodwin:1994jh,Brambilla:2004jw,Vairo:2009tn,Eichten:2007qx,Danilkin:2009hr}, a subject which is particularly topical in light of the sizeable differences which have recently been observed between the expectations of NLO NRQCD and the current data on the $J/\psi$ and $\Upsilon$ polarization in hadroproduction\cite{Braaten:2014ata}. Second, heavy quarkonium CEP exhibits characteristic features, based on Regge theory, that depend on the particle spin and parity $J^P$~\cite{Kaidalov03}, and these are altered by both the loop integration around the internal gluon momentum $Q_\perp$ and non--zero outgoing proton $p_\perp$ effects as well as by screening corrections, as we will see in section~\ref{numerchi}. A measurement of these effects, in particular the distributions of the outgoing proton momenta~\cite{Khoze:2002nf}, would provide a valuable source of spin--parity information about the centrally produced system as well as constituting an important test of the overall theoretical formalism.

In 2009\cite{Aaltonen:2009kg} the observation of $65\pm 10$ candidate exclusive $\chi_c$ events was reported by CDF, occurring via the radiative $\chi_c \to J/\psi \gamma$ decay chain. These signal events have a 
limited $M(J/\psi\gamma)$ resolution and were collected in a restricted area of final--state kinematics  (due to cuts and event selection criteria). It was in particular not possible to distinguish experimentally which of the three $\chi_{cJ}$ spin states were produced, and in which amounts; rather a broad mass peak about the $\chi_{cJ}$ mass region was observed. In order to determine the $\chi_c$ yield the dominance of $\chi_{c0}$ production
was {\it assumed}, and the CHIC Monte Carlo\footnote{CHIC is a publicly available Monte Carlo implementation of $\chi_{c0}$ CEP~\cite{Khoze:2004yb}.}, based on the $\chi_{c0} \to J/\psi\gamma\to \mu^+\mu^- \gamma$ decay, was used for the conversion of the observed events into a cross section. This assumption is based on general theoretical considerations: for the $\chi_{c1}$ case due to the Landau--Yang theorem~\cite{LY1,Yang50} for on--mass--shell gluons and for the $\chi_{c2}$ because in the non--relativistic approximation the $\chi_{c2}(2^{++})$ meson cannot be produced in a $J_z=0$ state, which is dominant in the case of CEP, see Section~\ref{select}. Under the assumption of $\chi_{c0}$ dominance, the corresponding cross section was found to be in good agreement with the first Durham estimate~\cite{Khoze:2004yb}.

However, while it is true that the $\chi_{c0}$ CEP cross section is expected to be strongly dominant, it was not possible experimentally to rule out the possibility that the $\chi_{c(1,2)}$ state may contribute to the CDF data~\cite{Aaltonen:2009kg}. This is because of the significantly ($\sim$ an order of magnitude) higher $\chi_{cJ} \to J/\psi \gamma$ branching ratio in the case of the $\chi_{c(1,2)}$ states~\cite{HarlandLang:2009qe,Pasechnik:2009bq}. Moreover, we will see that the eikonal survival probability, $\langle S_{\rm eik}^2\rangle$, is  larger for the $\chi_{c1}$ and $\chi_{c2}$  since, due to their spin structure, they are produced more peripherally. We will explicitly see later how we would indeed expect the higher spin $\chi_{c(1,2)}$ states to contribute to the CDF data. The predictions for this can be compared to the CDF measurement, and to the  more recent preliminary LHCb results on the CEP of $\chi_{c}$ mesons in the $\chi_c\to J/\psi\,+\,\gamma$ channel~\cite{LHCb,moranthesis}, and the agreement is found to be  reasonable, given the uncertainties in the theory and possible experimental issues related to exclusivity.

It is also worthwhile recalling that the reconstruction of the bottomonium spectroscopy is still incomplete and, despite a good deal of valuable information on the $b\overline{b}$ states and transitions, various issues  remain so far unresolved. Although the $\Upsilon({}^3S_{1})$ state was discovered in 1977~\cite{Herb:1977ek}, its spin--singlet partner  $\eta_{b}({}^1S_{0})$ was found  more than thirty years later~\cite{Aubert:2008ba}, while the spin assignments of the $P$--wave $\chi_{bJ}$ states still need experimental confirmation~\cite{Beringer:1900zz}. The CEP mechanism, with its spin--parity analyzing capability, could therefore potentially provide a way to establish the spin--parity assignments of the $C$--even  $b\overline{b}$ states. Moreover, we note that due to the higher $\chi_b$ mass, it suffers less from the sizeable theoretical uncertainties that are present in the $\chi_c$ case. For these reasons the CEP of $\chi_b$ and $\eta_b$ states was considered previously in the literature~\cite{HarlandLang:2009qe,HarlandLang:2010ep}, and we discuss this here.

We also note that a new area of experimental studies of CEP with tagged forward protons at c.m.s. energies up to 500~GeV is now being explored by STAR at RHIC~\cite{Guryn:2008ky,Leszek}. A capability to trigger on and to measure the outgoing forward protons provides an excellent means to extend the physics reach in studying CEP processes in exceptionally clean conditions. The encouraging preliminary results collected in 2009 during Phase I are already available~\cite{meson} and, hopefully, the large data sample expected from the measurements in Phase II~\cite{Lee:2010zzp,meson} should provide some very interesting exclusive physics results: recently, the first measurements of exclusive $\pi^+\pi^-$ production with tagged protons by the STAR collaboration have been reported~\cite{Leszek}. Motivated by this, we also discuss the potential for observing exclusive charmonium ($\chi_{cJ}$ and $\eta_c$) production at RHIC with tagged forward protons\cite{HarlandLang:2010ys}, paying particular attention to the new and interesting information that the forward proton distributions can provide. As discussed in the Introduction, there are also a range of possibilities for CEP measurements in this mass region with tagged protons at the LHC, in particular with the ATLAS+ALFA~\cite{Staszewski:2011bg} and TOTEM+CMS~\cite{Oljemark:2013wsa,CMSeds} detectors. The results we discuss here will remain qualitatively unchanged when going to higher LHC energies, so that they can also serve as a case study for such measurements.

Finally, as well as the conventional quarkonium states discussed above, the CEP of  `exotic'  charmonium--like states, which have been discovered over the past 10 years~\cite{Brambilla:2010cs,Braaten:2013oba,Chiochia:2014qva}, represents a very interesting and so far relatively unexplored topic of study. In addition, a number of new bottomonium--like states have been observed, and are the subject of much ongoing investigation~\cite{Chen:2008xia,Bondar:2011ev,Navarra:2011xa}, for example in the case of the new meson~\cite{Aad:2011ih,Abazov:2012gh} currently interpreted as the $\chi_b(3P)$, the origin of which is still the subject of discussion~\cite{Ferretti:2014xqa}.  Here we will discuss one particularly topical charmonium--like state, the $X(3872)$, and show how the CEP mechanism may shed light on the nature of this poorly understood state.

\subsection{Theory}\label{chitheory}

The original extension of the CEP formalism to $\chi_{(c,b)0}$ production\cite{Khoze:2004yb} was achieved by assuming that the $\chi_{(c,b)0}$ coupled to the gluons as a pure scalar, with any effects from its internal structure neglected. More recently~\cite{HarlandLang:2009qe,HarlandLang:2010ep} we went beyond this approximation, modelling the internal structure of the $\chi_{(c,b)}$ mesons for all three $J$ states and in particular their coupling to two gluons. This is done by a simple extension a previous calculation~\cite{Kuhn79}, where the coupling of ${}^3 P_J$ quarkonium states to two off--mass--shell photons is considered within the non--relativistic quarkonium approximation discussed previously; as the gluons are in a colour--singlet state the only difference will be constant prefactors resulting from colour algebra. The relevant calculation for $\chi_b$ production then proceeds in exact analogy to the $\chi_c$ case, the only difference being the input masses $M_\chi$ and widths $\Gamma(\chi \to gg)$. We can also consider pseudoscalar $\eta_{(c,b)}$ production, which can be calculated using the same formalism as for $\chi_{(c,b)}$ production. The $gg\to\chi,\eta$ vertices, defined as in (\ref{Vnorm}), are given by~\cite{HarlandLang:2009qe,HarlandLang:2010ep}

\begin{align}\label{V0}
&\mathcal{M}_{0^+}=\sqrt{\frac{1}{6}}\frac{c_\chi}{M_\chi}(3M_\chi^2(q_{1_{\perp}}q_{2_{\perp}})-(q_{1_{\perp}}q_{2_{\perp}})(
q_{1_{\perp}}^2+q_{2_{\perp}}^2)-2q_{1_{\perp}}^2q_{2_{\perp}}^2)  \; ,\\ \label{V1}
&\mathcal{M}_{1^+}=-\frac{2ic_\chi}{s} p_{1,\nu}p_{2,\alpha}((q_{2_\perp})_\mu(q_{1_\perp})^2-(q_{1_\perp})_\mu(q_{2_\perp})^2)\epsilon^{\mu\nu\alpha\beta}\epsilon^{*\chi}_\beta  \; ,\\
\label{V2}
&\mathcal{M}_{2^+}=\frac{\sqrt{2}c_\chi M_\chi}{s}(s(q_{1_\perp})_\mu(q_{2_\perp})_\alpha+2(q_{1_\perp}q_{2_\perp})p_{1\mu}p_{2\alpha})\epsilon_\chi^{*\mu\alpha}  \; ,\\ \label{V0m}
&\mathcal{M}_{0^-}=ic_\eta(q_{1_\perp}\times q_{2_\perp})\cdot n_0\;,
\end{align}
where $q_{i_\perp}$ are the incoming gluon momenta (\ref{q1pdef}--\ref{q2pdef}) and $n_0$ is a unit vector in the direction of the colliding hadrons (in the c.m.s. frame). $c_\chi$, $c_\eta$ are normalisation factors~\cite{HarlandLang:2010ep}, given by
\begin{equation}
c_\chi=\frac{1}{2\sqrt{N_C}}\frac{16 \pi \alpha_S}{(q_1 q_2)^2}\sqrt{\frac{6}{4\pi M_\chi}}\phi'_P(0), \qquad
c_\eta=\frac{1}{\sqrt{N_C}}\frac{4 \pi \alpha_S}{(q_1 q_2)}\frac{1}{\sqrt{\pi M_\eta}}\phi_S(0)\;,
\end{equation}
where $\phi_{S(P)}(0)$ is the $S?$--wave wavefunction at the origin. Thus the $\chi_{cJ}$ production amplitudes are proportional to $\phi_c'(0)$, the spatial derivative of the bound--state wave function at the origin: this corresponds to the non--perturbative probability amplitude for the formation of these states, and can be normalized to the $\chi_{c0}$ total width $\Gamma^{\rm tot}(\chi_{c0})\approx 10.4$ MeV~\cite{Beringer:1900zz}, with similar results for the $\eta_{c,b}$ and $\chi_{bJ}$ states\cite{HarlandLang:2010ep}

If we consider the $\chi_{c1}$ vertex, we can immediately see that it vanishes for on--shell gluons, that is when $q_i^2=q_{i\perp}^2=0$, as dictated by the Landau--Yang theorem~\cite{LY1,Yang50}. Furthermore, in the forward proton limit ($p_\perp=0$) we have $q_{1\perp}=-q_{2\perp}=Q_\perp$ and so
\begin{align}\label{M0}
\mathcal{M}_0 &\to -\sqrt{\frac{3}{2}}c M_\chi Q_\perp^2\; ,\\ \label{M1}
\mathcal{M}_1 &\to \frac{4ic}{s} Q_\perp^2 p_{1,\nu}p_{2,\alpha}Q_{\perp \mu} \epsilon^{\mu\nu\alpha\beta}\epsilon^{*\chi}_\beta\; ,\\  \label{M2}
\mathcal{M}_2 &\to -\frac{\sqrt{2}cM}{s}(sQ_{\perp\mu}Q_{\perp\alpha}+2Q_{\perp}^2p_{1\mu}p_{2\alpha})\epsilon_\chi^{*\mu\alpha} \; .
\end{align}
We see that $\mathcal{M}_1$ is odd in $Q_\perp$, and will therefore vanish upon the loop integration (\ref{bt}) over $Q_\perp$. For $\mathcal{M}_2$ we make use of the identity
\begin{equation}\label{Qang}
\int d^2Q_\perp Q_{\perp\mu}Q_{\perp\alpha}=\frac{\pi}{2}\int dQ^2_\perp Q_\perp^2 g_{\mu\alpha}^{_T} \; ,
\end{equation}
where $g_{\mu\sigma}^{_T}$, the transverse part of the metric, which can be written in the covariant form
\begin{equation}
g_{\mu\alpha}^{_T}=g_{\mu\alpha}-\frac{2}{s}(p_{1\mu}p_{2\alpha}+p_{1\alpha}p_{2\mu}) \; .
\end{equation}
We then find $\mathcal{M}_2 \propto \epsilon_{\phantom{\mu}\mu}^{\mu}$ which vanishes due to the tracelessness of the $\chi_2$ polarization tensor\footnote{It is worth mentioning that in the QED case the absence of the transition of a spin--2 positronium state into two-photons in a $J_z=0$ state was discovered in mid--fifties~\cite{alekseev}.} . We can see that the $\chi_{c(1,2)}$ production amplitudes vanish in the forward proton limit, and we will therefore expect the corresponding rates to be suppressed relative to $\chi_{c0}$ production, see the discussion in Section~\ref{select}. In fact we can give a very rough estimate for the level of suppression we will expect. Squaring and summing over polarization states gives
\begin{equation}\label{vcomp}
|\mathcal{M}_0|^2:|\mathcal{M}_1|^2:|\mathcal{M}_2|^2 \sim 1:\frac{\left\langle \mathbf{p}_{\perp}^2\right\rangle}{M_\chi^2}:\frac{\left\langle \mathbf{p}_{\perp}^2\right\rangle^2}{\left\langle \mathbf{Q}_\perp^2\right\rangle^2} \; ,
\end{equation}
where the factor of $\left\langle \mathbf{p}_{\perp}^2\right\rangle\sim 0.25$  ${\rm GeV}^2$ comes from integrating over the assumed exponential form ($\sim \exp(bt)$) of the proton vertex in (\ref{bt}). We note the suppression in the $\chi_2$ state is exactly as expected from the discussion in Section~\ref{select}. Although a full treatment will be described in Section~\ref{numerchi}, if for simplicity we assume $\langle \mathbf{Q}_\perp^2 \rangle \approx 2\, {\rm GeV}^2$, $M_{\chi_c}^2\approx 10\, {\rm GeV}^2$ and  $M_{\chi_b}^2\approx 100\, {\rm GeV}^2$ we then obtain
\begin{align}\label{roughc}
|\mathcal{M}_{0^+}|^2:|\mathcal{M}_{1^+}|^2:|\mathcal{M}_{2^+}|^2 &\sim 1:\frac{1}{40}:\frac{1}{64}\quad (c\overline{c}) \;,\\ \label{roughb}
|\mathcal{M}_{0^+}|^2:|\mathcal{M}_{1^+}|^2:|\mathcal{M}_{2^+}|^2 &\sim 1:\frac{1}{400}:\frac{1}{64}\quad (b\overline{b}) \; .
\end{align}
While  we will therefore expect a quite sizeable suppression of the $\chi_{c1}$ and $\chi_{c2}$ CEP cross sections, it is not clear that they are guaranteed to give negligible contributions to the $\chi_c \to J/\psi \gamma$ CEP cross section, for which the $\chi_{c(1,2)}$ branching ratios are a factor of $\sim 10$ higher than the $\chi_{c0}$. On the other hand, we can see that we would expect the $\chi_{b1}$ state to give a very small contribution to the overall $\chi_b$ CEP rate, because of the larger $\chi_b$ mass. However, as with $\chi_c$ CEP, there remains the possibility that $\chi_{b2}$ states may contribute to $\chi_b$ production via the $\chi_b \to \Upsilon\gamma$ decay chain. We will consider both these cases explicitly later on.

Turning now to pseudoscalar $\eta_{(c,b)}$ production, we first observe that the $\eta$ vertex $\mathcal{M}_{0^-}$ vanishes in the forward proton limit, as we expect from the $J^P_z=0^+$ selection rule, see section~\ref{select}. At small $p_\perp$ we will have~\cite{Kaidalov03}
\begin{equation}
|\mathcal{M}_{0^-}|^2 \sim p_{1_\perp}^2 p_{2_\perp}^2 \sin^2\phi\;,
\end{equation}
where $\phi$ is the azimuthal angle between the outgoing protons. The $\eta$ CEP cross section will therefore be heavily suppressed relative to the $\chi_0$ rate by a factor of $\sim \langle \mathbf{p}_{\perp}^2\rangle^2/\langle \mathbf{Q}_{\perp}^2\rangle^2$, i.e. roughly two orders of magnitude, as expected from the discussion in Section~\ref{select}. However, the $\chi$ cross section normalisation depends on the value of $\phi'_P(0)$ while the $\eta$ cross section depends on $\phi_S(0)$, and in fact this somewhat compensates the suppression, due to the higher value of $\phi_s(0)$, as evaluated from the $\eta_c$ width~\cite{HarlandLang:2010ep}.

Finally we note, see Section~\ref{durt}, that the applicability of pQCD to the CEP process is justified by the effect Sudakov factor, which assumes $M_X \gg Q_\perp$, where $Q_\perp$ is the transverse loop momentum. However, in the case of $\chi_{c}$ production, where the `hard' scale $\mu \sim M_X$ is of order a few GeV, we may expect that a non--negligible part of the cross section comes from the IR unstable low $Q_\perp$ region, and this will be even more true in the case of higher spin $\chi_{c(1,2)}$ and odd--parity $\eta_c$ states, where the corresponding pQCD amplitudes vanish at $p_\perp=0$, so that the integrand (\ref{bt}) will have a lower saddle point. Thus there may be important `non--perturbative' corrections to such predictions. In previous work~\cite{Khoze:2004yb,HarlandLang:2009qe,HarlandLang:2010ep} a pragmatic approach was taken, in which the main analysis was based on the perturbative contribution only, assuming a smooth matching between the perturbative regime and the `soft' regime, and at the same time considering which features (for example, the relative contributions of the various $J^P$ states, distributions of final state particles, etc.) are likely to be shared by a possible non--perturbative contribution; however other approaches~\cite{Pasechnik:2007hm,Pasechnik:2009if} are certainly possible. On the other hand we note that for the $\chi_{b}$, $\eta_b$ states, for which $M_X$ is higher, this should be much less of an issue. Combined with the other sources of uncertainty, from the gluon PDF at low $x$ and $Q^2$ (recalling from (\ref{bt}) that the CEP cross section depends on the gluon PDF to the fourth power), the soft survival factors $S^2_{\rm eik}$ and $S^2_{\rm enh}$ and the contribution from higher--order corrections, we may conservatively expect the total uncertainty in the predictions for charmonium CEP cross sections to be as large as $\sim {}^{\times}_{\div} 5$. On the other hand, this uncertainty is significantly reduced when we consider the ratios of the $\chi_{c0}$ to $\chi_{c(1,2)}$ perturbative contributions, where in particular the PDF uncertainty and that due to the survival factors $S^2_{\rm enh}$, $S^2_{\rm eik}$, largely cancels out. The Regge--based phenomenological models of~\cite{Stein93,Peng95}, which can give some estimate of any `non--perturbative' contribution, also give similar results for these ratios.

\subsection{Numerical predictions}\label{numerchi}

In this section we show some presented numerical predictions for quarkonium CEP, originally given elsewhere\cite{HarlandLang:2009qe,HarlandLang:2010ep,HarlandLang:2010ys}. We note that these predictions were calculated using GRV94HO partons~\cite{Gluck94}, as these extend down to quite low $Q^2$, at a lower c.m.s. energy $\sqrt{s}=60$ GeV, with a simple Regge--like scaling assumed to extrapolate to higher energies, in order to try and bypass some of the large uncertainty in the PDFs at low $x$ and $Q^2$ discussed above. However, this choice is certainly not the only possibility, and indeed in more recent estimates for $\gamma\gamma$ and meson pair production~\cite{HarlandLang:2010ep,HarlandLang:2011qd}, for example, more up--to--date PDF sets at the relevant $\sqrt{s}$ values are taken, see Sections~\ref{cha:gam} and~\ref{cha:meson}.

Our calculation requires an explicit inclusion of the ${\bf p}_\perp$--dependent eikonal survival factor $S^2_{\rm eik}$, as discussed in Section~\ref{durt}, which will alter the predicted `bare' (i.e. unscreened) distributions. In this way, we can give predictions for the ${\bf p}_\perp$ distributions of the outgoing protons, which depend on soft survival effects as well as the quantum numbers of the produced state~\cite{Kaidalov03}. Such predictions will be particularly important when considering measurements with tagged protons.

\begin{figure}[t]
\begin{center}
\includegraphics[scale=0.5]{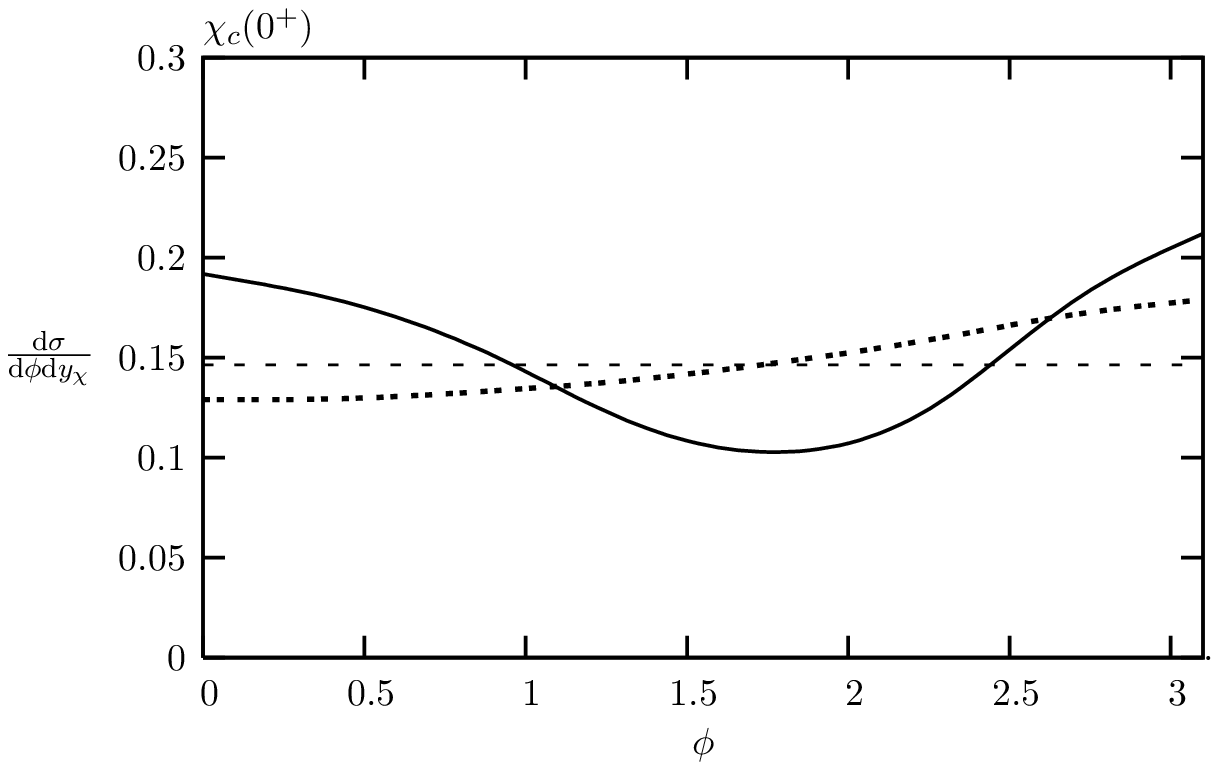}
\includegraphics[scale=0.5]{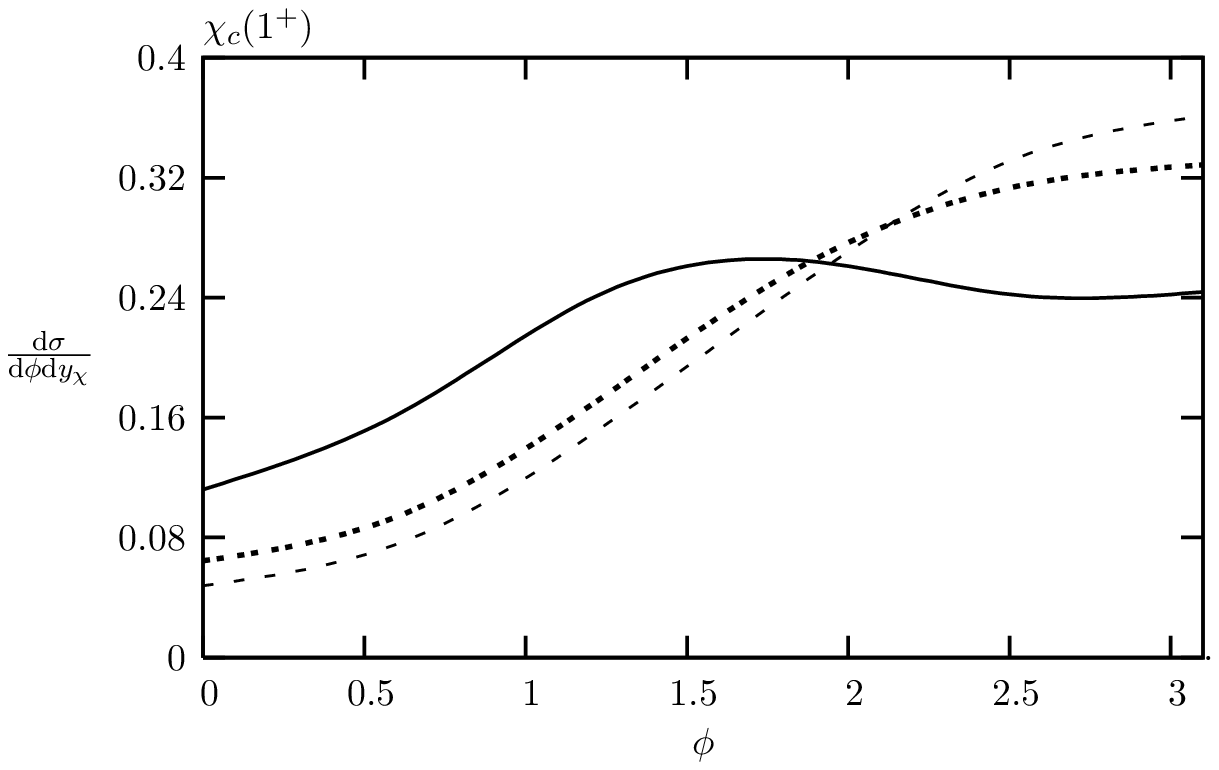}
\includegraphics[scale=0.5]{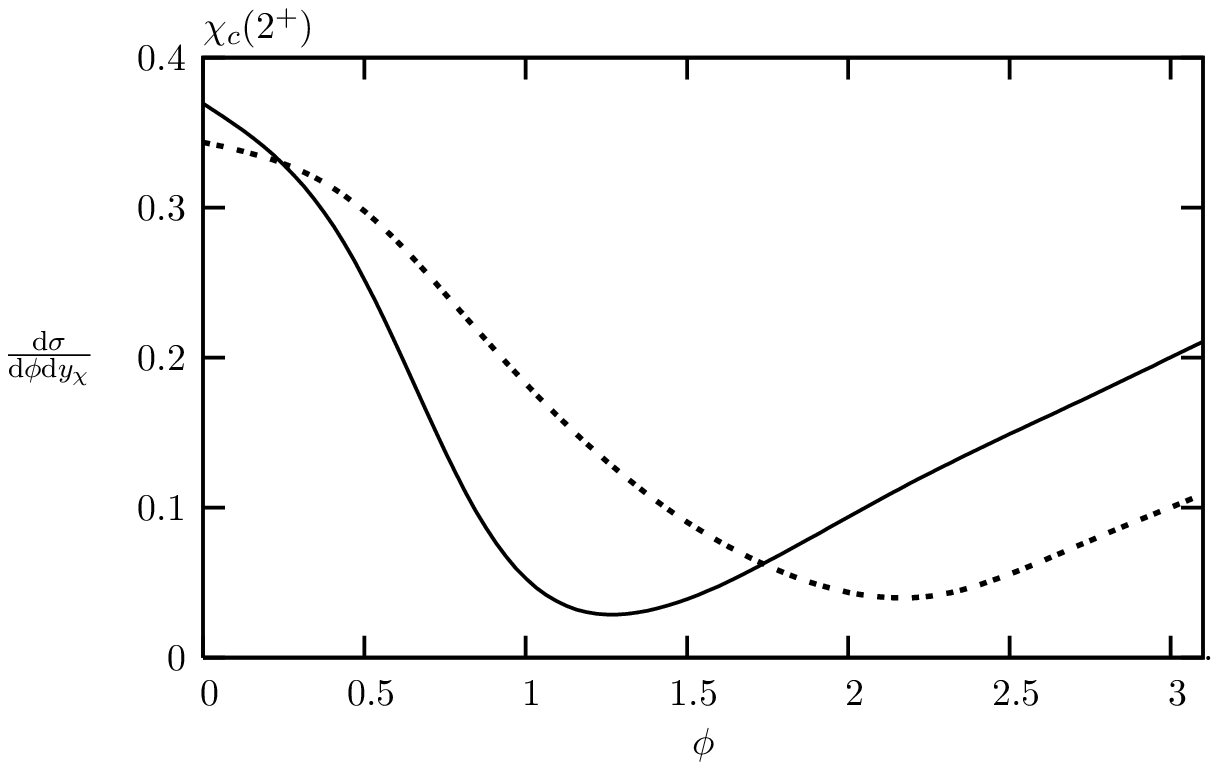}
\includegraphics[scale=0.5]{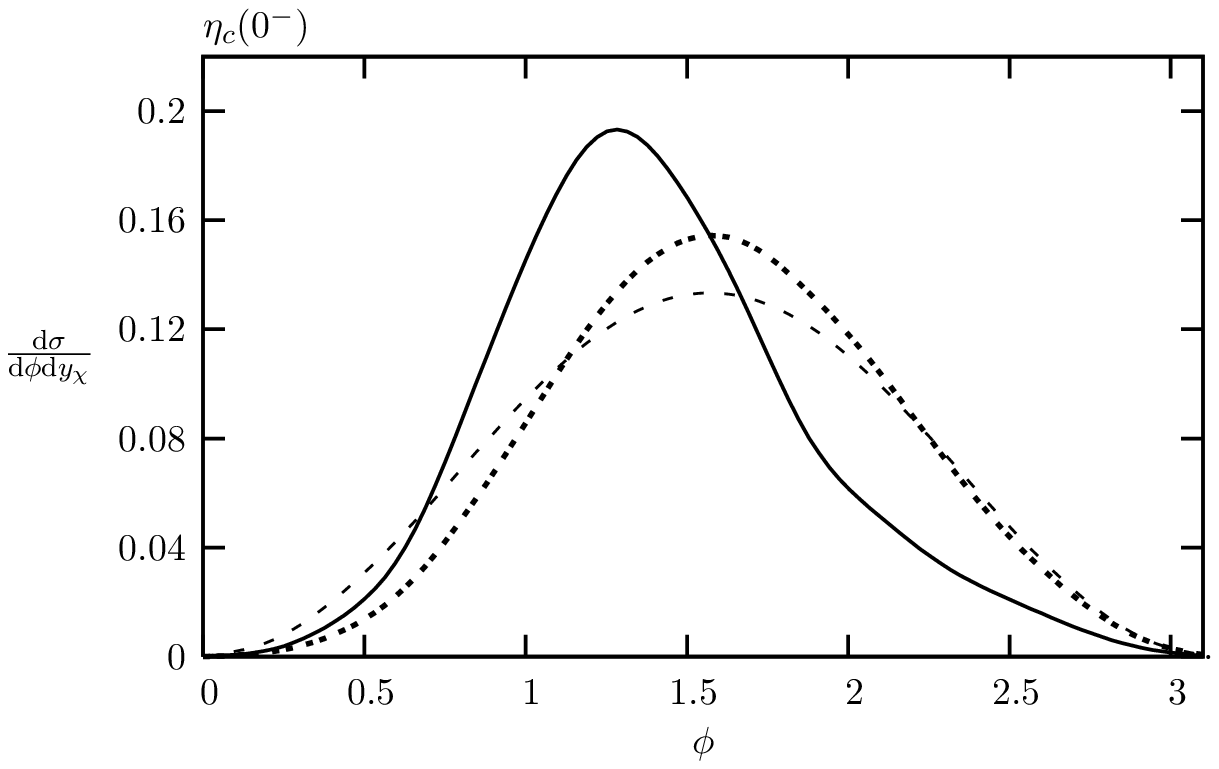}
\caption{Distribution (in arbitrary units) within the perturbative framework of the difference in azimuthal angle of the outgoing protons for the CEP of different $J^P$ $c\overline{c}$ states at $\sqrt{s}=14$ TeV and rapidity $y_X=0$. The solid (dotted) line shows the distribution including (excluding) the survival factor, calculated using the two--channel eikonal model~\cite{KMRsoft}, while the dashed line shows the distribution in the small $p_\perp$ limit, using the vertices of (\ref{R0p}--\ref{R0m}) and excluding the survival factor.}\label{surv1}
\end{center}
\end{figure}

In Fig.~\ref{surv1} we show the ${\rm d}\sigma/{\rm d}\phi$ distribution, where $\phi$ is the difference in azimuthal angle between the outgoing protons, for $\chi_{c(0,1,2)}$ and $\eta_c$ production at the LHC, while the following conclusions remains true for $\chi_b$/$\eta_b$ production and different c.m.s. energies, with the shape of the distributions only depending quite weakly on $\sqrt{s}$ and the central object mass $M_X$. The difference between the $J^P$ states, and the effect of including soft survival effects, is clear. We also see in Fig.~\ref{surv1} the effect of the $Q_\perp$ integral on the $\phi$ distributions. In the limit that ${\bf p}_\perp^2\ll {\bf Q}_\perp^2$, we expect~\cite{HarlandLang:2009qe,Kaidalov03} the $gg \to \chi/\eta$ vertices to have the form
\begin{align}\label{R0p}
|\mathcal{M}_{0^+}|^2 &\sim {\rm const.}\;,\\ \label{R1p} 
|\mathcal{M}_{1^+}|^2 &\sim ({\bf p}_{1_\perp}-{\bf p}_{2_\perp})^2\;,\\ \label{R0m}
|\mathcal{M}_{0^-}|^2 &\sim {\bf p}_{1_\perp}^2{\bf p}_{2_\perp}^2\sin^2{\phi}\;,
\end{align}
while there does not exist a simple closed form for the $\chi_2$. In the $\chi_0$ case we therefore expect a flat $\phi$ distribution as $p_\perp\to0$, but the inclusion of the $p_\perp$--dependent $gg \to \chi$ vertex factor and gluon propagators in (\ref{bt}) leads to corrections of the type $\sim{\bf p}_{1_\perp}\cdot{\bf p}_{2_\perp}/\langle Q_\perp^2\rangle$ which alter this. In the $\chi_{1}$ and $\eta$ cases, while the expected $\phi$ dependence is roughly the same as that predicted by (\ref{R1p}) and (\ref{R0m}), the $Q_\perp$ integral has again had some non--trivial effect on the original distributions. For $\chi_2$ production we can see that the $Q_\perp$ loop integral has induced a strong $\phi$ dependence, which cannot be predicted from general considerations, and is therefore specific to the pQCD--based model of CEP.

\begin{table}[h]
\begin{center}
\tbl{$p_\perp$--averaged survival factor $\langle S^2_{\rm eik}\rangle$ for $\chi_c$ and $\eta_c$ production at the Tevatron and LHC and $\chi_c$ production at RHIC ($\sqrt{s}=500$ GeV).}{
\begin{tabular}{|l|c|c|c|c|}
\hline
&$\chi_{c0}$&$\chi_{c1}$&$\chi_{c2}$&$\eta_c$ \\
\hline
Tevatron &0.058&0.15&0.11&0.18 \\
LHC (7~TeV) &0.037&0.11&0.084&0.13 \\
LHC (14~TeV) &0.029&0.091&0.072&0.10 \\
\cline{5-5}
RHIC &0.092&0.23&0.15& \multicolumn{1}{c}{}\\
\cline{1-4}
\end{tabular}\label{surv4}}
\end{center}
\end{table}

In Table~\ref{surv4} we show the $p_\perp$--averaged survival factors, calculated using (\ref{ampnew}), for $\chi_c$ and $\eta_c$ production at RHIC, Tevatron and LHC energies; the results for bottomonium production are approximately the same. We can see that the suppression factors for the higher spin and odd parity states are higher than for the scalar $\chi_{c0}$: due to the spin/parity structure of the vertices $gg \to \chi_{1,2},\eta$, the corresponding amplitudes vanish as $b_t\to 0$, in impact parameter space. These processes are therefore more `peripheral', occurring with a larger average impact parameter, where the probability of additional soft interactions is smaller, and therefore the survival factor $\langle S^2_{\rm eik}\rangle$ is larger. This is a clear example of the fact that the survival factors are in general process dependent and must therefore be considered carefully for each channel.

\begin{table}[ht]
\begin{center}
\tbl{Differential cross section (in nb) at rapidity $y_\chi=0$ for $\chi_{cJ}$ CEP via the $\chi_{cJ} \to J/\psi\gamma$ decay chain, summed over the $J=0,1,2$ contributions, at RHIC, Tevatron and LHC energies.}
{
\begin{tabular}{|l|c|c|c|c|}
\hline
$\sqrt{s}$ (TeV)&0.5&1.96&7&14\\
\hline
$\frac{{\rm d}\sigma}{{\rm d}y_{\chi_c}}(pp\to pp(J/\psi\,+\,\gamma))$&0.57&0.73&0.89&1.0\\
\hline
$\frac{{\rm d}\sigma(1^+)}{{\rm d}\sigma(0^+)}$&0.59&0.61&0.69&0.71\\
\hline
$\frac{{\rm d}\sigma(2^+)}{{\rm d}\sigma(0^+)}$&0.21&0.22&0.23&0.23\\
\hline
\end{tabular}\label{chires1}}
\end{center}
\end{table}
\begin{table}[ht]
\begin{center}
\tbl{Differential cross section (in nb) at rapidity $y_{\chi}=0$ for central exclusive $\chi_{(b,c)0}$ production at RHIC, Tevatron and LHC energies.}
{
\begin{tabular}{|l|c|c|c|c|}
\hline
$\sqrt{s}$ (TeV)&0.5&1.96&7&14\\
\hline
$\frac{{\rm d}\sigma}{{\rm d}y_\chi}(\chi_{c0})$&27&35&42&45\\
\hline
$\frac{{\rm d}\sigma}{{\rm d}y_\chi}(\chi_{b0})$&0.022&0.029&0.036&0.038\\
\hline
\end{tabular}\label{chires2}}
\end{center}
\end{table}

We next consider the predictions for quarkonium CEP cross sections at different collider energies.  In Table~\ref{chires1} we show the differential cross section for the central exclusive $\chi_c \to pp(J/\psi\gamma)$ process at RHIC, Tevatron and LHC energies. We can see that, as discussed above, a significant fraction of the $\chi_c$ events are predicted to correspond to the higher spin $\chi_{c(1,2)}$ states. In Table~\ref{chires2}  we show predictions for the $\chi_{c0}$ (and $\chi_{b0}$) CEP cross sections, which for example would be relevant for the observation of $\chi_c$ CEP via two--body decay channels (e.g. $\chi_c \to \pi\pi,K\overline{K}$). In Table~\ref{chires3} we show predictions for the differential cross section for the central exclusive $\chi_b \to \Upsilon\gamma$ process at Tevatron and LHC energies. While the cross section is smaller, $\chi_b$ CEP remains a potential observable at the LHC. We can see, as discussed above, that $\chi_{b1}$ states are expected to give a negligible contribution to the overall rate, while the relative $\chi_{b2}/\chi_{b0}$ contribution is reduced in comparison to the $\chi_c$ case. This suppression is largely due to the slightly higher $\chi_{b0}$ branching ratio ${\rm Br}(\chi_{b0}\to\Upsilon\gamma)\approx 1.76 \%$, when compared to the $\chi_{c0}\to J/\psi \gamma$ branching. Thus, we can predict with some certainty that the $\chi_{b0}$ contribution to any future observed $\chi_b$ events (via the $\chi_b \to \Upsilon\gamma$ decay chain) should be strongly dominant. We note that all of these results are in rough agreement with the expectations of (\ref{roughc}) and (\ref{roughb}).

\begin{table}[h]
\begin{center}
\tbl{Differential cross section (in pb) at rapidity $y_\chi=0$ for central exclusive $\chi_{bJ}$ production via the $\chi_{bJ} \to \Upsilon\gamma$ decay chain, summed over the $J=0,1,2$ contributions, at Tevatron and LHC energies.}
{
\begin{tabular}{|l|c|c|c|c|}
\hline
$\sqrt{s}$ (TeV)&1.96&7&14\\
\hline
$\frac{{\rm d}\sigma}{{\rm d}y_{\chi_b}}(pp\to pp(\Upsilon\,+\,\gamma))$&0.60&0.75&0.79\\
\hline
$\frac{{\rm d}\sigma(1^+)}{{\rm d}\sigma(0^+)}$&0.050&0.055&0.059\\
\hline
$\frac{{\rm d}\sigma(2^+)}{{\rm d}\sigma(0^+)}$&0.13&0.14&0.14\\
\hline
\end{tabular}\label{chires3}}
\end{center}
\end{table}

\begin{table}[h]
\begin{center}
\tbl{Differential cross section (in pb) at rapidity $y_{\eta}=0$ for central exclusive $\eta_{b,c}$ production at Tevatron and LHC energies.}
{
\begin{tabular}{|l|c|c|c|c|}
\hline
$\sqrt{s}$ (TeV)&1.96&7&14\\
\hline
$\frac{{\rm d}\sigma}{{\rm d}y_\eta}(\eta_c)$&200&200&190\\
\hline
$\frac{{\rm d}\sigma}{{\rm d}y_\eta}(\eta_b)$&0.15&0.14&0.12\\
\hline
\end{tabular}\label{chires4}}
\end{center}
\end{table}

In both cases, we can see that the predicted $\chi_{(c,b)}$ CEP cross sections depend only weakly on the c.m.s. energy, an effect which is reasonably generic to CEP processes. While the higher gluon density at lower $x$ will lead to an increase in the cross section, this growth is tamed by the eikonal and enhanced soft survival factors, which decrease with $\sqrt{s}$ due to the increase in proton opacity $\Omega(s,b)$ and increase in the size of the rapidity gaps $\sim \ln(s/M_X^2)$ available for enhanced absorption, respectively. Clearly the exact energy dependence due to both the gluon parton density and the soft survival factors carry their own uncertainties and as a consequence the predicted energy dependence of the CEP rates should be considered as an estimate only: in particular, if we do not assume the Regge scaling in the cross section that was taken in previous studies~\cite{HarlandLang:2010ep}, the cross section predictions will increase somewhat more with energy. However this reasonably weak dependence still represents a qualitative prediction, the validity of which could be probed by observations of these processes at RHIC, Tevatron and/or different LHC running energies.

Finally, we show in Table~\ref{chires4} predictions for the differential cross section for central exclusive $\eta_c$ and $\eta_b$ production at Tevatron and LHC energies. In both cases, the expected rates are roughly two orders of magnitude smaller than the associated $\chi_{(c,b)0}$ cross sections, consistent with the expected $\sim \langle \mathbf{p}_{\perp}^2\rangle^2/(\langle \mathbf{Q}_{\perp}^2\rangle^2)$ suppression described above. We can also see a similar, even decreasing, trend with energy.

\subsection{Comparison with data}\label{chiccomp}
  
As discussed in Section~\ref{chicintro}, in 2009 CDF reported~\cite{Aaltonen:2009kg} the observation of $65\pm 10$ candidate exclusive $\chi_c$ events, produced via the radiative $\chi_c \to J/\psi \gamma$ decay chain. Under the assumption that these correspond to purely $\chi_{c0}$ events, this corresponds to a cross section of
\begin{equation}\label{chi0cs}
\frac{{\rm d}\sigma_{\chi_{c0}}^{\rm exp}}{{\rm d}y_{\chi}}\bigg|_{y_\chi=0} = (76 \pm 14)\,{\rm nb}\;.
\end{equation}
If on the other hand we consider the possibility that the $\chi_{c(1,2)}$ states contribute to the CDF data, then we can translate (\ref{chi0cs}) into a total $\chi_{cJ} \to J/\psi\gamma$ cross section
\begin{equation}\label{CDFcj}
\frac{{\rm d}\sigma_{\chi_{cJ}\to J/\psi\gamma}^{\rm exp}}{{\rm d}y_{\chi}}\bigg|_{y_\chi=0} = (0.97 \pm 0.18)\,{\rm nb}\;,
\end{equation}
using the value for the $\chi_{c0}$ branching ratio taken in the CDF analysis. This is in excellent agreement with the theoretical expectation of $0.73$ nb given in Table~\ref{chires1}, although we recall from the discussion above that such a prediction caries very large theoretical uncertainties. On the other hand, these uncertainties are reduced when we consider the ratios of the $\chi_{c0}$ to $\chi_{c(1,2)}$ cross sections (we would estimate this to be of order $\sim {}^{\times}_{\div} 2$), where in particular the PDF uncertainty and that due to the survival factors $S^2_{\rm enh}$, $S^2_{\rm eik}$, largely cancels out: the results in Section~\ref{numerchi} therefore suggest that a non--negligible fraction of the observed $\chi_c$ events at the Tevatron are in fact $\chi_{c1}$ and $\chi_{c2}$ events. 

\begin{table}[h]
\begin{center}
\tbl{Comparison of preliminary candidate exclusive LHCb $\chi_{cJ}$ data~\cite{LHCb} with theory predictions, made using the \texttt{SuperCHIC} MC~\cite{SuperCHIC} .}
{
\begin{tabular}{|l | c| c|}
\hline
& $\sigma(pp \to pp(J/\psi+\gamma))$ LHCb (pb) & SuperCHIC prediction (pb)\\
\hline
$\chi_{c0}$ & $9.3\pm4.5$ & 14 \\
$\chi_{c1}$ & $16.4 \pm 7.1$ & 10 \\
$\chi_{c2}$ & $28\pm 12.3$ & 3 \\
\hline
\end{tabular}\label{lhcbcs}}
\end{center}
\end{table}

More recently, based on an analysis of 2010 data, LHCb have reported preliminary results on potentially exclusive $\chi_{c}$ meson production in the $\chi_c\to J/\psi\,+\,\gamma$ channel~\cite{LHCb}, where vetoing was imposed on additional activity in the rapidity region $1.9<\eta<4.9$, and charged particles in the backwards region $-4<\eta<-1.5$. Crucially, in this case LHCb were able to approximately distinguish the three $\chi_c$ spin states. The comparison of this data with our predictions, made using the \texttt{SuperCHIC} MC~\cite{SuperCHIC}, are shown in Table~\ref{lhcbcs}. We can see most significantly that all three $\chi_{cJ}$ states are observed to give non--negligible contributions, qualitatively supporting the results of the previous sections. Moreover, in the case of $\chi_{c0}$ and $\chi_{c1}$ production, there is good agreement between the data and theory, within the (fairly large) theoretical and experimental  uncertainties. 

On the other hand, we can see from Table~\ref{lhcbcs} in the case of the $\chi_{c2}$ that there is a clear excess of data events, even within the theoretical uncertainties. From the theoretical side, this may be due to possible non--perturbative contributions discussed above, which could enhance the $\chi_{c2}$ rate above the (strongly suppressed) amount predicted by the Durham model~\cite{HarlandLang:2009qe}. Such effects may be particularly important for the $\chi_c$ case, where we do not necessarily have $M_\chi \gg Q_\perp$, as discussed in Section~\ref{chitheory}. These dynamically suppressed cross sections may also be particularly sensitive to higher--order QCD corrections, which could also enhance the $\chi_{c(1,2)}$ rates somewhat. Alternatively, we recall that the $gg \to \chi_{c2}$ coupling only vanishes for $J_z=0$ gluons in the non--relativistic limit, that is ignoring higher--order corrections in the relative quark velocity $v$. Any, even small, correction to this limit which allows a $J_z=0$ contribution, could enhance the cross section. However an analysis of the $\chi_{c2} \to \gamma\gamma$ decay distribution~\cite{Ablikim:2012xi} has shown that such a contribution is very small and even consistent with zero, giving
\begin{equation}\label{f02}
f_{0/2}\equiv \frac{\Gamma_{\gamma\gamma}^{\lambda=0}(\chi_{c2})}{\Gamma_{\gamma\gamma}^{\lambda=2}(\chi_{c2})}=0.00\pm0.02\pm0.02\;,
\end{equation}
where $\lambda$ is the $J_z$ projection of the $\gamma\gamma$ state, while the first error is statistical and the second systematic. However, recalling the $\sim 2$ orders of magnitude suppression of the ($|J_z|=2$) non--relativistic $\chi_{c2}$ CEP cross section (see Table~\ref{chires1} and (\ref{roughc})), a non--zero value of $f_{0/2}\lesssim 5\%$, which is still consistent with (\ref{f02}) could enhance the $\chi_{c2}$ cross section by a factor of a `few'.

Experimentally, at the LHC, without forward proton detectors exclusivity must instead be selected by vetoing on additional hadronic activity in a large enough rapidity region, but in this case there will also be some contribution from non--exclusive events, in particular where one or both of the protons dissociates\cite{HarlandLang:2012qz},
\begin{equation}\label{dd}
pp(\overline{p}) \to Y+X+Z\;,
\end{equation}
where $X=\chi_c$. In the present case, the crucial point is that the veto region at LHCb is fairly limited and so such a contribution may be quite large. In particular, as well as low mass dissociation ($M_{Y,Z}\lesssim\, 2\, {\rm GeV}$), which we expect to enhance the observed cross section by a factor of $\lesssim 30\%$ but not to alter the particle distributions or, importantly, relative $\chi_{cJ}$ fractions significantly, we must also consider the contamination from higher mass dissociation where $M_{Y,Z}\gtrsim\, 2\, {\rm GeV}$. Here, while the theoretical uncertainties are quite large, we would expect a similar level of contribution to the case of low mass dissociation at LHCb, but due to the higher mass of the proton dissociative system, the $p_\perp$ of the outgoing proton systems (and hence of the central system, $X$) can be much larger than in the exclusive case. Recalling (\ref{roughc}) that the $\chi_{c2}$ cross section is suppressed by the ratio $\langle p_\perp^2\rangle^2/\langle Q_\perp^2\rangle^2$, we can see that the $\chi_{c2}$ cross section (and to a lesser extent the $\chi_{c1}$ cross section) will be particularly sensitive to such a contribution, with larger values of $p_\perp$ allowing an increasing violation of the $J_z=0$ selection rule which operates for pure CEP. This may therefore also explain, at least in part, why the observed LHCb $\chi_{c2}$ cross section is higher than the CEP prediction, although event selection techniques, including cuts on the central $\mu^+\mu^-$ system $p_\perp <0.9$ GeV are imposed to suppress the dissociative contribution. 

Finally, we note in passing that recently~\cite{Albrow:2013mva,Mikeeds} the CDF collaboration have reported a new preliminary limit on the $\chi_{c0}$ CEP cross section at $\sqrt{s}=0.9$ and 1.96 TeV, via the $\chi_{c0} \to \pi^+\pi^-$, $K^+K^-$ channels, of ${\rm d}\sigma/{\rm d}y|_{y=0}(\chi_{c0}) \lesssim 20 $ nb at 90\% confidence. This measurement, which should only contain a small contribution from proton dissociation (in particular in the $\sqrt{s}=0.9$ TeV case), appears to suggest a somewhat larger contribution from the higher spin $\chi_{c(1,2)}$ states to the $\chi_{cJ} \to J/\psi \gamma$ combined CDF cross section measurement~\cite{Aaltonen:2009kg} than that predicted in Section~\ref{numerchi}, possibly supporting the LHCb result. With the application of additional cuts on the pions\cite{Harland-Lang:2013dia} this limit should be further reduced and may even be translated into an observation. This may suggest, along with the LHCb data, that some further theoretical work may be needed to account for these measurements, and that perhaps the limit of the perturbative approach is being reached at these masses; as discussed above, it may be that some correction from the `non--perturbative' (i.e. low gluon $Q_\perp$) region could increase the $\chi_{c(1,2)}$ rates beyond the purely perturbative expectations. These effects should on the other hand be much less pronounced in the case of $\chi_b$ production, which would for this reason represent a particularly interesting measurement. Hopefully then the arrival of further quarkonium data may shed further light on this interesting issue. In particular, in the 2011-2012 data set LHCb have collected a much higher event rate, by a factor of about 80, which would allow a much more accurate study of  $\chi_c$ CEP, including via two--body decay channels (e.g. $\pi^+\pi^-$, $K^+K^-$, $p\overline{p}$, $\Lambda\overline{\Lambda}$), and help clarify the question of the dissociation background. Moreover, it is expected that data collected during the LHC Run 2 would nearly double usable LHCb luminosity.

\subsection{Quarkonium CEP with tagged protons}\label{rhic}

As discussed in the Introduction, previous work on quarkonium CEP has included a study\cite{HarlandLang:2010ys} of the possibility of measuring charmonium CEP with tagged protons at RHIC. This may also be relevant in the future at the LHC, where CEP measurements in the the quarkonium mass region with tagged protons is a realistic possibility~\cite{Oljemark:2013wsa,Staszewski:2011bg}. We therefore present a selection of these previous theoretical results\cite{HarlandLang:2010ys} here.

\begin{figure}
\begin{center}
\includegraphics[trim=30 0 0 0,scale=0.5]{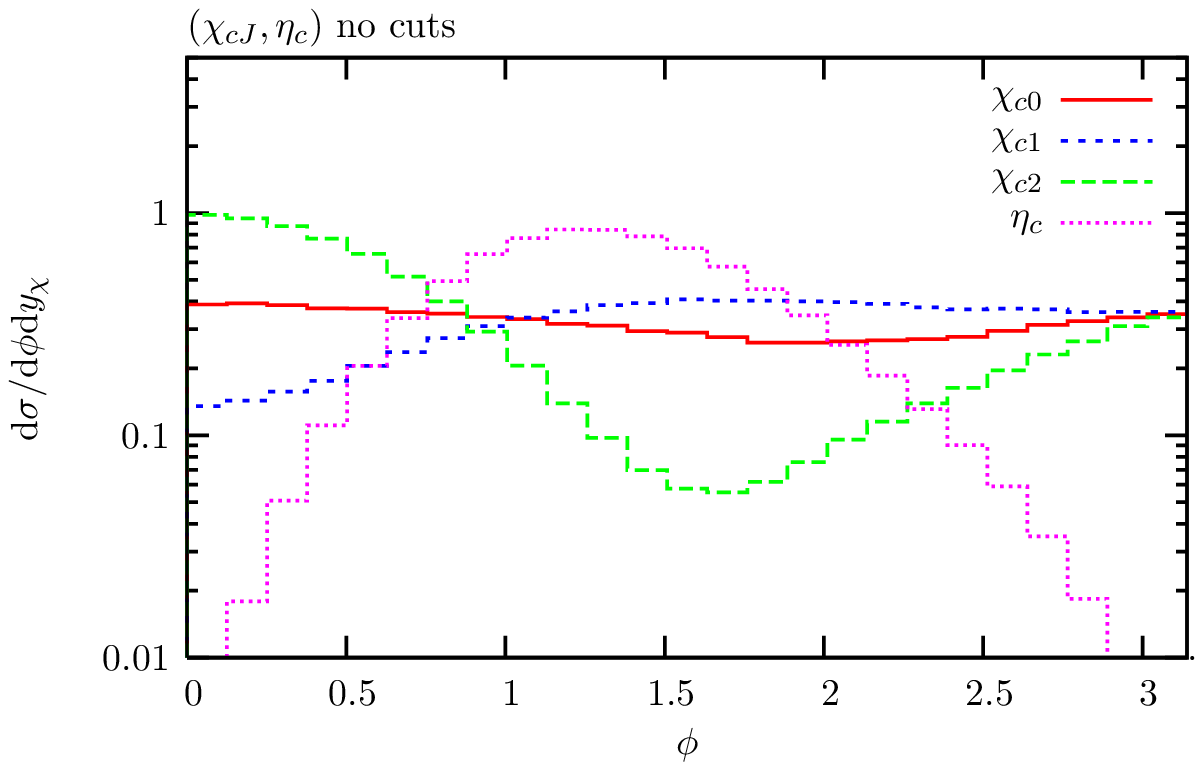}
\includegraphics[scale=0.5]{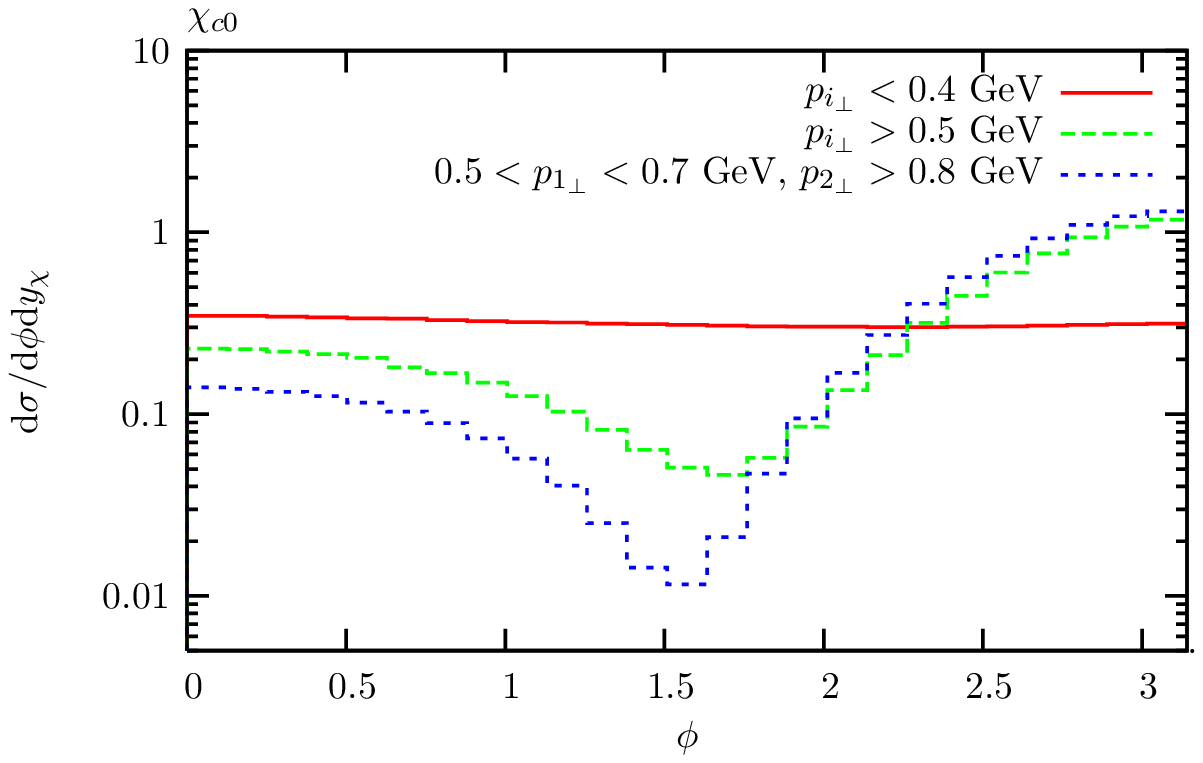}
\includegraphics[trim=30 0 0 0,scale=0.5]{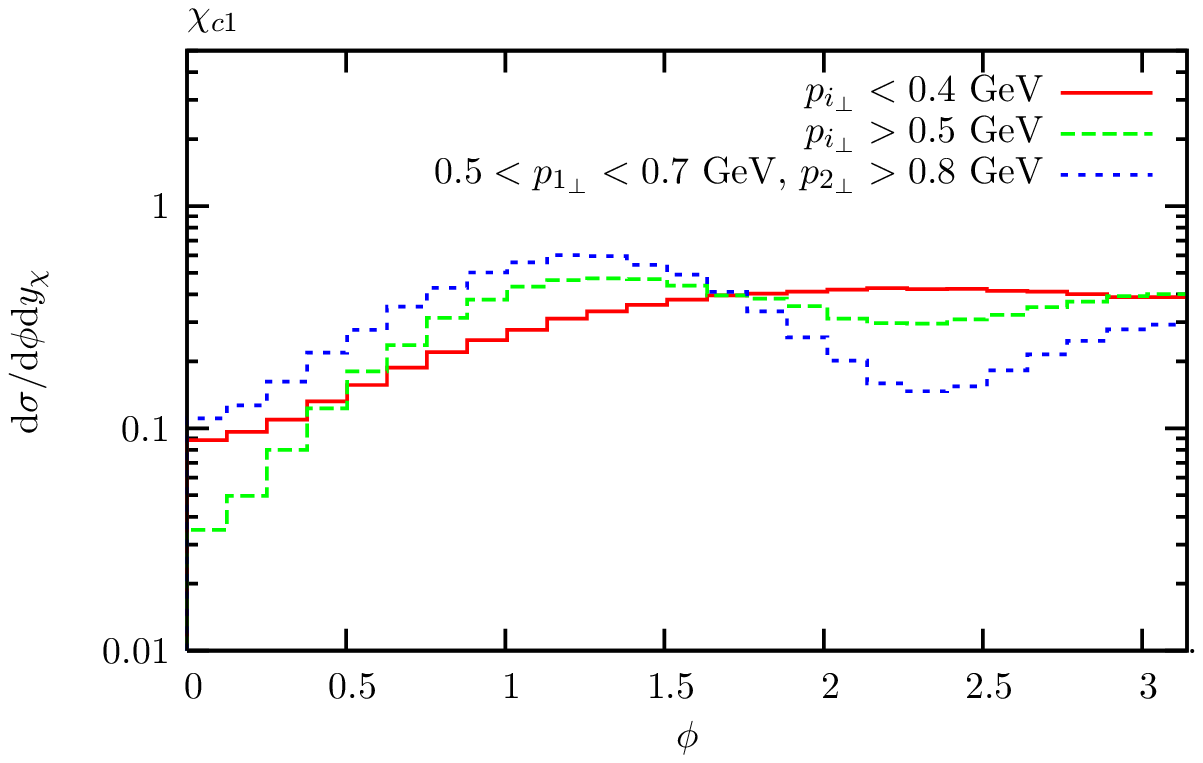}
\includegraphics[scale=0.5]{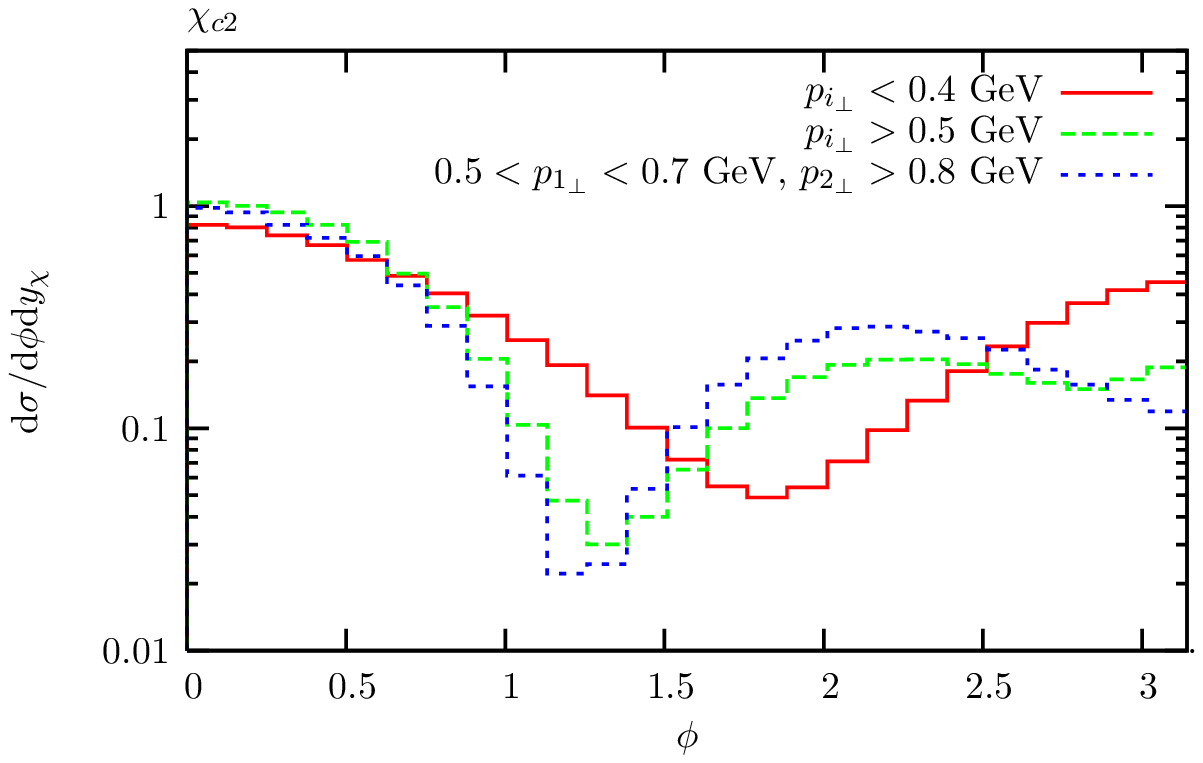}
\includegraphics[clip=true,trim=0 0 0 20,scale=0.3]{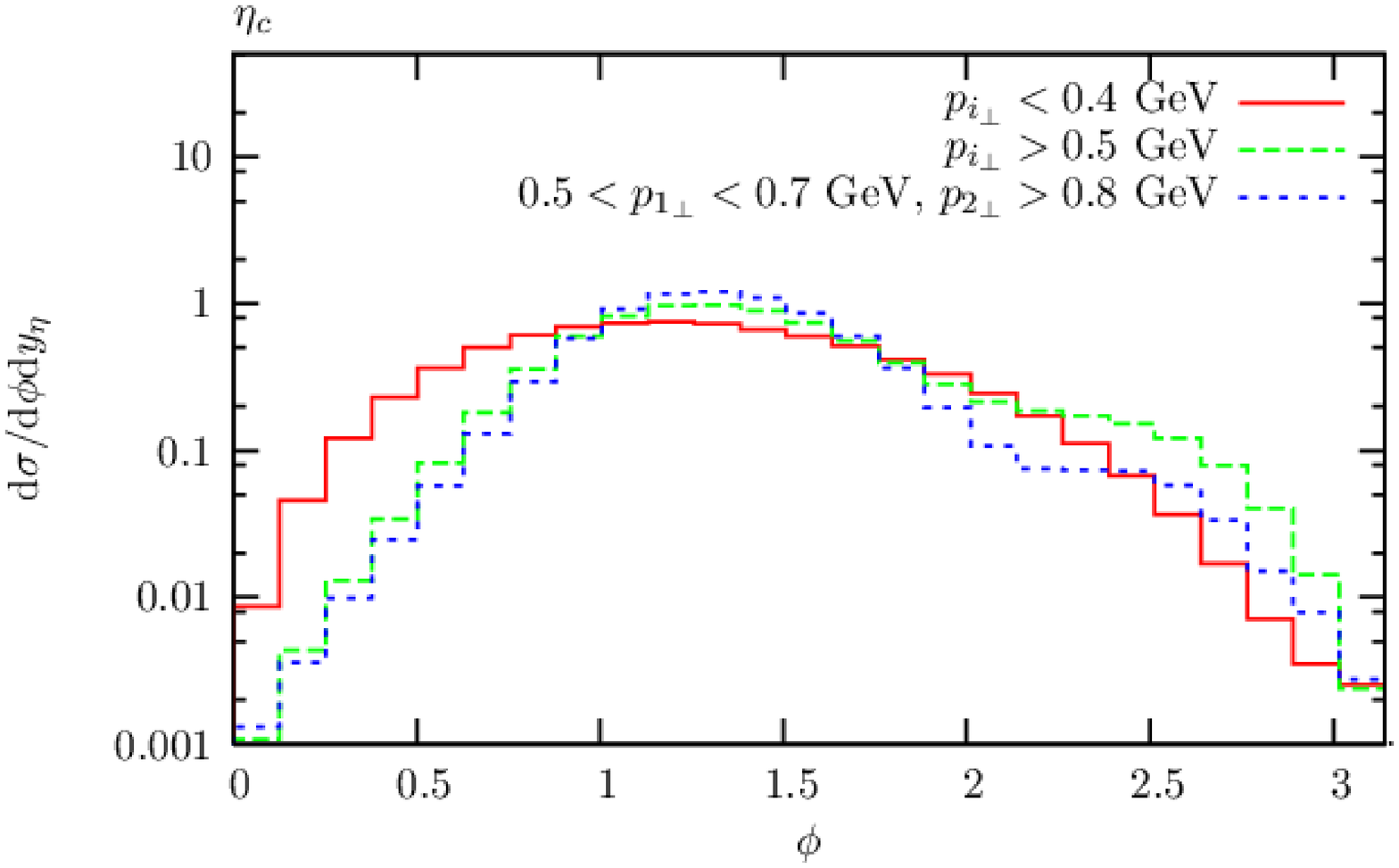}
\caption{Normalised distributions (in arbitrary units) of the difference in azimuthal angle between the outgoing protons for the CEP of $\chi_{c(0,1,2)}$ and $\eta_c$ states at $y=0$ and $\sqrt{s}=500$ GeV, and for a range of cuts on the proton $p_\perp$.}\label{dist1}
\end{center}
\end{figure}

\begin{figure}
\begin{center}
\includegraphics[trim=30 0 0 0,scale=0.5]{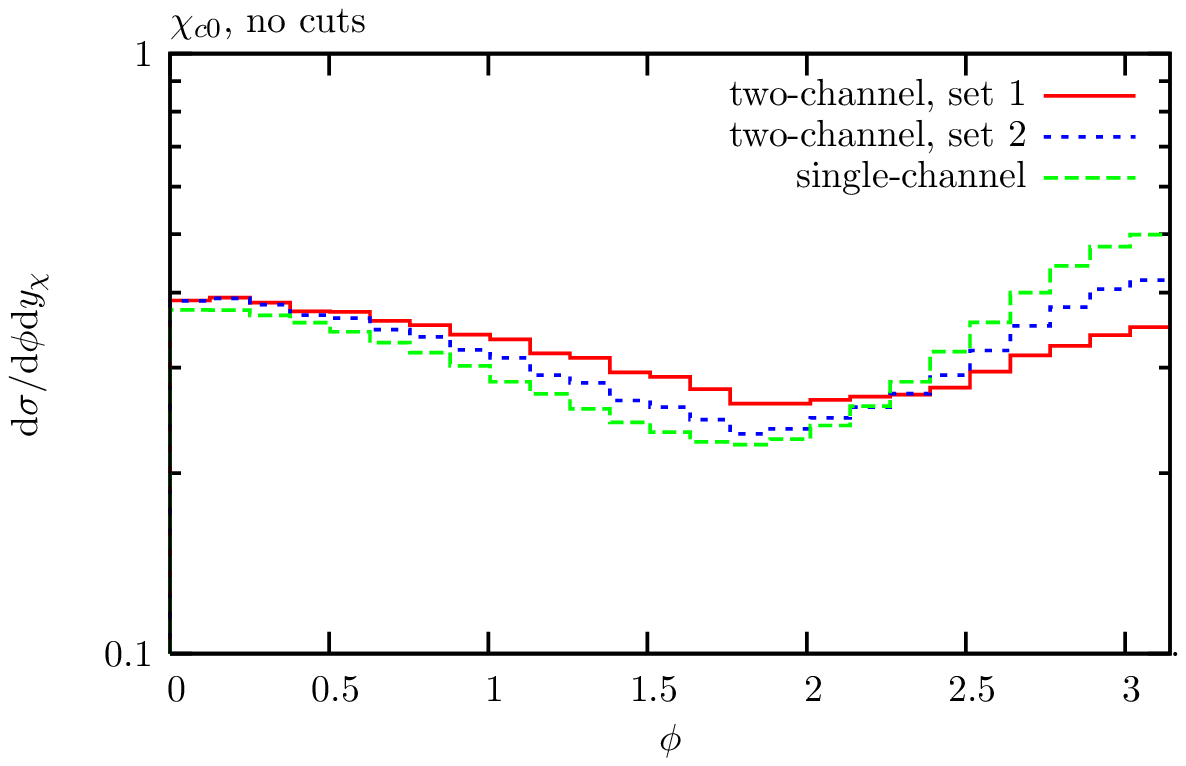}
\includegraphics[scale=0.5]{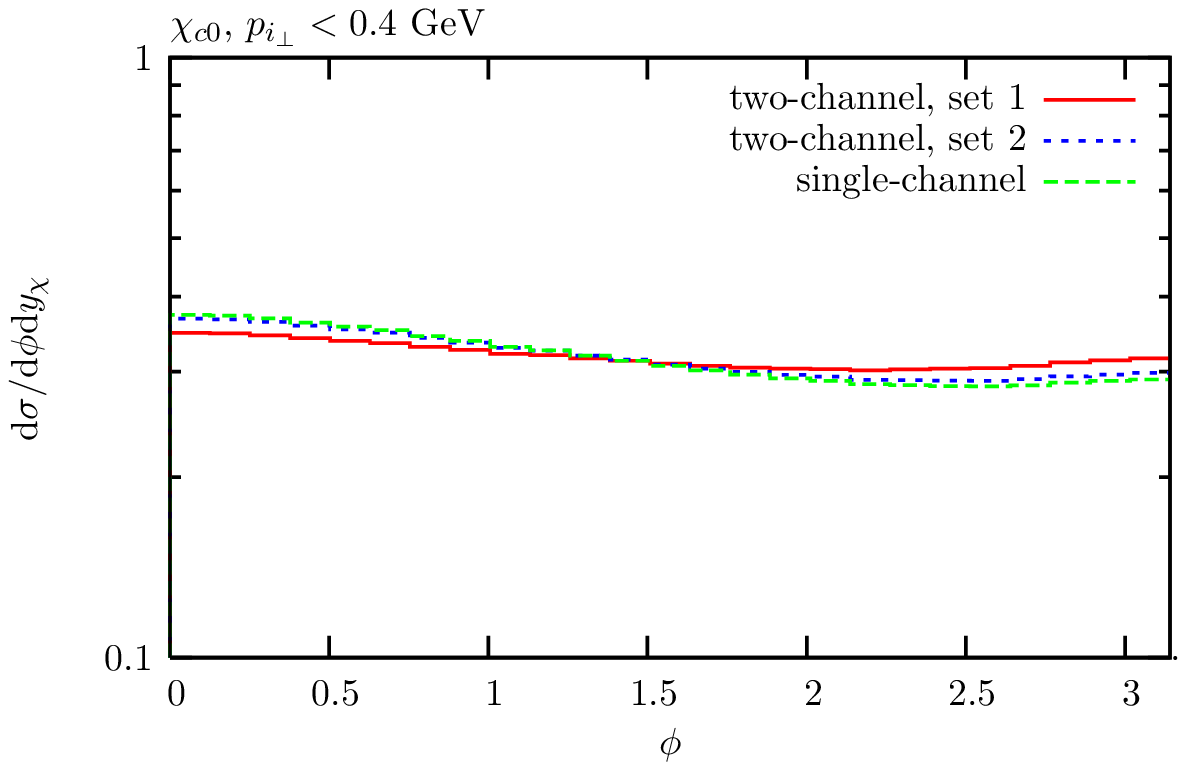}
\includegraphics[scale=0.5]{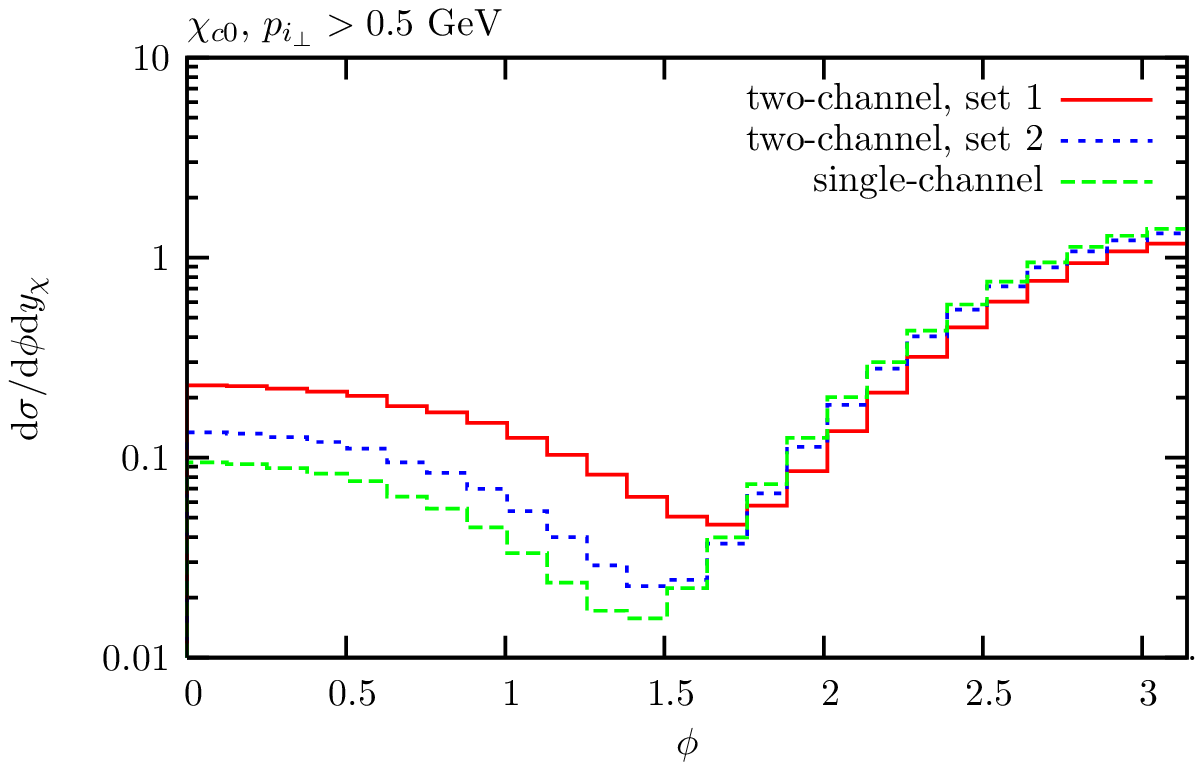}
\caption{Normalised distributions (in arbitrary units) of the difference in azimuthal angle between the outgoing protons for $\chi_{c0}$ CEP at $y=0$, with the survival factor $S^2_{\rm eik}$ calculated using the two--channel eikonal model~\cite{KMRsoft}, with two different choices of model parameters.  Also shown is the result of using the simplified single--channel eikonal approach~\cite{Khoze:2002nf}. Note that the `two--channel, set 1' $\chi_{c0}$ distributions are the same as those 
plotted in Fig.~\ref{dist1}.} \label{dist2}
\end{center}
\end{figure}

The predicted cross section at $\sqrt{s}=500$ GeV were presented in Section~\ref{numerchi}, and are not repeated here: we recall that they are not expected to be greatly different from the LHC case. In Fig.~\ref{dist1} we plot the expected ${\rm d}\sigma/{\rm d}\phi$ distributions with respect to the relative azimuthal angle $\phi$ between the outgoing protons, for $\chi_{cJ}$ and $\eta_c$ production, calculated using the SuperCHIC MC~\cite{SuperCHIC}. We can see, as described above (see in particular (\ref{R0p}--\ref{R0m})), that these are highly sensitive to the spin and parity of the produced object. By applying different cuts to the outgoing proton $p_\perp$, we can also in principle probe the underlying theory in a more detailed way, and to illustrate this we plot the $\phi$ distributions for three different sets of $p_\perp$ cuts: $p_{(1,2)_\perp}<0.4$ GeV, $p_{(1,2)_\perp}>0.6$ GeV and $0.5<p_{1_\perp}<0.7$ GeV, $p_{2_\perp}>0.8$ GeV. 
For low $p_\perp$ the screening corrections do not change this `bare' behaviour too much, however in the case of a relatively large $p_\perp$ (the green and blue lines in Fig.~\ref{dist1}) the role of absorptive effects becomes quite visible: starting from $\phi=0$ the absorptive correction increases with $\phi$ producing a dip in the region of $\phi\sim \pi/2$ for the cases of the $\chi_{c0}$ and $\chi_{c2}$ and about $\phi\sim 2.3$ for the $\chi_{c1}$. We note that these characteristic `diffractive dip' structures have the same physical origin as the proton azimuthal distribution patterns seen previously in the literature~\cite{Khoze:2002nf}. For the $\eta_c$ these effects are less significant, although some non--negligible dip structure around $\phi=2-2.5$ can be seen.

As well as depending on the spin--parity of the centrally produced particle, the proton distributions will also be affected non--trivially by soft--survival effects~\cite{Khoze:2002nf}, and on the particular model of soft diffraction that is used to calculate these. We show in Fig.~\ref{dist2} the predicted $\phi$ distribution for $\chi_{c0}$ CEP using the `two--channel' eikonal formalism~\cite{KMRsoft} for a range of cuts on the proton $p_\perp$. To give some indication of any model variation, we show the distributions using two different choices of model parameters as well as using a simplified single--channel model~\cite{Khoze:2002nf}. While the differences seen between the models in Fig.~\ref{dist2} are fairly small, they are not necessarily negligible, and it would certainly be of great interest to observe this dip structure, which is completely driven by soft survival effects, experimentally. We note that while these soft physics models have now been updated and adjusted\cite{Khoze:2013dha,Khoze:2014aca}, to account for recent theoretical developments as well as the TOTEM Run I data, and so are not to be taken completely literally, similar effects are also seen using these latest models~\cite{Harland-Lang:2013dia}, and so these distributions can still serve as an reliable indication of the effects described above.

\subsection{Exotic charmonium--like states: the $X(3872)$}

The CEP mechanism could also provide a complementary way to shed light on the nature of the large number of `exotic' XYZ charmonium--like states which have been discovered over the past 10 years~\cite{Brambilla:2010cs,Braaten:2013oba,Chiochia:2014qva}. In some cases the $J^{PC}$ quantum numbers of these states has not been determined experimentally, and often a range of interpretations are available: a $D^0\overline{D}^{*0}$ molecule, tetraquarks, $c\overline{c}g$ hybrids, the conventional $c\overline{c}$ charmonium assignment, and more generally a mixture of these different possibilities. Considering the CEP of such objects, then the effect of the $J_z^{P}=0^{+}$ selection rule which~\cite{Khoze:2001xm,Khoze:2000jm} strongly suppresses the CEP of non--$J_z^{P}=0^{+}$ states, as well as a measurement of the distribution  in the azimuthal angle $\phi$ between the transverse momenta of the outgoing protons (as in e.g.~\cite{Kaidalov03,HarlandLang:2010ep}), may help to fix the quantum numbers of the centrally produced system. Moreover, since the original $c\overline{c}$ pair is in the exclusive case necessarily produced at rather short distances, the CEP process can probe the wavefunction of the corresponding charmonium at the origin.

One interesting case is the  CEP of the $Y(3940)$, in particular via the $J/\psi \omega$ channel~\cite{Uehara:2009tx,Abe:2004zs,Aubert:2007vj}, which could help to resolve current uncertainties~\cite{Albuquerque:2013owa,Sreethawong:2013qua} in the interpretation of this state. Another particularly interesting example is the well--known $X(3872)$: this was the first such `exotic' state to be discovered (by BELLE in 2003~\cite{Choi:2003ue}), with a concrete interpretation  for it still remaining elusive. It has become even more topical with the establishment of its quantum numbers to be $J^{PC}=1^{++}$ by LHCb~\cite{Aaij:2013zoa}, an assignment which leaves both the more exotic and the conventional $\chi_{c1}(2{}^3 P_1)$ interpretations in principle available, as well a combination of, for example, the core $c\overline{c}$ and molecular $D$ meson states~\cite{Voloshin:2007dx}.

The $X(3872)$ has been seen in prompt inclusive production at both the Tevatron and LHC, and this raises the interesting possibility of observing its production in the exclusive channel. Such an observation would first of all probe the direct (i.e. not due to feed--down from the decay of higher mass states) production channel $gg \to X$ of this state. If the $X(3872)$ is a $D^0\overline{D}^{*0}$ molecule, then the binding energy of this would have to be very small, and so such a loosely bound system would have to be produced with a very small relative $k_\perp$ in the $D^0\overline{D}^{*0}$ rest frame, corresponding to a large separation between the mesons. The hadroproduction of such a state with the size of cross section observed in the $X(3872)$ case~\cite{Chatrchyan:2013cld} if possible at all, should in general take place in an environment where additional particles are emitted~\cite{Artoisenet:2009wk,Bignamini:2009sk}, so that the initially produced short--distance $c\overline{c}$ pair can form a loosely--bound, $D^0\overline{D}^{*0}$ state, at long distances. We would expect such a transition to be quite rare in the exclusive case, where no additional particles can be present, and so the observation of $X(3872)$ CEP would on general grounds disfavor such a purely molecular interpretation.

For a conventional $\chi_{c1}(2{}^3 P_1)$ state, the ratio of the CEP cross sections $\sigma(\chi_{c1}(2P))/\sigma(\chi_{c1}(1P))$ is predicted to first approximation (ignoring reasonably small corrections due to the different masses, relativistic effects etc) to be simply given by the ratio of the respective squared wave functions at the origin $|\phi_P'(0)|^2$. That is, we will expect them to be of comparable sizes. Moreover, we have seen in Section~\ref{chiccomp} that the CEP of the ground--state $\chi_{c1}(1P)$ has already been observed by LHCb~\cite{LHCb}, thus raising the possibility of such a measurement in the same experimental conditions.  The $X \to J/\psi \pi^+\pi^-$ decay is of particular interest, and in this case the final state is the same as for $\psi(2S)$ photoproduction, which may be used to give a handle on experimental efficiencies.  This result of course depends on the conventional charmonium interpretation for the $X(3872)$ being valid. If, as may be more realistic, it is a mixture of a $\chi_{c1}(2P)$ and a molecular $D^0\overline{D}^{*0}$ state, then the size of this ratio will also  be driven by the probability weight of the purely $c\overline{c}$ component; if this is small, that is the molecular component is dominant, then the $X(3872)$ cross section will be suppressed relative to the $\chi_{c1}(1P)$. In this way, the CEP mechanism could shed light on the nature of this puzzling state.

\section{Diphoton CEP}\label{cha:gam}

A further process which is of much interest, in particular as a `standard candle' with which to test the Durham framework, is the CEP of a pair of photons\cite{Khoze:2004ak,HarlandLang:2010ep,HarlandLang:2012qz}, produced via an intermediate quark loop, as shown in Fig.~\ref{gggam}. Such a process is experimentally very clean, and as we can access much higher central masses than in the case of $\chi_c$ CEP, it is less sensitive to the sort of theoretical uncertainties which are such a significant issue there. It also in principle allows a more detailed investigation of the underlying theory, as we can compare the $\gamma\gamma$ invariant mass distribution with the theoretical prediction, in a similar manner to the analysis performed for exclusive dijet production, where good agreement between theory and experiment was found, see~\cite{Khoze:2007td,Martin:2009ku}. Such a measurement is in particular sensitive to both the gluon PDF and the Sudakov factor, as well as to the theoretically challenging `enhanced' survival factor $S^2_{\rm enh}$. 

Already exclusive $\gamma\gamma$ data have been taken, with hopefully more to come in the near future. In 2007 CDF published a search for $\gamma\gamma$ CEP~\cite{Aaltonen:2007am} at the Tevatron, with $E_T(\gamma) >$ 5~GeV. Three candidate events were observed, in agreement with the Durham model expectations~\cite{Khoze:2004ak}. Subsequently, to increase statistics the $E_T(\gamma)$ threshold has been decreased to 2.5 GeV, and in 2011~\cite{Aaltonen:2011hi} CDF reported the  observation of 43 $\gamma\gamma$ events in $|\eta(\gamma)|<1.0$ with no other particles detected in $-7.4<\eta<7.4$. More recently~\cite{CMSgam} CMS have presented a search for exclusive $\gamma\gamma$ events at $\sqrt{s}=7$ TeV, and while no candidate events were observed the corresponding limits were fairly close to the theoretical predictions. We may therefore expect an observation of $\gamma\gamma$ CEP at the LHC to be forthcoming in the future.

\subsection{Theory}\label{gamCEP}

\begin{figure}
\begin{center}
\includegraphics[scale=0.8]{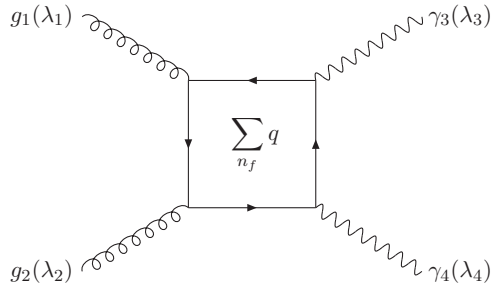}
\caption{Feynman diagram for $gg \to \gamma\gamma$ subprocess, mediated via a virtual quark loop. The sum is over all active quark flavours.}\label{gggam}
\end{center}
\end{figure}

The cross section for $\gamma\gamma$ CEP was originally calculated~\cite{Khoze:2004ak} assuming a simple factorized form, with the CEP amplitude (\ref{bt}) calculated in the limit of forward outgoing protons. However, as discussed in section~\ref{durt}, in general the skewed PDFs, the survival factor $S^2_{\rm eik}$, and the hard subprocess amplitude $\hat{\sigma}(gg \to \gamma\gamma)$ will depend on the outgoing proton $p_\perp$ and so this simple multiplicative factorization, used in earlier theoretical estimates~\cite{Khoze:2001xm,Khoze:2004ak}, will not hold. A proper treatment of these effects will not only in general lead to a more reliable cross section estimate in the $J_z=0$ case, but also allow for the inclusion of contributions that violate the $J_z=0$ selection rule~\cite{Khoze:2000jm}, which we recall is only exact in the limit that the proton $p_\perp=0$. We have seen in Section~\ref{chicintro} that in the case of $\chi_c$ production, these have been observed to have a significant effect.

The procedure for separating out the different $J_z$ contributions is outlined in Section~\ref{select}, see (\ref{pol}) and (\ref{Agen}). The $\mathcal{M}_{\lambda_1\lambda_2}$ in (\ref{Agen}) are the $g(\lambda_1)g(\lambda_2)\to \gamma(\lambda_3)\gamma(\lambda_4)$ helicity amplitudes (with the photon helicity labels implicit), corresponding to the process shown in~Fig.~\ref{gggam}. These are given by a simple generalisation of the lowest order $\gamma(\lambda_1)\gamma(\lambda_2)\to \gamma(\lambda_3)\gamma(\lambda_4)$ helicity amplitudes, which can be found in the literature~\cite{Bern:2001dg}. For completeness we give the $gg\to \gamma\gamma$ amplitudes here. The colour--averaged (as in (\ref{Vnorm})), amplitudes can be written as
\begin{equation}
\mathcal{M}_{\lambda_1\lambda_2\lambda_3\lambda_4}=4\left(\sum_f Q_{f}^2\right)\alpha\alpha_s\mathcal{M}^{(1)}_{\lambda_1\lambda_2\lambda_3\lambda_4}\;,
\end{equation}
where the sum is over the fractional charges $Q_f$ of all active quark flavours. The functions $\mathcal{M}^{(1)}_{\lambda_1\lambda_2\lambda_3\lambda_4}$ are given by
\begin{align}\label{gamp1}
\mathcal{M}_{--++}^{(1)}&=1\;,\\ \label{gamp2}
\mathcal{M}_{-+++}^{(1)}&=1\;,\\ \label{gamp3}
\mathcal{M}_{++++}^{(1)}&=-\frac{1}{2}\frac{\hat{t}^2+\hat{u}^2}{\hat{s}^2}\left(\ln^2\left(\frac{\hat{t}}{\hat{u}}\right)+\pi^2\right)-\frac{\hat{t}-\hat{u}}{\hat{s}}\ln\left(\frac{\hat{t}}{\hat{u}}\right)-1\;,\\
\end{align}
where $\hat{s}, \hat{t}, \hat{u}$ are the usual Mandelstam variables, and the other amplitudes follow from crossing symmetry, parity and time inversion~\cite{Jikia:1993tc}.

\subsection{Numerical Results}\label{gamnumer}

\begin{figure}[h]
\begin{center}
\includegraphics[scale=0.55]{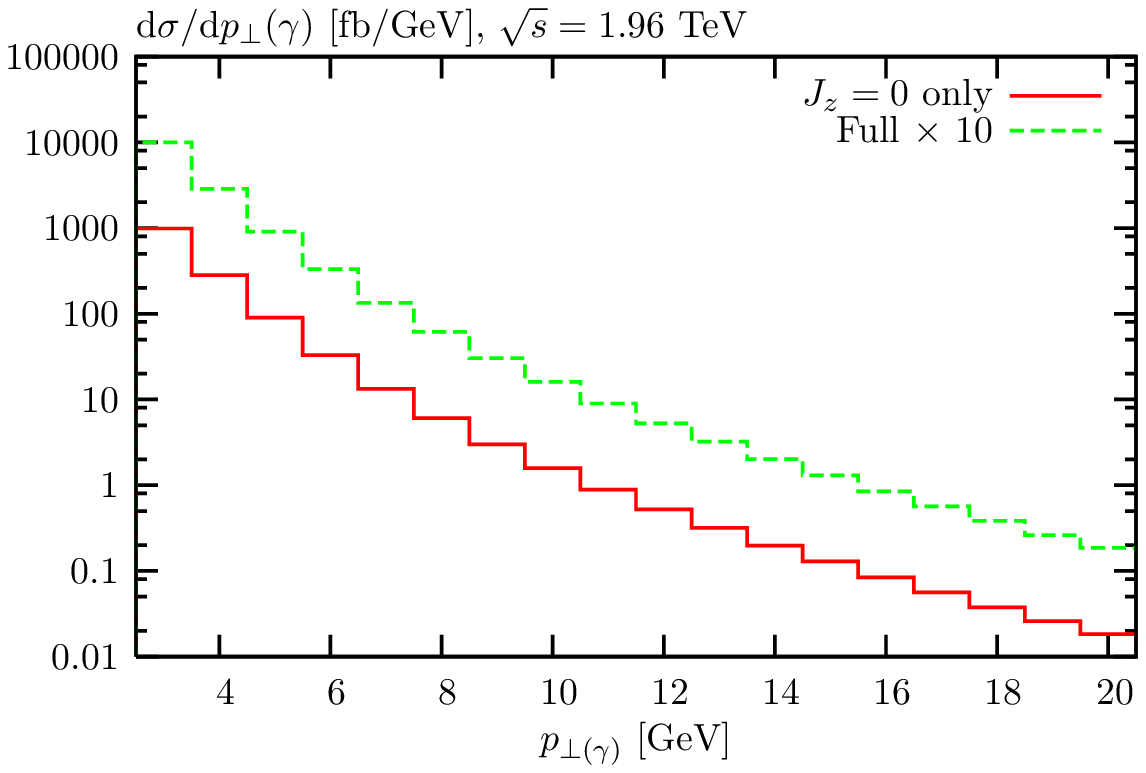}
\includegraphics[scale=0.55]{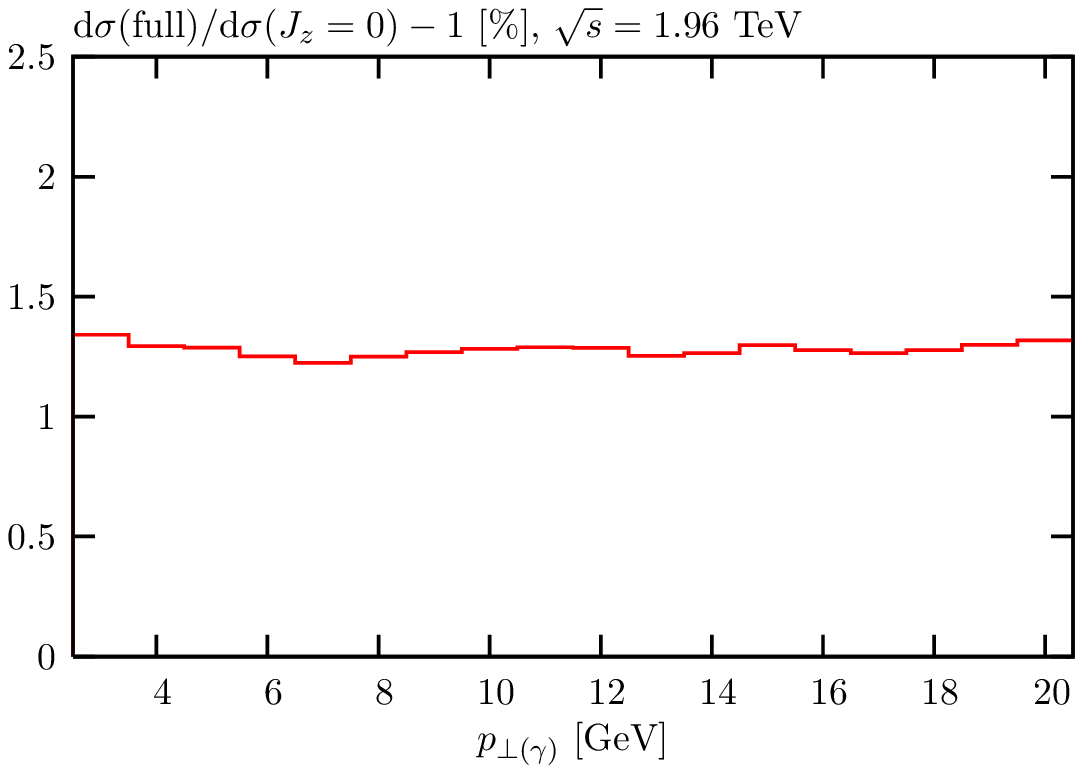}
\caption{Differential $\gamma\gamma$ CEP cross section ${\rm d}\sigma/{\rm d}p_\perp(\gamma)$ including and excluding the $|J_z|=2$ contribution, as discussed in the text. MSTW08LO PDFs~\cite{Martin:2009iq} are used, and the cuts $|\eta|<1$ and $E_\perp>2.5$ GeV are imposed on the photons. Also shown is the ratio ${\rm d}\sigma({\rm full})/{\rm d}\sigma(J_z=0)-1$ in percent, which shows the relative (small) size of the contribution from the $|J_z|=2$ piece.}\label{jzcomps}
\end{center}
\end{figure}

\begin{figure}
\begin{center}
\includegraphics[scale=0.6]{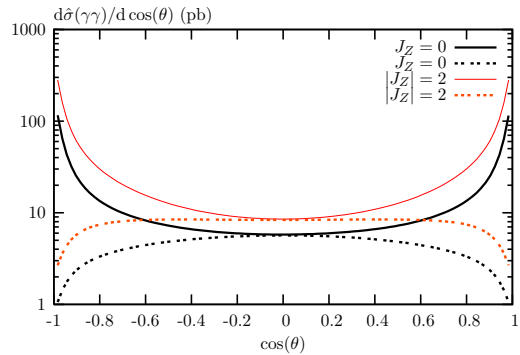}
\caption{Centre--of--mass scattering angle dependence of the hard subprocess $gg\to\gamma\gamma$ cross section, averaged over incoming gluon polarizations at the amplitude level, for a $J_z=0$ and $|J_z|=2$ incoming $gg$ system. The continuous curve represents production for a fixed $M_{\gamma\gamma}=10$~GeV, and the dashed for fixed $E_{\perp_\gamma}=5$~GeV.}\label{gamsig}
\end{center}
\end{figure}

We begin this section by considering the size of the $|J_z|=2$ contribution to the $\gamma\gamma$ CEP cross section; as well as being relevant to this specific process, this result will be important for other CEP processes where such a contribution may enter. Using the theory described above and in Section~\ref{durt}, we can give an initial rough estimate for the expected contribution by considering for example the production of a $(++)$ $\gamma\gamma$ state. In this case the $|J_z|=2$ contribution simplifies to
\begin{equation}\label{qjz2n}
T(|J_z|=2) \sim (q_{1_\perp}^x q_{2_\perp}^x-q_{1_\perp}^y q_{2_\perp}^y) \;.
\end{equation}
After performing the $Q_\perp$ integral and squaring, this will be of size
\begin{equation}\label{simjz2n}
|T|^2 \sim p_{1_\perp}^2p_{2_\perp}^2 \to \frac{\langle p_\perp^2 \rangle^2}{\langle Q_\perp^2\rangle^2}\;,
\end{equation}
as expected from (\ref{simjz2}). To give an exact evaluation of the expected suppression we should include all contributing amplitudes. While the $|J_z|=2$ subprocess cross section $\hat{\sigma}$ is a factor of $\sim 2$ larger than the $J_z=0$ cross section, this level of suppression is quite significant, and an explicit calculation of the full contribution, calculated as described in Section~\ref{gamCEP} using all contributing helicity amplitudes and summed over the final--state photon polarizations, gives
\begin{equation}
\frac{|T(|J_z|=2)|^2}{|T(J_z=0)|^2} \sim 1\%\;,
\end{equation}
again consistent with (\ref{simjz2}), with some variation depending on the particular PDF set used and the $x$ values probed. While this will receive some compensation from a larger survival factor, as in the $\chi_{c2}$ case (see section~\ref{numerchi}), this nonetheless represents a small correction to the overall cross section, which is well within other theoretical uncertainties. We show this explicitly in Fig.~\ref{jzcomps}, where we plot the differential $\gamma\gamma$ CEP cross section ${\rm d}\sigma/{\rm d}p_\perp(\gamma)$ at $\sqrt{s}=1.96$ TeV including and excluding the $|J_z|=2$ contribution, as well as the ratio ${\rm d}\sigma({\rm full})/{\rm d}\sigma(J_z=0)-1$ in percent. We can see that this ratio is indeed of order 1\% and is relatively flat in the photon $p_\perp$ and pseudorapidity $\eta$, although we do not show the latter explicitly here. In Fig.~\ref{gamsig} we plot the $J_z=0$ and $|J_z|=2$ subprocess differential  cross sections, and we can see clearly that the angular distributions for the two spin cases do not differ significantly, as they are dominated in both cases by double logarithmic singularities in the amplitudes as $u,t \to 0$. The $|J_z|=2$ contribution to the $\gamma\gamma$ CEP cross section is therefore expected to represent a very small normalisation correction, with the particle distributions being almost unchanged. 

\begin{figure}[t]
\begin{center}
\includegraphics[scale=0.5]{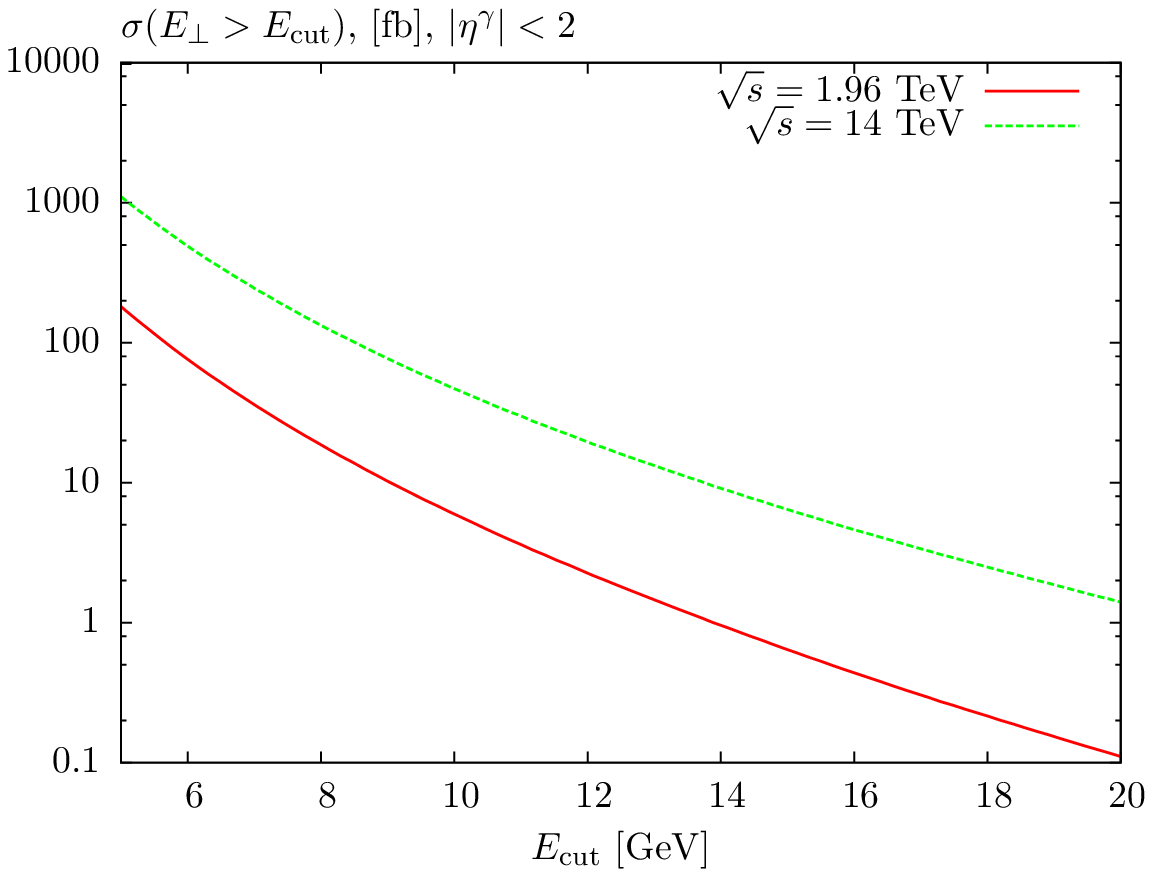}
\includegraphics[scale=0.5]{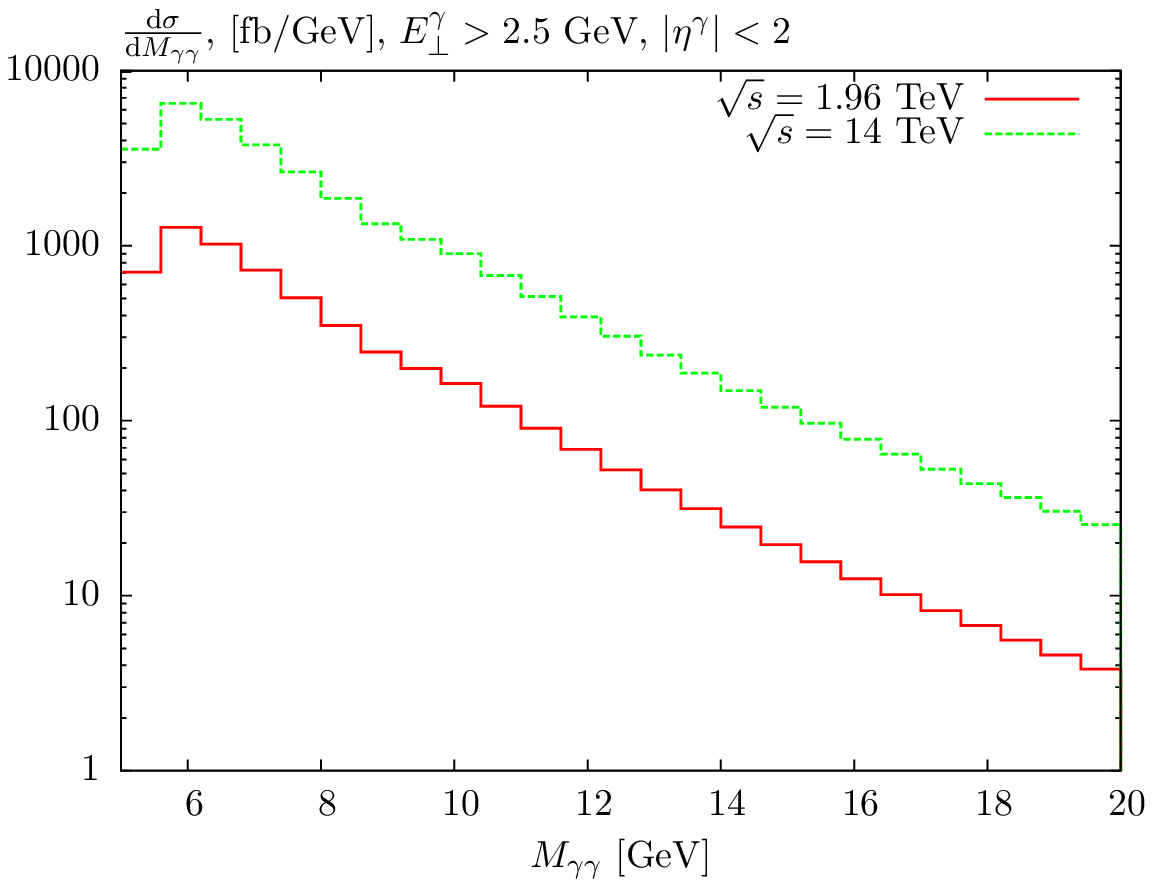}
\caption{Cross sections for $\gamma\gamma$ CEP at $\sqrt{s}=1.96$ and $14$ TeV and using MSTW08LO PDFs~\cite{Martin:2009iq}, as a function of the cut on the photon transverse energy $E_\perp>E_{\rm cut}$, and invariant mass distribution ${\rm d}\sigma/{\rm d}M_{\gamma\gamma}$, for $E_\perp>2.5$ GeV. In both cases the photon pseudorapidity is required to lie within $|\eta^\gamma|<2$. Predictions made using the \texttt{SuperCHIC}~\cite{SuperCHIC} MC}\label{gam1}
\end{center}
\end{figure}

We may therefore concentrate on the $J_z^P=0^+$ contribution to the $\gamma\gamma$ CEP cross section. In Fig.~\ref{gam1} we show the predicted $\gamma\gamma$ CEP cross section at $\sqrt{s}=$1.96 and 14~TeV, using MSTW08LO PDFs~\cite{Martin:2009iq}, as a function of the cut, $E_{\rm cut}$, on the photon transverse energy, $E_\perp$. As we have $M_{\gamma\gamma}/2 \sim E_\perp \sim E_{\rm cut}$, this cut effectively controls the invariant mass of the $\gamma\gamma$ system being produced. We also show in Fig.~\ref{gam1} the invariant mass distribution for a particular cut $E_\perp>2.5$ GeV on the photon transverse energy. The steep fall--off with $M_{\gamma\gamma}$ coming in part from the Sudakov factor, is clear. 

Although some caution is needed, recalling the theoretical uncertainties, we can also in principle use measurements of $\gamma\gamma$ CEP to shed some light on the gluon PDF in this low $x$ and low $Q^2$ region, where it is poorly determined. We note in particular the significant difference between the current LO and NLO PDF fits: while the LO PDFs have quite a steep low $x$ dependence, the NLO PDFs are much smaller and can even be negative at small $x$ and $Q^2$ for the more modern fits. We note that while the CEP cross section is calculated at LO in perturbation theory, and so formally should be used with LO PDFs, we may take both LO and NLO PDFs as input to provide an estimate of the range of predictions, given the (large) uncertainty in the gluon PDF at the low $x$ and $Q^2$ values relevant to CEP~\cite{HarlandLang:2012qz}. With this in mind, in Table~\ref{pdft} we show cross section predictions for $\gamma\gamma$ CEP at the LHC ($\sqrt{s}=1.96$ TeV) c.m.s. energies, for a range of LO and NLO PDF sets. These use a fairly recent fit to the survival factor~\cite{Ryskin:2012az}, which accounts for early LHC data, and were first presented elsewhere\cite{HarlandLang:2012qz}. We can see the wide variation in predictions, which future measurements, in particular at different collider energies, could shed light on. As we will see below, the CDF data~\cite{Aaltonen:2011hi} favour the higher predictions which use LO PDFs. We also show predictions at Tevatron energies, using slightly older models~\cite{Ryskin:2011qe,Ryskin:2009tk} to calculate the soft survival factors, which were used in earlier publications~\cite{HarlandLang:2010ys,HarlandLang:2011qd}, consistently with the cross section predictions quoted in the CDF publication~\cite{Aaltonen:2011hi}. However, we note that an updated calculation, with a more recent evaluation of the survival factor~\cite{Khoze:2013dha}, predicts a somewhat lower $\gamma\gamma$ cross section, by up to a factor of 2 depending on the specific soft model used, although this appears to be somewhat disfavoured by the CDF data. We also note that this more recent fit can predict somewhat larger cross sections, again by up to a factor of 2 depending on the specific soft model used, than those presented for the LHC in Table~\ref{pdft}. This sort of variation is unfortunately an inevitable result of the current uncertainty in the models of soft survival, and this issue will hopefully be clarified by future data.

\begin{table}[h]
\begin{center}
\tbl{$\gamma\gamma$ CEP cross sections (in pb) for different choices of gluon PDFs (GRV94HO~\cite{Gluck94}, MSTW08LO~\cite{Martin:2009iq}, CTEQ6L~\cite{Pumplin:2002vw}, MRST99~\cite{Martin:1999ww}, CT10~\cite{Lai:2010vv} and NNPDF2.1~\cite{Ball:2010de}), at $\sqrt{s}=1.96$ and 7 TeV, and for different cuts on the photon pseudorapidity, $\eta$. The photons are restricted to have transverse energy $E_\perp>2.5$ GeV at $\sqrt{s}=1.96$ TeV and $E_\perp>5.5$ GeV at $\sqrt{s}=7$ TeV.}
{
\begin{tabular}{|l|c|c|c|c|c|c|}
\hline
&\footnotesize{MSTW08LO}&\footnotesize{CTEQ6L}&\footnotesize{GJR08LO}&\footnotesize{MRST99}&\footnotesize{CT10}&\footnotesize{NNPDF2.1}\\
\hline
$\sqrt{s}=1.96$ TeV ($|\eta|<1$) &1.4&2.2&3.6&0.35&0.47&0.29\\
\hline
$\sqrt{s}=7$ TeV ($|\eta|<1$)&0.061&0.069&0.16&0.013&0.0094&0.0057\\
\hline
$\sqrt{s}=7$ TeV ($|\eta|<2.5$)&0.18&0.20&0.45&0.039&0.027&0.017\\
\hline
\end{tabular}\label{pdft}}
\end{center}
\end{table}

\subsubsection{Comparison with data}\label{gamcomp}

\begin{figure}[h]
\begin{center}
\includegraphics[scale=0.25]{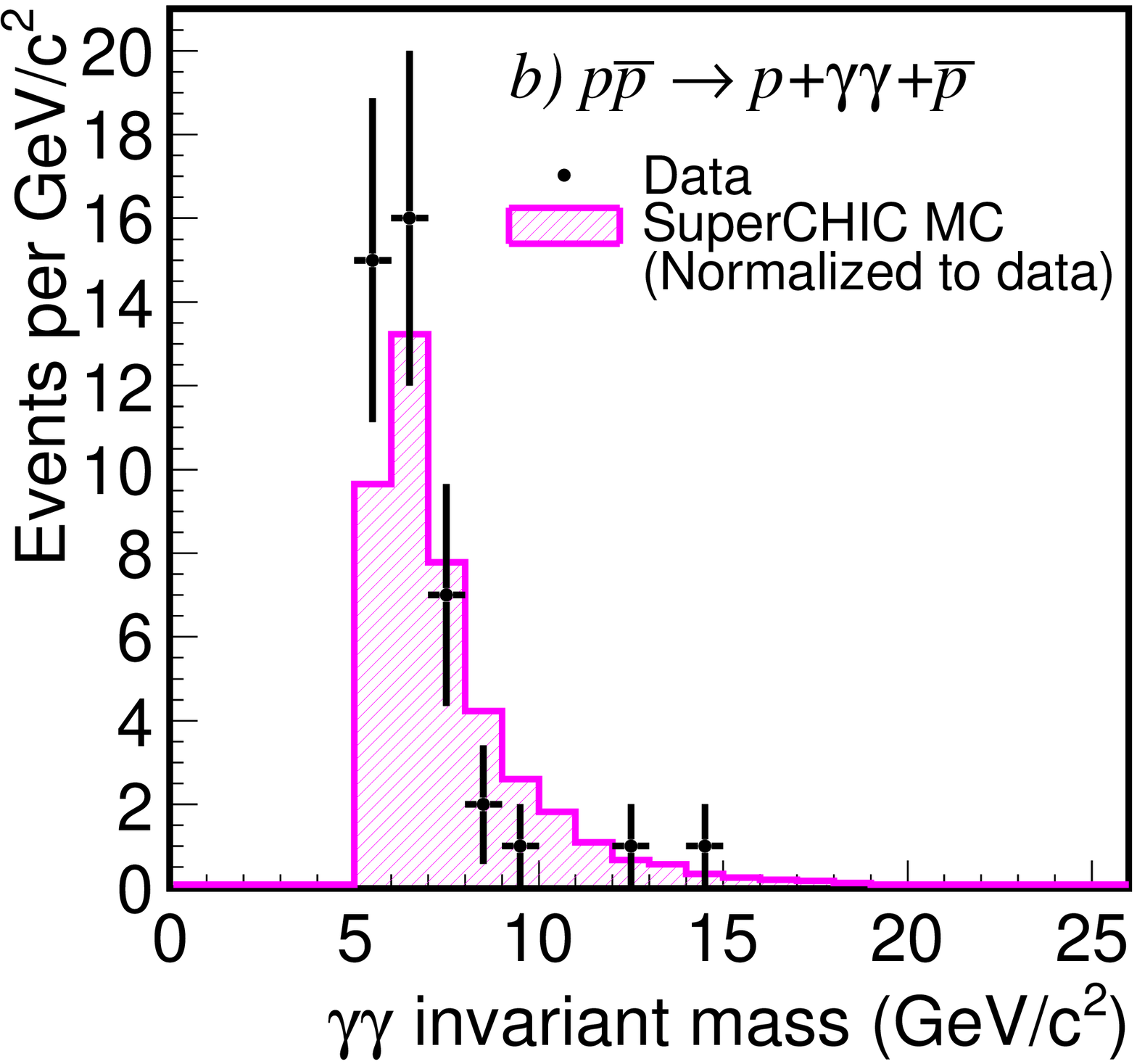}
\includegraphics[scale=0.25]{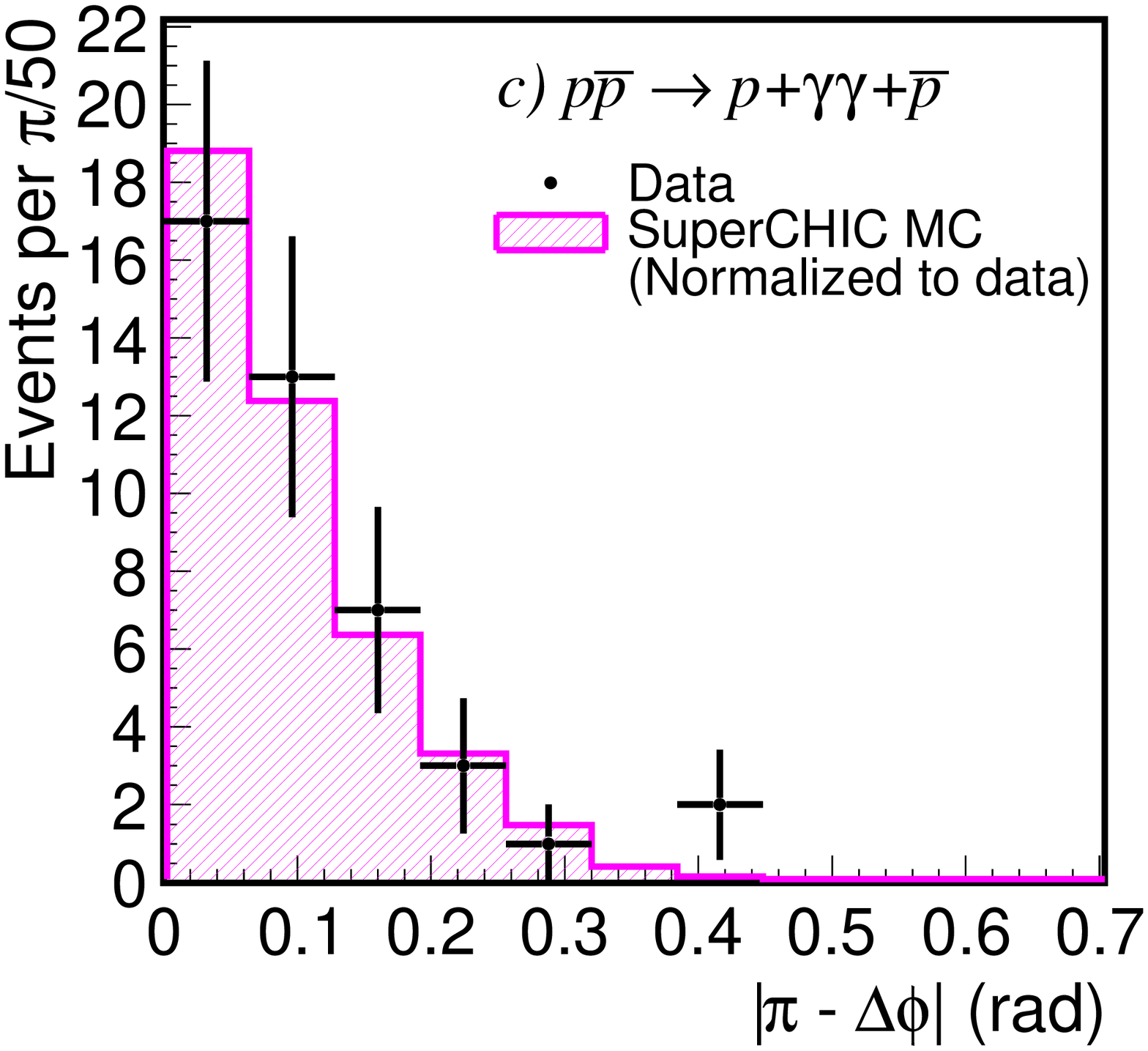}
\includegraphics[scale=0.25]{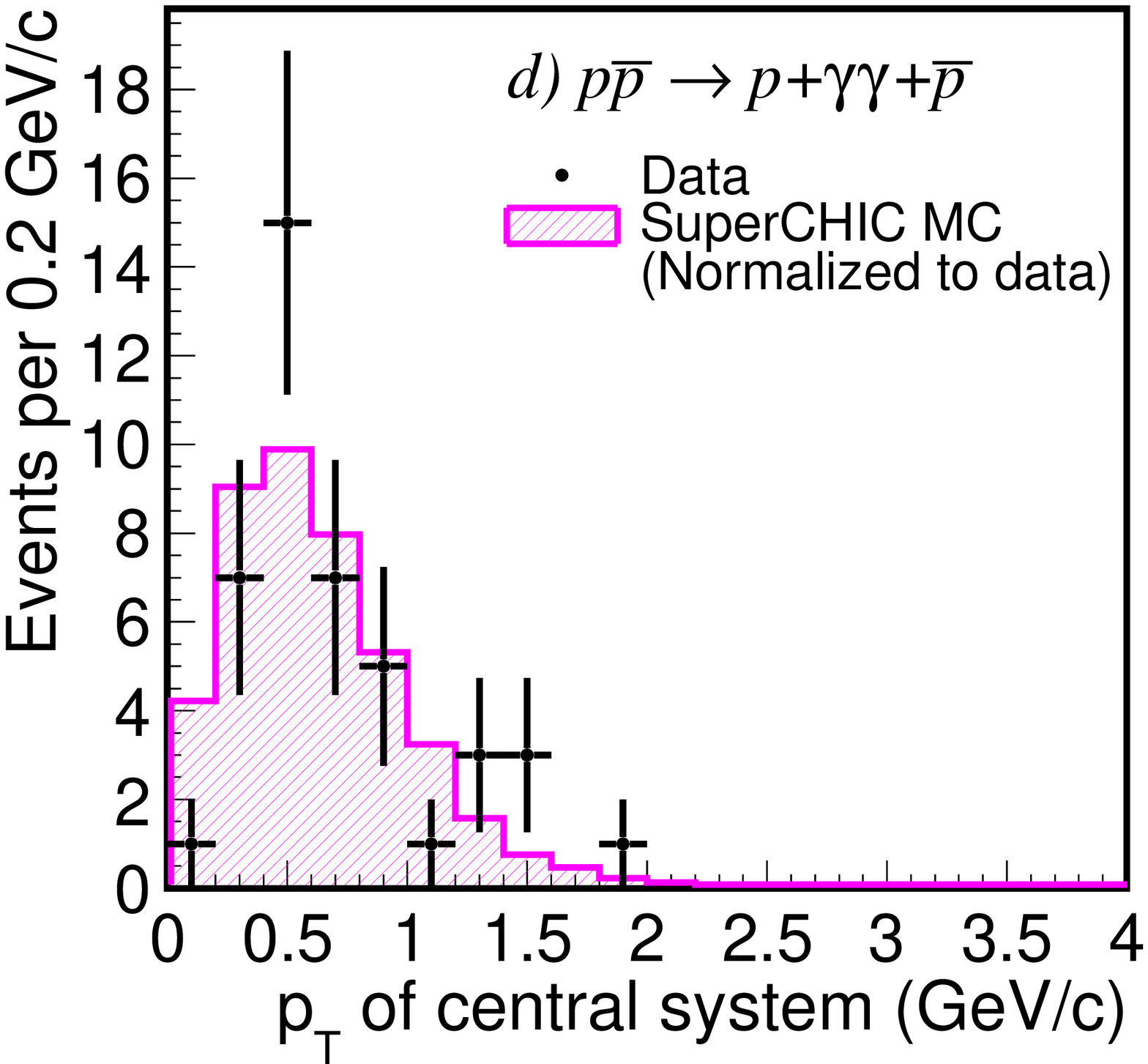}
\caption{Distributions of $\gamma\gamma$ invariant mass $M_{\gamma\gamma}$, difference in azimuthal separation from back--to--back configuration $|\pi-\Delta(\phi)_{\gamma\gamma}|$, and $\gamma\gamma$ transverse momentum $p_\perp(\gamma\gamma)$. Theory curves are calculated using the \texttt{SuperCHIC} MC~\cite{SuperCHIC}, and normalised to the data for comparison. Plots taken from CDF analysis~\cite{Aaltonen:2011hi} .}\label{gamdplots}
\end{center}
\end{figure}

As discussed in Section~\ref{cha:gam}, in 2011 CDF reported~\cite{Aaltonen:2011hi} the observation of 43 candidate $\gamma\gamma$ events in $|\eta(\gamma)|<1.0$ with no other particles detected in $-7.4<\eta<7.4$, which corresponds to a cross section of
\begin{equation}\label{gexp}
\sigma_{\gamma\gamma}^{\rm exp.}= 2.48\,{}^{+0.40}_{-0.35} \,({\rm stat}) \,{}^{+0.40}_{-0.51} \,({\rm syst}) \,{\rm pb}\;.
\end{equation}
This data was compared in the CDF analysis to the predictions of the Durham model~\cite{HarlandLang:2010ys}, made with MSTW08LO~\cite{Martin:2009iq} and MRST99 (NLO) PDFs~\cite{Martin:1999ww} (roughly representing an envelope of the different PDF predictions, following the logic described above), calculated using the formalism described in the preceding sections and implemented in the \texttt{SuperCHIC} MC generator~\cite{SuperCHIC}. These were
\begin{align}\label{gamtheory1}
\sigma_{\gamma\gamma}^{\rm theory}({\rm MRST99})&= 0.35\, {\rm pb}\;,\\ \label{gamtheory2}
\sigma_{\gamma\gamma}^{\rm theory}({\rm MSTW08LO})&=1.42\, {\rm pb}\;.
\end{align}
In the CDF publication~\cite{Aaltonen:2011hi}, these were quoted as having a factor of $\sim{}^{\times}_{\div}3$ uncertainties (in addition to that due to the PDF variation), although~\cite{HarlandLang:2012qz} it could be argued that these estimates are somewhat over--conservative. We can see that the prediction using the LO PDF set is consistent with the result within theoretical uncertainties, although both predictions lie below the observed cross section. As shown in Fig.~\ref{gamdplots}, the $p_\perp$, $\Delta \phi$ and invariant mass distributions of the $\gamma\gamma$ pair are described fairly well by the \texttt{SuperCHIC} MC.

It is natural to ask why the $\gamma\gamma$ CEP cross section predictions in (\ref{gamtheory1}) and (\ref{gamtheory2}) are somewhat lower than the data (\ref{gexp}). In fact, there are reasons why we may expect this to be the case. Most importantly\footnote{Experimentally we may also expect the observed $\gamma\gamma$ cross section to be enhanced by the small fraction of $\gamma\gamma$ events seen by CDF which are not truly exclusive, but rather due to diffractive production where one or both of the proton and antiproton dissociates, but where the proton dissociation products are not seen within the CDF acceptance; we estimate such a fraction  to be 10\% or lower.}, we recall that the predictions of (\ref{gamtheory1}, \ref{gamtheory2}) include only the LO perturbative QCD contribution to the $\gamma\gamma$ CEP process. In general, we may reasonably expect a numerically large NLO K--factor correction to the $gg \to X$ subprocess: for example, the higher--order corrections to Standard Model Higgs boson production via $gg \to H$ (see~\cite{Spira:1995rr,Kunszt:1996yp} and references therein) and P--wave quarkonia decay $\chi \to gg$~\cite{Barbieri:1981xz} are known to be quite large. 

Thus, in particular considering the possibility of such a large K--factor for the case of $\gamma\gamma$ CEP as well as the other theory uncertainties (due to for example the soft survival factors), we can see that the MSTW08LO prediction (\ref{gamtheory2}) is in good agreement with the data. We can see from Table~\ref{pdft} that for the predictions made using other representative LO PDF sets, the agreement is also generally good. This gives encouraging support for pQCD--based CEP framework, and in particular for the predictions for the topical example of light Higgs boson CEP, as discussed in Section~\ref{cha:higgs}.

Hopefully, further LHC $\gamma\gamma$ (and other) CEP data can shed further light on this: currently, we recall that CMS have presented a search for exclusive $\gamma\gamma$ events at $\sqrt{s}=7$ TeV~\cite{CMSgam}, and while no candidate events were observed the corresponding limits were fairly close to the theoretical predictions. In particular, the cross section for $E_\perp(\gamma)>5.5$ GeV, $|\eta(\gamma)|<2.5$ is found to be $<1.30$ pb at 95\% confidence, which we can see lies somewhat above\footnote{We should recall however that  accounting for the possibility of proton dissociation should lead to an increase of the theoretical CEP predictions by about a factor of 2, which is still consistent with the CMS upper limit~\cite{HarlandLang:2012qz}.} the theoretical predictions shown in Table~\ref{pdft}.
Hopefully, therefore, with the higher statistics that should come in the future an observation of $\gamma\gamma$ CEP at the LHC may be possible. Such data would be highly valuable, providing an additional and important constraint the model predictions, in particular in terms of the predicted $\sqrt{s}$ dependence from the Tevatron to the LHC, and with an extension out to higher values of $E_\perp(\gamma)$ allowing for a more differential test of the theory.

\section{Exclusive meson pair production}
\label{cha:meson}

Another particularly interesting CEP process is the production of light meson pairs ($X=\pi\pi, KK, \rho\rho, \eta(')\eta(')$É). At sufficiently high meson transverse momentum $k_\perp$ a perturbative approach, applying the Durham model and the `hard exclusive' formalism~\cite{Brodsky:1981rp,Benayoun:1989ng}, to evaluate the meson production subprocess, may be taken. 

There are two principle, related, reasons to look at such reactions. First, the  helicity amplitudes relevant to the CEP of meson pairs exhibit some remarkable theoretical features. Such `exclusive' $gg \to q\overline{q} q\overline{q}$, $q\overline{q}gg$, $gggg$ 6--parton amplitudes with fixed helicities of the incoming gluons are not relevant in a typical high--multiplicity inclusive process, and consequently they have not been studied in this context before. Second, they are of phenomenological interest, being experimentally realistic observables at hadron colliders from the Tevatron and RHIC to the LHC, with the wide variety of meson states available offering various channels with which to probe the non--trivial theory predictions for these different processes. Moreover the two--meson ($\pi\pi$, $KK$É) decays of the $\chi_{c(0,2)}$ states are of much interest in the CEP mode~\cite{HarlandLang:2010ep,HarlandLang:2012qz}, and so a proper understanding of the continuum background to these decays is important.

In the previous section we discussed the case of $\gamma\gamma$ CEP, including the CDF measurements~\cite{Aaltonen:2007am,Aaltonen:2011hi} of this process. However, an important possible background to $\gamma\gamma$ CEP is the exclusive production of a pair of $\pi^0$ mesons, with one photon from each $\pi^0$ decay undetected or the two photons merging. At first sight it would appear that the cross section for this purely QCD process may be much larger than the $\gamma\gamma$ cross section and so would constitute an appreciable background. However, as we will show in this section, we in fact expect the $\pi^0\pi^0$ CEP cross section to be significantly lower than in the case of $\gamma\gamma$ CEP~\cite{HarlandLang:2011qd}. This is a non--trivial result of the perturbative CEP formalism and as such this prediction represents a further important test of the framework. In fact, in the recent CDF measurement~\cite{Aaltonen:2011hi}, despite previous hints of a non--negligible $\pi^0\pi^0$ contribution in earlier data~\cite{Aaltonen:2007am}, of the 43 candidate $\gamma\gamma$ events, the contamination caused by $\pi^0\pi^0$ CEP was indeed observed to be very small ($< 15$ events, corresponding to a ratio $N(\pi^0\pi^0)/N(\gamma\gamma)<0.35$, at 95\% CL), lending support to this prediction. This encouraging result motivated a detailed investigation of the CEP of meson pairs~\cite{HarlandLang:2011qd,Harland-Lang:2013ncy,Harland-Lang:2013dia,Harland-Lang:2013qia}, which presents a new test of the perturbative formalism, with all its non--trivial ingredients.

A particularly interesting example of this is $\eta\eta$ and $\eta'\eta'$ CEP, which could allow a probe of the gluonic structure of $\eta(')$ mesons~\cite{Harland-Lang:2013ncy}. Currently, while different determinations of the $\eta$--$\eta'$  mixing parameters are generally consistent, the long--standing issue concerning the extraction of the gluon content of the $\eta'$ (and $\eta$) remains uncertain, in particular due to non--trivial theory assumptions and approximations that must be made, as well as the current experimental uncertainties and limitations~\cite{Thomas:2007uy,DiDonato:2011kr}. We will see in this section how the CEP process may provide a novel and potentially sensitive handle on this uncertain issue.

\subsection{Basic formalism}

\begin{figure}
\begin{center}
\includegraphics[scale=1.2]{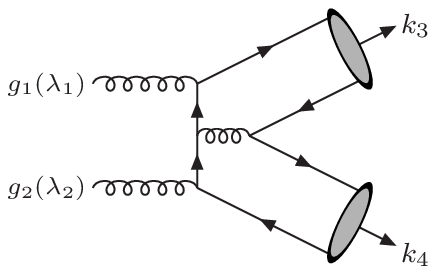}
\caption{Representative Feynman diagram for the $gg \to q\overline{q}q\overline{q}$ process.}\label{feyn1}
\end{center}
\end{figure}

As discussed above, the CEP of meson pairs has recently~\cite{HarlandLang:2011qd,Harland-Lang:2013qia,Harland-Lang:2013ncy}, been considered within a new approach, combining the Durham model and the `hard exclusive' formalism~\cite{Brodsky:1981rp,Benayoun:1989ng}, to calculate the parton--level helicity amplitudes relevant to the CEP of meson pairs, i.e. $gg \to M\overline{M}$  (where $M,\overline{M}$ is a meson, anti--meson).  The basic idea of this latter formalism is that the hadron--level amplitude can be written as a convolution of a (perturbatively calculable) parton--level amplitude, $T$, and a `distribution amplitude' $\phi$, which contains all the (non--perturbative) information about the binding of the partons in the meson. The $gg \to M\overline{M}$ amplitude can then be written as
\begin{equation}\label{amp}
\mathcal{M}_{\lambda\lambda'}(\hat{s},\theta)=\int_{0}^{1} \,{\rm d}x \,{\rm d}y\, \phi_M(x)\phi_{\overline{M}}(y)\, T_{\lambda\lambda'}(x,y;\hat{s},\theta)\;,
\end{equation}
 where $\sqrt{\hat{s}}$ is the $M\overline{M}$ invariant mass, $x,y$ are the meson momentum fractions carried by the partons and $\theta$ is the scattering angle in the $gg$ cms frame. $T_{\lambda\lambda'}$ is the hard scattering amplitude for the parton--level process \linebreak[4]$gg\to q\overline{q}(gg)\,q\overline{q}(gg)$, where each $q\overline{q}$ or $gg$ pair is collinear with the meson momentum\footnote{For a meson produced with large momentum, $|\vec{k}|$, we can to good approximation neglect the transverse component of the parton momentum, $\vec{q}$, with respect to $\vec{k}$.} and has the appropriate colour, spin, and flavour content projected out to form the parent meson. $\lambda$, $\lambda'$ are the gluon helicities: for our considerations there are two independent helicity configurations, $(\pm\pm$) and $(\pm \mp)$, which correspond to the incoming gluons being in a $J_z=0$ and $|J_z|=2$ state, respectively, along the incoming $gg$ direction. A representative diagram for purely $q\overline{q}$ 
valence components is shown in Fig.~\ref{feyn1}. Provided the meson $k_\perp$ is large enough, all intermediate quark and gluon propagators will be far off--shell and the amplitude can be calculated using the standard tools of pQCD.

The meson distribution amplitude depends on the (non--perturbative) details of hadronic binding and cannot be predicted in perturbation theory. However, the overall normalization of the $q\overline{q}$ distribution amplitude can be set by the meson decay constant $f_M$ via~\cite{Brodsky:1981rp}
\begin{equation}\label{wnorm}
\int_{0}^{1}\, {\rm d}x\,\phi_M(x)=\frac{f_M}{2\sqrt{3}}\;.
\end{equation}
It can also be show~\cite{Lepage:1980fj} that for very large $Q^2$ the meson distribution amplitude evolves towards the asymptotic form
\begin{equation}\label{asym}
\phi_M(x,Q^2)\underset{Q^2\to \infty}{\to} \,\sqrt{3} f_M\, x(1-x)\;.
\end{equation}
However this logarithmic evolution is very slow and at realistic $Q^2$ values the form of $\phi_M$ can in general be quite different. Indeed, a BABAR  measurement of the pion transition form factor $F_{\pi\gamma}(Q^2)$~\cite{Aubert:2009mc,Druzhinin:2009gq}, for example, strongly suggests that $\phi_\pi(x,Q)$ does not have the asymptotic form out to $Q^2\lesssim 40\,{\rm GeV}^2$, although more recent Belle data~\cite{Uehara:2012ag} are in conflict with this. Another possible choice is the `Chernyak--Zhitnisky' (CZ) form, which we will make use of later on~\cite{Chernyak:1981zz}
\begin{equation}\label{CZ}
\phi^{{\rm CZ}}_M(x,Q^2=\mu_0^2)=5\sqrt{3}f_M\, x(1-x)(2x-1)^2\;,
\end{equation}
where the starting scale is roughly $\mu_0\approx 1$ GeV. For the two--gluon distribution amplitude, $\phi_G(x)$, the normalization cannot be set as in (\ref{wnorm}), but an analogous formula can be written down~\cite{Ohrndorf:1981uz,Atkinson:1983yh,Baier:1985wv}
\begin{equation}
\int_0^1 {\rm d}y\, \phi_G(x,Q^2) (2x-1) \propto f_G(Q^2)\;,
\end{equation}
which serves to define $f_G$. While as $Q^2\to \infty$, it can be shown that the $gg$ distribution amplitude vanishes due to QCD evolution~\cite{Ohrndorf:1981uz,Baier:1981pm}
\begin{equation}
\lim_{Q^2 \to \infty} \phi_G(x)=0\;,
\end{equation}
there is no reason to assume this will be the case at experimentally relevant energies. More precisely, the meson distribution amplitudes can be expanded in terms of the Gegenbauer polynomials $C_n$~\cite{Lepage:1980fj,Ohrndorf:1981uz,Baier:1981pm}
\begin{align} \nonumber
\phi_{(1,8),M}(x,\mu_F^2)&=\frac{6 f_{(1,8)}^M}{2\sqrt{N_C}} x(1-x)[1+\sum_{n=2,4,\cdots} a_n^{(1,8)}(\mu_F^2)C_n^{3/2}(2x-1)]\;,\\ \label{waves}
\phi_{G,M}(x,\mu_F^2)&=\frac{f_1^M}{2\sqrt{N_C}}\sqrt{\frac{C_F}{2n_f}} x(1-x)\sum_{n=2,4,\cdots} a_n^G(\mu_F^2) C_{n-1}^{5/2}(2x-1)\;,
\end{align}
where $\mu_F$ is the factorization scale, taken as usual to be of the order of the hard scale of the process being considered, and $n_f=3$ for $\eta(')$ mesons. The $\phi_{(1,8)}$ are the flavour singlet(octet) quark wave functions, and the $f_{1,8}^M$ are given by (\ref{etafit}) for the case of the $\eta'$, $\eta$ mesons, while for other states these are given simply by their measured values, e.g. $f_{8}^\pi \equiv f_\pi=133$ MeV for the pion, see Section~\ref{mres} for more details. We note that for the $q\overline{q}$ distributions, the asymptotic (\ref{asym}) and CZ (\ref{CZ}) distribution amplitudes correspond to taking $a_2^{1}(\mu_0^2)=0$ and $a_2^{1}(\mu_0^2)=2/3$ in (\ref{waves}), respectively, with all higher $n$ terms being zero.  

To make contact with the physical $\eta$, $\eta'$ states we will consider later on, we introduce the flavour--singlet and non--singlet quark basis states
\begin{align} \nonumber
|q\overline{q}_1\rangle &=\frac{1}{\sqrt{3}}|u\overline{u}+d\overline{d}+s\overline{s}\rangle\;,\\ \label{qfock}
|q\overline{q}_8\rangle &=\frac{1}{\sqrt{6}}|u\overline{u}+d\overline{d}-2s\overline{s}\rangle\;,
\end{align}
and the two--gluon state
\begin{equation}\label{gfock}
|gg\rangle\;,
\end{equation}
with corresponding distribution amplitudes given by (\ref{waves}). Here, we take a general two--angle mixing scheme~\cite{Feldmann:1997vc,Kiselev:1992ms,Leutwyler:1997yr} for the $\eta$ and $\eta'$ mesons. That is, the mixing of the $\eta$, $\eta'$ decay constants is not assumed to follow the usual (one--angle) mixing of the states. This is most easily expressed in terms of the $\eta$ and $\eta'$ decay constants
\begin{align}\nonumber
f_8^\eta=f_8 \cos \theta_8\;,   \qquad\qquad f_1^\eta&=-f_1 \sin \theta_1 \;,\\ \label{etafit}
f_8^{\eta'}=f_8 \sin \theta_8\;,     \qquad\qquad f_1^{\eta'}&=f_1 \cos \theta_1 \;,
\end{align}
with~\cite{Feldmann:1998vh}
\begin{align}\nonumber
  f_8=1.26 f_\pi\;, \qquad & \qquad \theta_8   = -21.2^\circ\;,\\ \label{thetafit}
 f_1=1.17 f_\pi\;,  \qquad & \qquad \theta_1  = -9.2^\circ\;.
\end{align}
We then take the distribution amplitudes (\ref{waves}) with the decay constants given as in (\ref{etafit}), for the corresponding Fock components (\ref{qfock}) and (\ref{gfock}). A more in depth discussion is given elsewhere~\cite{Harland-Lang:2013ncy}.

\subsection{Parton--level amplitudes}

\subsubsection{Scalar flavour non--singlet mesons}\label{sfns}

The simplest case of scalar flavour--non--singlet mesons ($\pi\pi, KK$...) proceeds via the type of diagram shown in Fig.~\ref{feyn1}, where the $q\overline{q}$ pair forming the parent mesons are {\it not} connected by a quark line~\cite{HarlandLang:2011qd}. There are 31 Feynman diagrams which contribute to the leading--order amplitude, and after an explicit calculation we find
\begin{align}\label{T++}
T^{qq}_{++}=T^{qq}_{--}&=0\;,\\ \label{T+-}
T^{qq}_{+-}=T^{qq}_{-+}&=\frac{\delta^{\rm AB}}{N_C}\frac{64\pi^2\alpha_S^2}{\hat{s}xy(1-x)(1-y)}\frac{a-b^2}{a^2-b^2\cos^2{\theta}}\\ \nonumber
&\cdot \frac{N_C}{2}\bigg(\cos^2{\theta}-\frac{2 C_F}{N_C}a\bigg)\;,
\end{align}
where `$qq$' indicates that the final--state partons are $q\overline{q}$ pairs, `$A,B$' are the gluon colour indices and
\begin{align}\label{a}
a&=(1-x)(1-y)+xy\; ,\\ \label{b}
b&=(1-x)(1-y)-xy\; .
\end{align}
Considering first the $J_z=0$ amplitude, we can see that this completely vanishes at LO, a non--trivial result which follows from an overall cancellation between the many different non--zero contributing Feynman amplitudes, and is a direct generalization of the case of the equivalent $\gamma\gamma \to M \overline{M}$ amplitudes, which vanish for neutral mesons, when the photons are in a $J_z=0$ state~\cite{Brodsky:1981rp}. At leading order, the production of flavour non--singlet meson pairs, which must proceed via these diagrams, can therefore only occur when the incoming gluons are in a $|J_z|=2$ state. Recalling the `$J_z=0$' selection rule discussed in Section~\ref{select}, which strongly disfavours such a configuration, this gives the non--trivial prediction that the CEP of flavour non--singlet meson pairs ($\pi\pi, KK$...) in the perturbative regime will be strongly suppressed\footnote{However it should be noted that any NNLO corrections which allow a $J_z=0$ contribution may cause the precise value of the cross section to be somewhat larger than the leading--order, leading--twist $|J_z|=2$ estimate, although qualitatively the strong suppression should remain.}.

\begin{figure}[h]
\begin{center}
\includegraphics[scale=0.65]{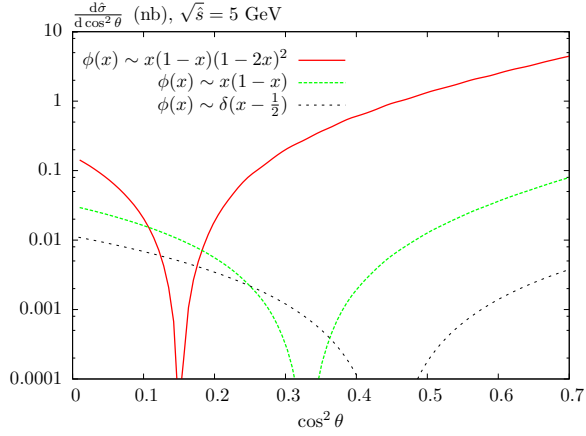}
\caption{Differential cross section ${\rm d}\sigma/{\rm d}|\cos \theta|$ at $\sqrt{\hat{s}}=5$ GeV, for the $gg\to M\overline{M}$ process for non--flavour singlet scalar mesons. For comparison, the distribution for three choices of meson wavefunction are shown, the asymptotic form $\phi_M(x)\propto x(1-x)$, the CZ form (\ref{CZ}), and a $\delta$--function $\phi_M(x)\propto\delta(x-\frac{1}{2})$.}\label{rad0}
\end{center}
\end{figure}

Considering now the $|J_z|=2$ amplitude, we can see that this is in general non--zero, but still vanishes for a particular value of $\cos^2\theta$. This behaviour, which at first sight may appear quite unusual, is in fact not completely unexpected: the vanishing of a Born amplitude for the radiation of massless gauge bosons, for a certain configuration of the final state particles, is a known effect, usually labeled a `radiation zero'~\cite{Heyssler:1997ng,Brown:1982xx}. It results from the complete destructive interference of the classical radiation patterns, leading to a vanishing of the amplitude, and the general conditions for the existence of these zeros have been written down elsewhere~\cite{Brown:1982xx}. . The position of the zero is determined by an interplay of both the internal (in the present case, colour) and space-time (the particle $4$-momenta) variables, as can be seen in (\ref{T+-}), where the position of the zero 
depends on the choice of meson wavefunction, $\phi(x)$, through the variables $a$ and $b$, as well as on the QCD colour factors, see Fig.~\ref{rad0}.  In particular, the zero occurs in (\ref{T+-}) when $\cos^2\theta \approx 2C_F\langle a \rangle/N_C$, where $\langle a \rangle$ is the average value of $a$ integrated over the meson wavefunctions $\phi_M(x),\phi_M(y)$. As $0\leq a \leq 1$ for all physical values of $x,y$ and $2 C_F/N_C<1$ for all $N_C$ (while the prefactors in (\ref{T+-}) are strictly positive), this will always occur in the physical region for any non--abelian SU(N) gauge theory. While this effect, which is present in all theories with massless gauge bosons, occurs in general in QCD, it is usually neutralised along with colour by the averaging of hadronization. The CEP process, for which the fusing gluons are selected to be in a colour--singlet state by the exclusivity of the event, therefore offers an in principle unique possibility to observe these zeros. 

\subsubsection{Scalar flavour--singlet mesons}\label{sfs}

\begin{figure}
\begin{center}
\subfigure[]{\includegraphics[scale=0.75]{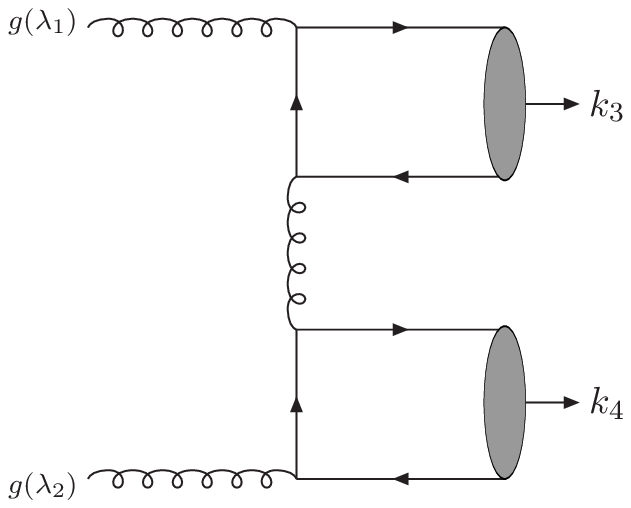}}
\subfigure[]{\includegraphics[scale=0.75]{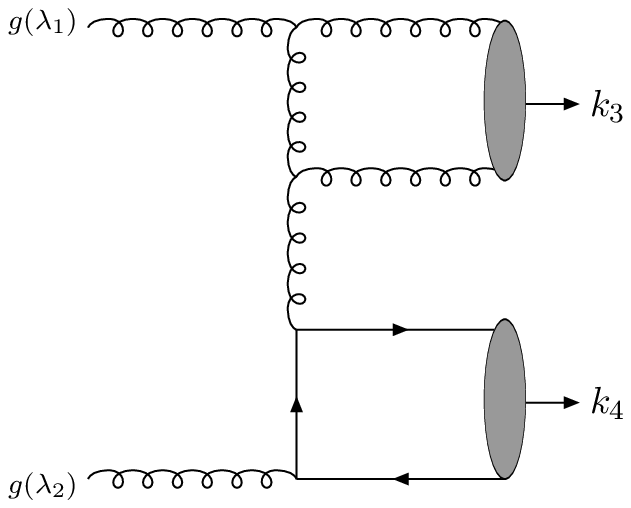}}
\subfigure[]{\includegraphics[scale=0.75]{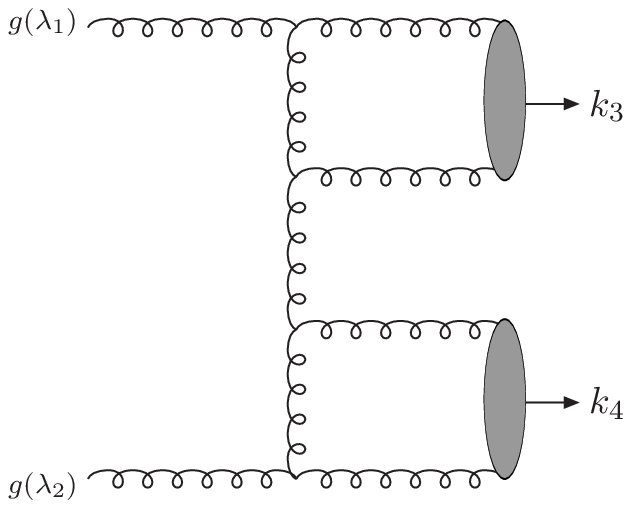}}
\caption{Representative Feynman diagrams for the $gg \to M\overline{M}$ process, where the $M$ are flavour--singlet mesons. There are 8 Feynman diagrams of type (a), and the corresponding helicity amplitudes are given by (\ref{lad0}, \ref{lad2}). There are 76 Feynman diagrams of type (b), and the corresponding $J_z=0$ helicity amplitude is given by (\ref{tgq0}). There are 130 Feynman diagrams of type (c), and the corresponding $J_z=0$ helicity amplitude is given by (\ref{tgg0}). In the case of the amplitudes for (b) and (c), all diagrams allowed by colour conservation are included, and not just diagrams of this ladder type.}\label{feyn2}
\end{center}
\end{figure}

As well as the configuration shown in Fig.~\ref{feyn1}, the outgoing $q\overline{q}$ pairs can also combine in a second way, with the $q\overline{q}$ pair forming each meson being connected by a quark line. A representative such `ladder--type' diagram is shown in Fig.~\ref{feyn2} (a); as each meson couples individually to two (isosinglet) gluons, only flavour--singlet states can be produced via such a diagram (for e.g. a $\pi^\pm$ state, this is clear, while for a $\pi^0$ the $u\overline{u}$ and $d\overline{d}$ components of the flavour Fock state interfere destructively). There are 8 Feynman diagrams which contribute to the amplitude, and we find that these give
\begin{align}\label{lad0}
T_{++}^{S,qq}=T_{--}^{S,qq}&=\frac{\delta^{ab}}{N_C}\frac{64\pi^2 \alpha_S^2}{\hat{s}xy(1-x)(1-y)}\frac{(1+\cos^2 \theta)}{(1-\cos^2 \theta)^2}\;,\\ \label{lad2}
T_{+-}^{S,qq}=T_{-+}^{S,qq}&=\frac{\delta^{ab}}{N_C}\frac{64\pi^2 \alpha_S^2}{\hat{s}xy(1-x)(1-y)}\frac{(1+3\cos^2 \theta)}{2(1-\cos^2 \theta)^2}\;,
\end{align}
where the label `$S,qq$' is used to distinguish these amplitudes, which only contribute for flavour--singlet mesons, from (\ref{T++}, \ref{T+-}), which contribute for both flavour--singlet and non--singlet states. Thus we can see that the amplitude does not vanish for  $J_z=0$ incoming gluons, in contrast to the flavour--non--singlet case (\ref{T++}). From this simple observation, we then have the highly non--trivial prediction that the CEP of flavour--singlet meson pairs, in particular $\eta'\eta'$, will not be suppressed by the $J_z=0$ selection rule, and so it is predicted to be strongly enhanced relative to e.g. $\pi\pi$ CEP. In the case of $\eta\eta'$ and $\eta\eta$ production, we also expect some enhancement, although the level of this is dependent on the specific $\eta$--$\eta'$ mixing parameters that are taken. We will examine this in more detail in Section~\ref{mres}.

As well as having valence $q\overline{q}$ components, it is well known that the dominantly flavour--singlet $\eta'$ (and also, through mixing, $\eta$) mesons should have a valence $gg$ component, which also carries flavour--singlet quantum numbers, and as discussed above, the size of such a gluonic component remains largely uncertain. With this in mind, we have previously extended~\cite{Harland-Lang:2013ncy}  the above calculation to include the case that one or both outgoing $q\overline{q}$ pairs forming the meson states are replaced by $gg$ pairs in a pseudoscalar state, as shown in Figs.~\ref{feyn2} (b) and (c). Such diagrams will contribute to the $gg \to \eta(')\eta(')$ processes in the presence of any non--zero $gg$ valence component. After a quite lengthy calculation, it was found that
\begin{align}\label{tgq0}
 T^{gq}_{++}&=T^{gq}_{--}=2\,\sqrt{\frac{N_C^3}{N_C^2-1}}(2x-1)\cdot\,T_{++}^{S,qq}\;,\\ \label{tgg0}
T^{gg}_{++}&=T^{gg}_{--}=4\,\frac{N_C^3}{N_C^2-1}(2x-1)(2y-1)\cdot\,T_{++}^{S,qq}\;,
\end{align}
where `$qg$' and `$gg$' correspond to the $q\overline{q}gg$ and $gggg$ final states, respectively.
We can find no simple form in the case of the $|J_z|=2$ ($\pm,\mp$) amplitudes, but the magnitude and angular dependences are found numerically to be similar.

We can see that again the $J_z=0$ amplitudes do not vanish. Moreover, they are in fact identical to the amplitude (\ref{lad0}) for a purely valence quark component in the mesons, up to overall colour and normalization factors. That is, they are predicted to have the same angular dependence in the incoming $gg$ rest--frame. We emphasize that in these cases, all diagrams allowed by colour conservation contribute to the total amplitude, and not just diagrams of this ladder type shown in Fig.~\ref{feyn2}: for each process (`$S,qq$', `$qg$' and `$gg$') the final amplitude receives contributions from a set of diagrams which are completely distinct and apparently unrelated between the three cases. That these final results should be so remarkably similar in form is therefore very surprising. An immediate phenomenological implication of this fact is that as the $gg$ valence contribution is not suppressed relative to the $q\overline{q}$ contribution, and so, as we will show in more detail in Section~\ref{mres}, any 
reasonable gluonic component of the $\eta'$ ($\eta$) may have a significant effect on the predicted $\eta(')\eta(')$ CEP cross sections.

\subsubsection{Vector mesons}\label{svect}

Finally, we give the amplitudes $T_{\lambda_1\lambda_2,\lambda_3\lambda_4}$ for the $g(\lambda_1)g(\lambda_2)\to V(\lambda_3)\overline{V}(\lambda_4)$ process, where $V(\overline{V})$ are helicity $\pm 1$ spin-1 mesons. Considering first the flavour--non--singlet case, for the helicity--0 state we find that the amplitudes are identical to those for scalar mesons -- see (\ref{T++}) and (\ref{T+-}). For the transverse polarisations we have
\begin{align}\label{Tv1}
T_{++,+-}&=T_{++,-+}=T_{--,+-}=T_{--,-+}=0\;,\\\label{Tv2}
T_{+-,+-}&=T_{-+,-+}=-\frac{\delta^{ab}}{N_C}\frac{64\pi^2\alpha_S^2}{\hat{s}xy(1-x)(1-y)}\bigg(C_F b^2-\frac{N_C}{2}a\bigg)\frac{\cos \theta (1+\cos \theta)}{a^2-b^2 \cos^2 \theta}\;,\\\label{Tv3}
T_{-+,+-}&=T_{+-,-+}=\frac{\delta^{ab}}{N_C}\frac{64\pi^2\alpha_S^2}{\hat{s}xy(1-x)(1-y)}\bigg(C_F b^2-\frac{N_C}{2}a\bigg)\frac{\cos \theta (1-\cos \theta)}{a^2-b^2 \cos^2 \theta}\;,
\end{align}
where it is clear (in the limit that the quark $q_t=0$ relative to the meson momentum) that $T_{\lambda_1\lambda_2,++}=T_{\lambda_1\lambda_2,--}=0$ (as well as those amplitudes in which both helicity--0 and 1 states are produced), as helicity must be conserved along the fermion line for massless quarks. This result immediately follows from the helicity conserving gluon--$q\overline{q}$ vertices which enter the perturbative calculation, and the fact that the meson helicity is given by the sum of the helicities of its valence quarks: this forms the basis of the so--called `hadronic helicity conservation' selection rule~\cite{Brodsky:1981kj}. 

Considering now the flavour--singlet case, we recall these may in principle receive an additional contribution from the `ladder--type' diagrams shown in Fig.~\ref{feyn2} (a). While it also immediately follows from helicity conservation along the quark lines that the production of transversely polarized vector mesons cannot proceed via these diagrams, for longitudinally polarised vector mesons we find
\begin{align}\label{long0}
T_{++,00}^{\rm S,qq}=T_{--,00}^{\rm  S,qq}&=(2a-1)\,T_{++}^{\rm  S,qq}\;,\\ \label{long2}
T_{+-,00}^{\rm  S,qq}=T_{-+,00}^{\rm  S,qq}&=(1-2a)\,T_{+-}^{\rm  S,qq}\;,
\end{align}
which are antisymmetric under the interchange $x\leftrightarrow (1-x)$ (or $y\leftrightarrow (1-y)$), and will therefore vanish upon integration over the (symmetric) meson wavefunction, $\phi_M(x)$. This result for these `ladder' diagrams, where each meson state couples separately to two gluons, recalls the well--known Landau--Yang theorem~\cite{LY1,Yang50}, which states that a spin--1 particle cannot couple to two on--shell massless vector bosons: although the intermediate $t$--channel gluon will in general be far off--shell, the overall amplitude nevertheless vanishes. It therefore follows from (\ref{Tv1}) and  (\ref{long0})--(\ref{long2}) that the CEP of light vector pairs will be strongly suppressed in the perturbative regime, irrespective of their flavour structure, and indeed the $\rho\rho$, $\omega\omega$ and $\phi\phi$ rates are predicted within this formalism to be the same up to small (higher--order, higher--twist) corrections. 

\subsection{Calculation within a MHV approach}

It is well known~\cite{Mangano:1990by} that the tree level $n$--gluon scattering amplitudes, in which ($n-2$) gluons have the same helicity, the so--called `maximally helicity violating' (MHV), or `Parke--Taylor', amplitudes, are given by remarkably simple formulae~\cite{Parke:1986gb,Berends:1987me}. These results have been extended using supersymmetric Ward identities to include amplitudes with one and two quark--antiquark pairs~\cite{Mangano:1990by}, where `MHV' refers to the case where ($n-2$) partons have the same helicity. In these cases, simple analytic expressions can again be written down for the MHV amplitudes, while for greater than 2 fermion--anti--fermion pairs (recalling that the helicities of a connected massless fermion--anti--fermion pair must be opposite) no MHV amplitudes exist. More recently, it has been shown~\cite{Britto:2004ap,Georgiou:2004by} that the $n$--parton scattering amplitude for {\it any} helicity configuration can be calculated with this formalism; in particular they can be constructed from tree graphs in which the vertices are the usual tree-level MHV scattering amplitudes continued off-shell in a specific way.

With this in mind, it can be shown that the quite simple results in the previous sections follow from the observation that the corresponding $J_z=0$ helicity amplitudes are MHV, with $n-2=4$ partons (the two incoming gluons, and two outgoing partons) having the same helicity; in particular, the vanishing of the $J_z=0$ amplitude for the production of scalar flavour--non--singlet mesons described in Section~\ref{sfns} and the identical form of the flavour--singlet amplitudes (\ref{lad0}, \ref{tgq0}, \ref{tgg0}) for quark and gluon valence final states described in Section~\ref{sfs} both follow from a quite simple application of the MHV formalism~\cite{HarlandLang:2011qd,Harland-Lang:2013ncy}.

These results follow by considering the relevant MHV 6--parton amplitude for general particle momenta and colours, and making the assignments that the final--state $q\overline{q}$ and/or $gg$ pairs are collinear with the outgoing mesons and are in a  colour--singlet state. More precisely, we recall that within the MHV approach the full $n$--parton amplitude $\mathcal{M}_n$ can be written in the form of a `dual expansion', as a sum of products of colour factors $T_n$ and purely kinematic partial amplitudes $A_n$
\begin{equation}\label{mhv}
\mathcal{M}_n(\{p_i,h_i,c_i\})=ig^{n-2}\sum_\sigma T_n(\sigma\{c_i\})A_n(\sigma\{1^{\lambda_1},\cdots,n^{\lambda_n}\})\;,
\end{equation}
where $c_i$ are colour labels, $i^{\lambda_i}$ corresponds to the $i$th particle ($i=1\cdots n$), with momentum $p_i$ and helicity $\lambda_i$, and the sum is over different particle orderings in the amplitude, i.e. over appropriate and simultaneous non--cyclic permutations $\sigma$ of colour labels and kinematics variables. The purely kinematic part of the amplitude $A_n$ encodes all the non--trivial information about the full amplitude, $\mathcal{M}_n$, while the factors $T_n$ are given by known colour traces. 

While each particle ordering in general has a different colour factor $T_n$, the assignment in our case that the outgoing partons must form colour--singlet pairs tends to lead to a substantial simplification in this, with certain colour factors vanishing or factorizing for a set of particle orderings. For example~\cite{HarlandLang:2011qd}, in the case of flavour--non--singlet meson pair production the only non--vanishing colour factors are given by a universal factor of ${\rm Tr}(\lambda^a \lambda^b)=\delta^{ab}/2$, where $a,b$ are the incoming gluon colour labels. Moreover, the kinematic partial amplitudes can also take very simple forms when the collinearity assignment is made for the outgoing $q\overline{q}$ and/or $gg$ pairs. The combination of this vanishing and/or factorisation of the colour factors and simplifications of the kinematic terms can lead to very simple results for the full amplitudes, exactly as we have found in the preceding sections. In the case of  flavour--non--singlet meson pair production, after making the colour--singlet assignment there is in fact a full--scale cancelation between the surviving kinematic partial amplitudes, with
\begin{align}\nonumber
M&\propto \delta^{ab}\left( \frac{\langle k_3\, k_4 \rangle}{ \langle k_4\, k_1 \rangle \langle k_1\, k_3 \rangle \langle k_3\, k_2 \rangle \langle k_2\, k_4 \rangle}+\frac{1}{\langle k_3\, k_1 \rangle\langle k_1\, k_2 \rangle
\langle k_2\, k_4 \rangle}+\frac{1}{\langle k_3\, k_2 \rangle \langle k_2\, k_1 \rangle \langle k_1\, k_4 \rangle}\right)\\ \label{canc}
&\propto \delta^{ab}\left(\langle k_3\, k_2 \rangle\langle k_1\, k_4 \rangle+\langle k_1\, k_3 \rangle\langle k_2\, k_4 \rangle-\langle k_3\, k_4 \rangle\langle k_1\, k_2 \right)\rangle=0\;,
\end{align}
where `$\langle k,l\rangle$'  is the standard spinor contraction (for particles of momenta $k,l$) and the last line corresponds to the well--known Schouten identity, which holds for any set of four $4$--momenta $k_i$~\cite{HarlandLang:2011qd}. This is exactly as was found above (\ref{T++}), after a quite involved calculation involving 31 separate (and in general non--zero) Feynman diagrams. This result (\ref{canc}), on the other hand, follows after just a few lines of calculation, giving some indication of the power of this approach. The $J_z=0$ amplitudes for vector meson production given in Section~\ref{svect}, and the purely valence quark flavour--singlet amplitude (\ref{lad0}) follow in an equally simple way.

In the case of flavour--singlet mesons, and the identical forms (\ref{lad0}, \ref{tgq0}, \ref{tgg0}) for valence quark and gluon final--states described in Section~\ref{sfs}, it takes a little more work to arrive at this result within the MHV approach, but the principle of applying the colour--singlet and collinearity assignments to simplify the known MHV amplitudes, is the same. Moreover, this result certainly follows much more simply than from the very complicated Feynman diagram calculation (we recall that for example in the case of the purely gluonic amplitude there are 130 contributing Feynman diagrams). For more details of these calculations we refer the interested reader to previous detailed treatments~\cite{HarlandLang:2011qd,Harland-Lang:2013ncy}.

\subsection{Numerical results}\label{mres}

Using the amplitudes calculated above we can show some representative predictions for the corresponding meson pair CEP cross sections. We take $f_\pi=133$ MeV and $f_\rho^\perp=f_\rho^0=200$ MeV~\cite{Benayoun:1989ng}, and we assume a universal $q\overline{q}$ wavefunction given by (\ref{CZ}) throughout. Although the correct form of the quark distribution amplitude remains an open question, this choice is found~\cite{Chernyak:2009dj} to describe the $\gamma\gamma \to M\overline{M}$ data quite well, and we take it as our benchmark choice here. 

\begin{figure}
\begin{center}
\includegraphics[scale=0.6]{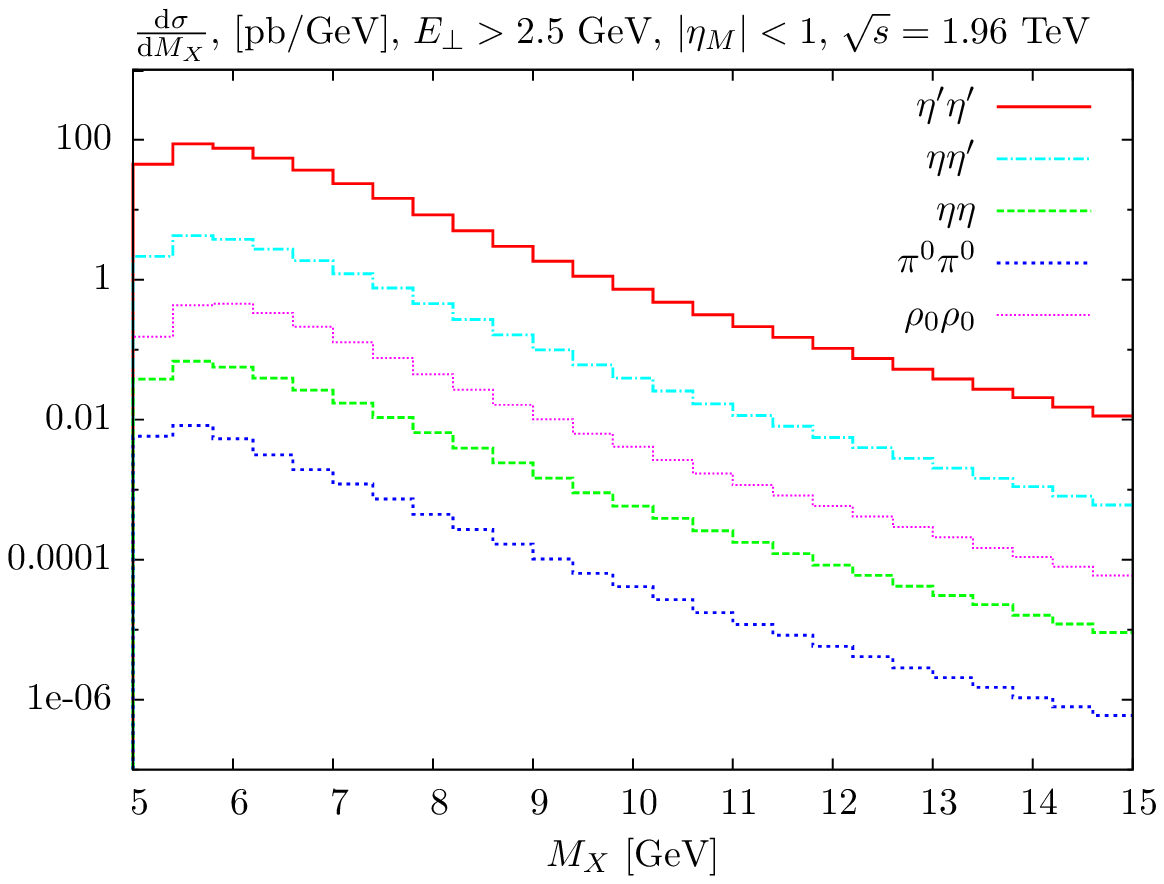}
\includegraphics[scale=0.6]{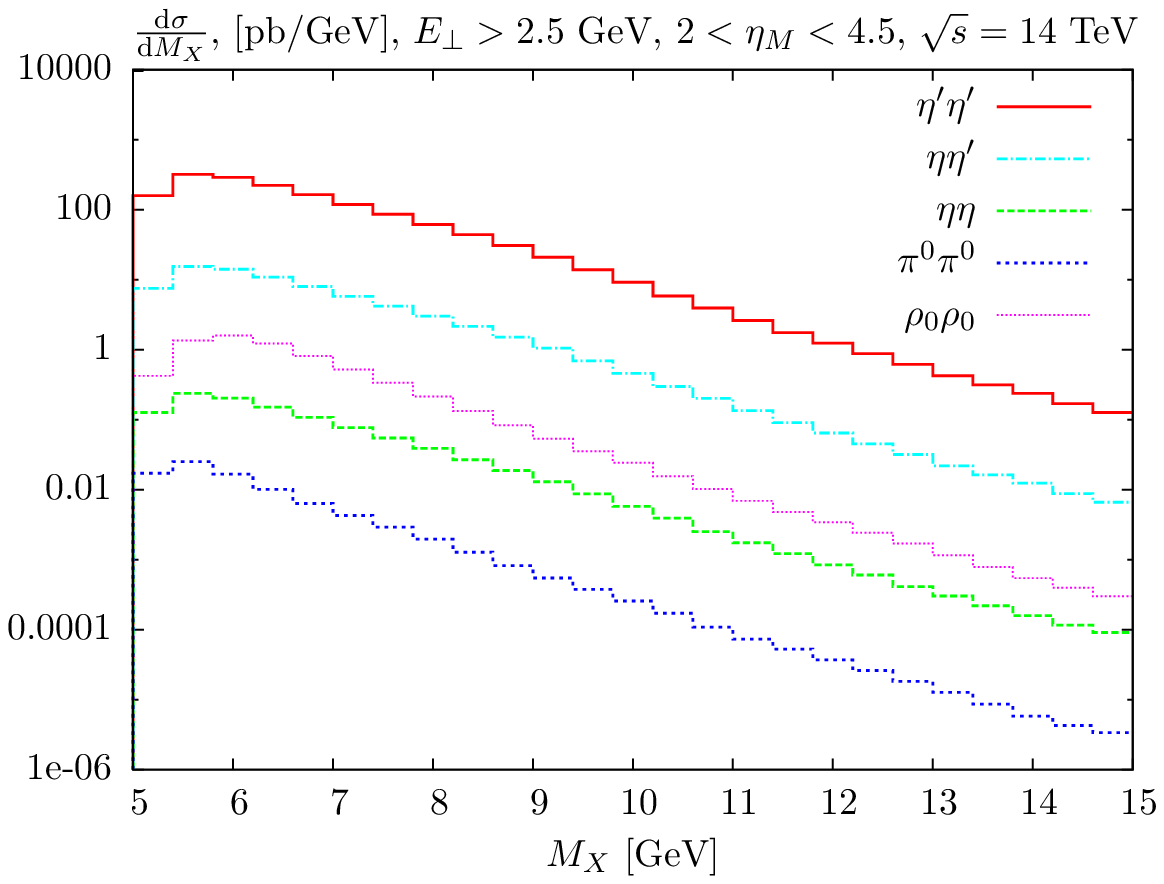}
\caption{Differential cross section ${\rm d}\sigma/{\rm d} M_X$ for the CEP of meson pairs, for meson transverse energy $E_\perp>2.5$ GeV, and for different c.m.s. energies and cuts on the meson pseudorapidities. Predictions made using \texttt{SuperCHIC}~\cite{SuperCHIC} MC.}\label{vsm}
\end{center}
\end{figure}

We begin by considering the purely valence quark contribution. We recall from the results above that we expect a strong enhancement in the cross section for the dominantly flavour--singlet $\eta'\eta'$ states relative to the flavour non--singlets ($\pi^+\pi^-$, $\pi^0\pi^0$É). The $\eta\eta$ and $\eta\eta'$ CEP cross sections are strongly dependent on the precise level of $\eta-\eta'$ mixing, through which a $J_z=0$ component can enter, but are also expected to be enhanced.  In Fig.~\ref{vsm} we show the CEP cross sections ${\rm d}\sigma/{\rm d}M_X$, where $M_X$ is the invariant mass of the meson pair, for the production of various scalar and vector states. We consider different c.m.s. energies and cuts on the meson pseudorapidity, corresponding to the experimentally relevant situations at the Tevatron and LHCb, for illustration. The suppression of the $\pi^0\pi^0$ and vector meson cross sections is clear\footnote{The charged $\pi^+\pi^-$ and $\rho^+\rho^-$ CEP cross sections are expected to be a factor of 2 larger from isospin symmetry and the non--identity of the final--state particles.}, in particular in the $\pi^0\pi^0$ case where the radiative zero in the $|J_z|=2$ will tend to further reduce the cross section. The $\eta'\eta'$ cross section, on the other hand, is predicted to be much larger. The vector meson $\rho\rho$ cross section is also shown: within the perturbative formalism, the $\phi\phi$ and $\omega\omega$ rates are to lowest order expected to be identical to this, see section~\ref{svect}. In the vector meson case the cross sections are also suppressed by the $J_z=0$ selection rule, see (\ref{Tv1}--\ref{Tv3}), however this is less severe than for the scalar $\pi^0\pi^0$, due in part to the larger decay constant $f_\rho$ ($\sim 200$ MeV), and in part the particular form of the $|J_z|=2$ production amplitudes, for which there is no additional suppression from a radiation zero as in (\ref{T+-}). Unfortunately, as discussed in Section~\ref{cha:gam} for the low $x$ and $Q^2$ values probed in the CEP of lighter mass objects, there is a large degree of uncertainty in the single gluon PDFs. Here we present estimates here using MSTW08L0~\cite{Martin:2009iq} PDFs, which as we have seen in Section~\ref{gamcomp} give a prediction for $\gamma\gamma$ CEP that is in reasonable agreement with the CDF data.

We recall that the calculation of the $\pi^0\pi^0$ CEP cross section has important consequences for the possible $\pi^0\pi^0$ background to $\gamma\gamma$ CEP, when one photon from each $\pi^0$ decay is undetected or the two photons merge. At first sight it would appear that the cross section for this purely QCD process could be much larger than for $\gamma\gamma$ production and so would constitute an appreciable background, but in fact this is not the case. Firstly, we have seen that the amplitude to form an exclusive pion with transverse momentum, $k_\perp$, is proportional to the ratio $f_\pi/\sqrt{\hat{s}}\sim f_\pi/k_\perp$ (see (\ref{wnorm})), that is the cross section of the $gg\to \pi^0\pi^0$ hard subprocess contains the numerically small factor $(f_\pi/k_\perp)^4$ which in the region of interest is comparable to (or even smaller than) the QED suppression, $(\alpha_{QED}/\alpha_S)^2$, of the $gg\to\gamma\gamma$ cross section. Secondly, and crucially, the vanishing of the LO amplitude $gg\to\pi^0\pi^0$ with $J_z=0$ initial--state gluons leads to a further $\sim$ two orders of magnitude suppression in the CEP cross section, see Section~\ref{select}. We therefore expect the $\pi^0\pi^0$ background contribution to $\gamma\gamma$ CEP to be small, a prediction which is supported by the recent CDF $\gamma\gamma$ data~\cite{Aaltonen:2011hi}, see Section~\ref{mesoncomp}.

Our results can readily be extended to the kaon sector, although we do not consider this numerically here. In particular, under the assumption of exact $SU(3)$ flavour symmetry, the $K^0\overline{K}^0$ and $K^+K^-$ CEP cross sections can be calculated in the same way as the $\pi^0\pi^0$ and $\pi^+\pi^-$ cross sections, with the replacement $f_\pi \to f_K$. In fact, $SU(3)$ flavour symmetry breaking effects can be non--negligible, and to precisely estimate the $K^0\overline{K}^0$ and $K^+K^-$ cross sections, a modified narrower form of the meson wavefunction, which accounts for asymmetry between the $s$ and $(u,d)$ quark masses, should be taken~\cite{Benayoun:1989ng}. Without the inclusion of this modified wavefunction, the perturbative formalism tends to overestimate the $\gamma\gamma \to K^+ K^-$ cross section when compared to BELLE data~\cite{Nakazawa:2004gu}.

\begin{figure}
\begin{center}
\includegraphics[scale=0.6]{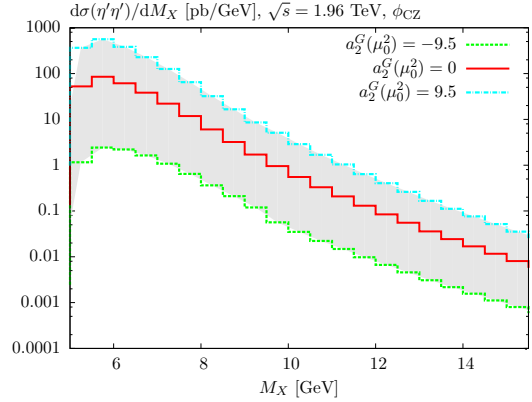}
\caption{Differential cross section ${\rm d}\sigma/{\rm d}M_X$ for $X=\eta'\eta'$ production  at $\sqrt{s}=1.96$ TeV with MSTW08LO PDFs~\cite{Martin:2009iq}, taking the CZ form~\cite{Chernyak:1981zz} for the quark distribution amplitude, and for a band of $a_2^G(\mu_0^2=1\,{\rm GeV}^2)$ values, corresponding to different normalizations of the $gg$ distribution amplitude $\phi_G(x,Q^2) \propto a_2^G(Q^2)$. The mesons are required to have transverse energy $E_\perp>2.5$ GeV and pseudorapidity $|\eta|<1$.}\label{etamcz}
\end{center}
\end{figure}

Considering now the effect of a $gg$ valence component to the $\eta'$, $\eta$ cross sections, we have previously~\cite{Harland-Lang:2013ncy} made use of a fit~\cite{Kroll:2013iwa} to the $\gamma\gamma^* \eta(')$ form factor, which finds 
 \begin{equation}\label{b2g}
a^G_{2,{\rm fit}}(\mu_0^2)=19 \pm 5\;.
\end{equation}
While this may give some rough guidance for the expected size of the $gg$ component of the $\eta(')$, we note that this fit contains important uncertainties, in particular because the gluonic contribution to the $\eta(')$ transition form factor $F_{\eta(')\gamma}(Q^2)$ only enters at NLO, and so is a relatively small effect, thus requiring  a precision fit to the data in regions where other theoretical uncertainties are not necessarily under such good control\footnote{We also recall, for example, that the situation with regards to the $\chi_{c(0,2)}$ decays into $\eta$ and $\eta'$ pairs appears to be somewhat puzzling. Experimentally, no enhancement in the decays to $\eta,\eta'$ pairs relative to pions is observed (after taking trivial phase space effects into account). This may indicate that there is some destructive interference between the $q\overline{q}$ and the $gg$ components of the pseudoscalar bosons, or that the $gg$ component is small~\cite{Kroll:2013iwa,Ochs:2013gi}.}. Therefore, to give a conservative evaluation of the sensitivity of the CEP process to the size of this $gg$ component, we have considered a band of cross section predictions, guided by (\ref{b2g}), corresponding to the range~\cite{Harland-Lang:2013ncy}
\begin{equation}\label{b2g1}
 a_2^G(\mu_0^2)\in (-a^G_{2,{\rm fit}}/2,+a^G_{2,{\rm fit}}/2)=(-9.5,9.5)\;,
\end{equation}
 where $a_2^G$ is defined in (\ref{waves}), and all higher  ($n=4,6$...) order terms are neglected for simplicity. Even with this quite narrow and conservative range of values, we find that the predicted CEP cross section changes considerably.  We show this in Fig.~\ref{etamcz}, where we plot the $M_X$ distribution for $X=\eta'\eta'$ CEP at $\sqrt{s}=1.96$ TeV for this band of possible $gg$ components. At the LHC, we expect the cross section (for the same event selection) to be roughly a factor of $\sim 3$--5 larger for $\sqrt{s}=7$--14 TeV, with the particle distributions almost unchanged. We can see that a reasonable  $gg$ component of the $\eta'$ (and $\eta$) can have a strong effect on the CEP cross section, increasing (or decreasing) it by up to $\sim$ an order of magnitude, depending on the specific size and sign of the $gg$ component\footnote{Depending on the sign of the $gg$ component, the quark and gluon valence contributions will interfere destructively or constructively.}. 

More precisely~\cite{Harland-Lang:2013ncy}, due to the identical angular dependence of the amplitudes (\ref{lad0}), (\ref{tgq0}) and (\ref{tgg0}), the effect of including a non--zero $gg$ component  of the $\eta'$ ($\eta$) mesons on the $\eta(')\eta(')$ CEP amplitudes will to first approximation be to multiply them by an overall normalization factor. The ratio of the different $\eta(')\eta(')$ cross sections are then determined by the mixing parameters~\cite{Feldmann:1998vh}
\begin{align}\nonumber
\sigma(\eta'\eta'):\sigma(\eta\eta'):\sigma(\eta\eta)&=1:2\tan^2(\theta_1):\tan^4(\theta_1)\;,\\ \label{ratcross}
&\approx 1:\frac{1}{19}:\frac{1}{1450}\;,
\end{align}
 irrespective of the size of the $gg$ component. In Table~\ref{etarats} we show numerical results for the cross section ratios (\ref{ratcross}): due to the effect of the $|J_z|=2$ flavour non--singlet contribution, this scaling is only expected to be approximate (there is also in all cases a small effect due to the differing $\eta$ and $\eta'$ masses).

\begin{table}[h]
\begin{center}
\tbl{Ratios of $\eta(')\eta(')$ CEP cross sections at $\sqrt{s}=1.96$ TeV with MSTW08LO PDFs~\cite{Martin:2009iq}, for a $gg$ distribution amplitude with different choices of $a_2^G(\mu_0^2)$ and with the $q\overline{q}$ distribution amplitude given by the CZ form (\ref{CZ}). The mesons are required to have transverse energy $E_\perp>2.5$ GeV and pseudorapidity $|\eta|<1$.}{
\begin{tabular}{|l|c|c|c|}
\hline
$a_2^G(\mu_0^2)$&-9.5&0&9.5\\
\hline
$\sigma(\eta'\eta')/\sigma(\eta\eta)$&210&1300&1600\\
\hline
$\sigma(\eta'\eta')/\sigma(\eta\eta')$&20&20&20\\
\hline
$\sigma(\eta\eta')/\sigma(\eta\eta)$&11&66&78\\
\hline
\end{tabular}
\label{etarats}}
\end{center}
\end{table}

Thus, to first approximation we can only look at absolute value of the various $\eta(')\eta(')$ CEP cross sections to determine the size of the $gg$ component, $a_2^G(\mu_0^2)$. This is potentially problematic because of the other uncertainties in the CEP calculation, due primarily to the value of the survival factors $S^2_{\rm eik}$, $S^2_{\rm enh}$, which are not known precisely, and potential higher--order corrections in the hard process, which combined are expected to a give a factor of $\sim {}^\times_\div 2-3$ uncertainty, as well as a sizeable PDF uncertainty in the low--$x$, $Q^2$ regime relevant to such processes~\cite{HarlandLang:2012qz}. Nevertheless, given the sensitivity of the CEP cross section to the $gg\to gggg$ and $gg \to ggq\overline{q}$ subprocess, if the $gg$ component of the $\eta(')$ is sizeable enough, such a measurement may still provide useful information. However, it is more reliable to look at the ratio of the $\eta(')\eta(')$ cross section to other processes, in which case many of the uncertainties due to PDFs and survival factors largely cancel out and a potentially much cleaner measurement of the $gg$ component of the $\eta(')$ becomes possible. With this in mind we show for illustration in Table~\ref{etarats1} the ratio of the $\eta\eta$ and $\eta'\eta'$ to $\pi^0\pi^0$ and $\eta'\eta'$ to $\gamma\gamma$ cross sections. Recalling that the $\pi^0\pi^0$ CEP cross section is also predicted to be strongly suppressed, due to the vanishing at LO of the $gg \to \pi^0\pi^0$ amplitude for $J_z=0$ incoming gluons, a measurement of the ratios $\sigma(\eta(')\eta('))/\sigma(\pi^0\pi^0)$ would also represent as an important probe of the $J_z=0$ selection rule.

\begin{table}[h]
\begin{center}
\tbl{Ratios of $\eta(')\eta(')$ to $\pi^0\pi^0$ and $\gamma\gamma$ CEP cross sections at $\sqrt{s}=1.96$ TeV with MSTW08LO PDFs~\cite{Martin:2009iq}, for a $gg$ distribution amplitude with different choices of $a_2^G(\mu_0^2)$ and with the $q\overline{q}$ distribution amplitude given by the CZ form (\ref{CZ}). The meson/photons are required to have transverse energy $E_\perp>2.5$ GeV and pseudorapidity $|\eta|<1$.}{
\begin{tabular}{|l|c|c|c|}
\hline
$a_2^G(\mu_0^2)$&-9.5&0&9.5\\
\hline
$\sigma(\eta\eta)/\sigma(\pi^0\pi^0)$&2.7&12&66\\
\hline
$\sigma(\eta'\eta')/\sigma(\pi^0\pi^0)$&570&16000&100000\\
\hline
$\sigma(\eta'\eta')/\sigma(\gamma\gamma)$&3.5&100&660\\
\hline
\end{tabular}
\label{etarats1}}
\end{center}
\end{table}

\subsection{The non--perturbative regime}\label{nps}

\begin{figure}[h]
\begin{center}
\includegraphics[scale=1.0]{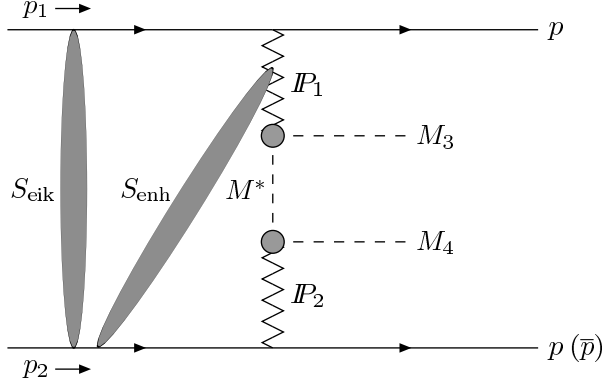}
\caption{Representative diagram for the non--perturbative meson pair ($M_3$, $M_4$) CEP mechanism, where $M^*$ is an intermediate off--shell meson of type $M$. Eikonal and (an example of) enhanced screening effects are indicated by the shaded areas.}\label{npip}
\end{center}
\end{figure}

Up until now we have only considered the CEP of meson pairs within a purely perturbative framework. However, the study of meson pair CEP in fact has a long history, which far predates this approach~\cite{Kaidalov:1974qi,Azimov:1974fa,Pumplin:1976dm,Desai:1978rh}. In these cases~\cite{HarlandLang:2012qz,Lebiedowicz:2009pj,Lebiedowicz:2012nk}, the production process was instead considered within the framework of Regge theory~\cite{Collins:1977jyp}, with the meson pair produced by the exchange of two Pomerons in the $t$--channel. Such a `non--perturbative' picture, see Fig.~\ref{npip}, should be relevant at lower values of the meson transverse momentum $k_\perp$, where the cross sections are largest, and may be particularly important for the case of flavour--non--singlet mesons ($\pi\pi$, $KK$...), for which the perturbative contribution is expected to be dynamically suppressed, as we have seen above. This has been considered in detail elsewhere~\cite{HarlandLang:2011qd,HarlandLang:2012qz}, with more recently a new \texttt{Dime} Monte Carlo (MC) implementation of this Regge--based approach being developed~\cite{Harland-Lang:2013dia}. 

It is crucial to consider such a contribution when comparing to measurements of exclusive meson pair production, which will lie dominantly in the lower mass region where this non--perturbative approach must be used. More generally, we may hope in the future to experimentally probe the transition between these two regimes, an issue which is still unclear, but may hopefully be clarified by future measurements of, for example, $\pi^+\pi^-$ CEP at the LHC. More details of this and the non--perturbative model used are considered in detail elsewhere~\cite{HarlandLang:2011qd,HarlandLang:2012qz}, and we will not consider this further here.

\subsection{Comparison with data}\label{mesoncomp}

Until recently, there was little existing data on meson pair CEP, in particular at high energies. This was limited to the ISR measurements~\cite{Breakstone:1989ty,Breakstone:1990at}, at reasonably low c.m.s energy ($\sqrt{s}=62$ GeV), which provide some constraint on the non--perturbative model of meson pair production but do not extend to sufficiently high values of meson transverse momentum, $k_\perp$, for the perturbative CEP formalism to be applicable. More recently a new CDF measurement of central $\pi^+\pi^-$ production at $\sqrt{s} = 900$ and 1960 GeV, which contain a large exclusive component has been reported~\cite{Albrow:2013mva,Mikeeds}. These data are in encouraging agreement with the non-perturbative model discussed above. We may expect further data on the CEP of meson pairs to be forthcoming from CMS~\cite{enterria}, CMS+Totem~\cite{CMSeds,Oljemark:2013wsa,Oljemarkeds}, ATLAS+ALFA~\cite{Staszewski:2011bg,Sykoraeds}, RHIC~\cite{Leszek}, and LHCb~\cite{Paula}.

It has been shown~\cite{Harland-Lang:2013dia} that the observation of exclusive meson pair production, in the presence of tagged protons, can act as a very sensitive test of the soft physics models used to calculate the survival factor, in a similar way to that described in Section~\ref{rhic}. In particular, a measurement of the distribution in azimuthal angle between the outgoing intact protons can provide a fully differential test of the soft survival factors. Such measurements are under consideration at the LHC, with the CMS+Totem~\cite{CMSeds,Oljemark:2013wsa,Oljemarkeds} and ATLAS+ALFA~\cite{Staszewski:2011bg,Sykoraeds} detectors, in particular during special low luminosity running conditions, and are already being made at RHIC by the STAR collaboration~\cite{Leszek} and by the COMPASS fixed--target experiment at CERN~\cite{fortheCOMPASS:2013vda}.

We may also hope that future measurements such as these will provide a further tests of the perturbative approach considered here. However, as discussed above, due to the suppression of the perturbative flavour--non--singlet meson ($\pi\pi$, $KK$...) cross sections, the dominantly `perturbative' phase space region is only expected to occur at relatively high values of $k_\perp$: in this case, the cross section are quite low and therefore the available statistics may be somewhat limited. A more promising observable may therefore be the CEP of flavour--singlet states ($\eta(')\eta(')$), which are not predicted to be suppressed in this way, and may therefore represent a more realistic experimental observable. As discussed above, the observation of this process, from a new analysis of the existing CDF data, as well as in forthcoming LHC data (in particular from CMS, CMS+TOTEM and LHCb) may also shed light on the important and uncertain issue concerning the size of the $gg$ component of the flavour--singlet $\eta,\eta'$ mesons.

\begin{figure}[h]
\begin{center}
\includegraphics[scale=0.3]{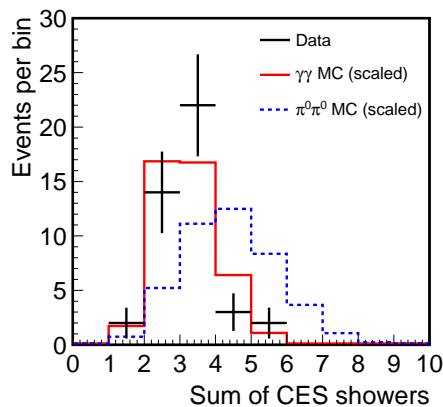}\qquad
\caption{Estimate of $\pi^0\pi^0$ background fraction in the candidate
data sample, taken from~\cite{Aaltonen:2011hi}. (a) Distribution of reconstructed CES showers per event for data
compared to $\gamma\gamma$ and $\pi^0\pi^0$ Monte Carlo simulations. 
}\label{pichit}
\end{center}
\end{figure}

One promising possibility is at the Tevatron, where we recall that CDF have set~\cite{Aaltonen:2011hi} the limit $N(\pi^0\pi^0)/N(\gamma\gamma)<0.35$, at 95\% CL on $\pi^0\pi^0$ CEP, by looking at the distribution of reconstructed proportional wire chamber (CES) showers per event, as shown in Fig.~\ref{pichit}. This lends support to the CEP framework, which we recall gives as a non--trivial prediction the strong suppression in the $\pi^0\pi^0$ cross section: without this $J_z=0$ suppression, we could certainly expect CDF to observe $\pi^0\pi^0$ events. In the future~\cite{Albrowpriv1} it is planned to extend this CDF study with increased statistics, and perhaps even to measure the $\pi^0\pi^0$ CEP cross section, as well as possibly the $\eta\eta$ cross section. While the $\eta\eta \to 4\gamma$ cross section is reduced by a factor of $15\%$ by the $\eta \to {\rm neutrals}$ branchings, and potentially further reduced by lower experimental efficiencies, it is still expected that the corresponding cross section, after accounting for this, will be of the same size and perhaps even larger than the corresponding $\pi^0\pi^0$ rate, see Table~\ref{etarats1}. In addition, we can expect further results from the LHC, with the  LHCb collaboration in particular having a promising CEP physics programme which includes the possibility of measuring these processes.

\section{Exclusive jet production}\label{cha:jets}

\noindent Exclusive jet production~\cite{Martin:1997kv,Khoze:2006iw}, and in particular the CEP of a dijet system
\begin{equation}
pp(\overline{p}) \to p\,+\,jj\,+\,p(\overline{p})\;,
\end{equation}
has been of great importance in testing the underlying Durham perturbative formalism. In 2008, the CDF collaboration reported~\cite{Aaltonen:2007hs} the observation and cross section measurement of this process using a data sample of $310$ ${\rm pb}^{-1}$, at $\sqrt{s}=1.96$ TeV and for $E_\perp^{\rm jet}>10$ GeV, selected by tagging the outgoing anti--proton and requiring a rapidity gap in the proton direction (which was not tagged). Crucially, they presented both dijet invariant mass $M_{jj}$ and jet transverse momenta $E_\perp^{\rm jet}$ distributions, out to quite high $M_{jj} \sim$ 130 GeV, and $E_\perp^{\rm jet} \sim 35$ GeV, and it was found that the perturbative approach of the Durham model, and in particular the characteristic fall--off with scale $\mu \sim M_{jj}$ produced by the Sudakov factor (\ref{tsp}), was essential to describe these distributions, with an overall good agreement between the Durham predictions and the CDF data. This observation was later supported by the measurement of the D0 collaboration~\cite{Abazov:2010bk}, which found evidence for exclusive dijet production with $M_{jj}>100$ GeV. The potential to probe a wide range of invariant masses, providing a differential test of the theory, in a similar fashion to $\gamma\gamma$ CEP but with much larger, $O({\rm nb})$, cross sections, therefore provides a strong motivation for studying this process. Moreover, as we will discuss, there are a range of interesting expectations for the gluon and quark jet cross sections in the exclusive mode which are quite different from those in the usual inclusive case~\cite{Khoze:2000jm,Khoze:2006um,Khoze:2006iw}.

On the other hand, although it is certainly an interesting channel, it is not without issues. In particular, an exclusive jet sample is in principle defined by requiring that a certain number of jets are present in the central detector, with no additional hadronic activity, and with outgoing intact protons (or anti--protons). However, it is not possible to uniquely assign particles to a given jet, and even for a genuinely exclusive event there will certainly in general be some remaining `unassigned' particles in the final state, due to additional radiation outside of the jets and experimental smearing effects, with the precise number depending on the details of the jet finding algorithm and experimental efficiencies. Moreover, there can be a significant background contribution from `inelastic' double Pomeron exchange (DPE), where the dijet system is produced by the collision of two Pomerons, but where there will in general be additional soft Pomeron remnants in the event~\cite{Khoze:2006iw}. However, if we consider the variable $R_{jj}=M_{jj}/M_X$, where $M_X$ is the invariant mass of the central system, then we will expect the exclusive signal to peak around $R_{jj} \sim 1$, while the inelastic background will populate a broad range of $R_{jj}$ values. This can then be used to extract the exclusive signal, in the high $R_{jj}$ region, provided the inelastic background can be suitably modelled, and this is precisely what was done in the CDF analysis.

Thus, exclusive jet production represents an interesting and novel QCD observable, and there is much potential to measure this process further at the LHC, in particular with both protons tagged using the installed and proposed forward proton spectrometers. The possibility of such measurements at the LHC with ATLAS+ALFA~\cite{Staszewski:2011bg} detectors is currently under consideration, while the first results on of a combined TOTEM+CMS measurement~\cite{Oljemark:2013wsa,CMSeds}  at 8 TeV, are expected to be available soon. In addition, a promising program of QCD studies, including jet production, is under discussion in the framework of the AFP~\cite{AFP,Tasevsky:2009zza,Royon:2013en,Royon:2013ala} and PPS~\cite{Albrow:2012oha,Albrow:2013yka} upgrade projects, which would allow an investigation of the region of centrally produced masses around 200--800 GeV, using proton detectors stationed at $\sim$220m  and $\sim$240m from the interaction points of ATLAS and CMS, respectively.

\subsection{Theory}

Exclusive dijet production is initiated by the colour--singlet $gg\to gg$ and $gg \to q\overline{q}$ subprocesses, for which the amplitudes are given by
\begin{align}\label{ggexc}
\mathcal{M}\left((g(\pm)g(\pm)\to g(\pm)g(\pm)\right) &= \delta^{CD}\frac{N_c}{N_c^2-1}\frac{32\pi\alpha_s}{(1-\cos^2\theta)}\;,\\ \label{qqexc0}
\mathcal{M}\left((g(\pm)g(\pm)\to q_h \overline{q}_{\bar{h}}\right) &=\frac{\delta^{cd}}{N_c}\frac{16\pi\alpha_s}{(1-\beta^2\cos^2\theta)}\frac{m_q}{M_X}(\beta h \pm 1)\delta_{h,\bar{h}}\;,\\ \label{qqexc2}
\mathcal{M}\left((g(\pm)g(\mp)\to q_h \overline{q}_{\bar{h}}\right) &=\pm h\frac{\delta^{cd}}{2N_c}8\pi \alpha_s \left(\frac{1\pm h \cos\theta}{1\mp h \cos\theta}\right)\delta_{h,-\bar{h}}\;,
\end{align}
for gluons of `$\pm$'  helicity and quarks of helicity $h$, while $c,d$ ($C,D$) are the outgoing quark (gluon) colour labels, $\beta=(1-4m_q^2/M_X^2)^{1/2}$ and $\theta$ is the scattering angle in the $gg$ rest frame. We can see that in the case of $q\overline{q}$ production the $J_z=0$ amplitude (\ref{qqexc0}) involves a helicity flip along the quark line, and vanishes as the quark mass $m_q \to 0$, and thus we expect a strong suppression in the CEP cross section for quark dijets. Considering for example the case of $b$--jets at $M_X=100$ GeV, we find
\begin{equation}
\frac{{\rm d}\sigma(b\overline{b})/{\rm d}t}{{\rm d}\sigma(gg)/{\rm d}t}\approx \frac{N_c^2-1}{4N_c^3}\frac{m_b^2}{M_X^2}\approx 10^{-4}\;,
\end{equation}
while for the inclusive case, this ratio is much larger. As we will discuss in Section~\ref{cha:higgs} this result is of great importance in the case of exclusive Higgs production via the $b\overline{b}$ mode, for which the direct QCD background is therefore expected to be suppressed~\cite{Khoze:2000jm,Heinemeyer:2007tu,Shuvaev:2008yn}. For lighter quark jets, the dominant contribution will come from the $|J_z|=2$ amplitude (\ref{qqexc2}), for which
\begin{equation}
\frac{{\rm d}\sigma^{|J_z|=2}(q\overline{q})/{\rm d}t}{{\rm d}\sigma(gg)/{\rm d}t}\approx\frac{N_c^2-1}{16 N_c^3}\frac{\left\langle p_\perp^2\right\rangle^2}{\left\langle Q_\perp^2 \right\rangle^2} \sim 10^{-4}\;,
\end{equation}
for typical values of the average proton transverse momentum $p_\perp$ and the loop momentum $Q_\perp$, see Section~\ref{durth}. Thus we expect a universal and strong suppression in quark jets relative to the gluon case, to a much greater extent than in inclusive production. Results consistent with such a suppression are indeed seen in a CDF study~\cite{Aaltonen:2007hs} of a 200 ${\rm pb}^{-1}$ sample of $b$--tagged jets. Thus the exclusive mode offers the possibility to study almost purely gluonic and, crucially, isolated jets (produced by the collision of a colour--singlet $gg$ state) in a hadronic environment, shedding light on the underlying properties of these jets (such as multiplicity, particle correlations etc) in a well--defined and comparatively clean exclusive environment. 

If we now consider the case of three jet production, that is $q\overline{q}g$ and $ggg$ jets, this suppression in the $q\overline{q}$ exclusive dijet cross section also leads to some interesting predictions~\cite{Khoze:2006um,Khoze:2009er}. In particular, we expect the behaviour of the $q\overline{q}g$ amplitude as the radiated gluon becomes soft to be governed by the corresponding Born--level, in this case $q\overline{q}$, amplitude. More precisely, the Low--Burnett--Kroll~\cite{Low:1958sn,Burnett:1967km} theorem tells us that for a soft gluon, carrying momentum fraction $x_g = 2 E_g/M_X\ll1$, the radiative amplitude $M_{q\overline{q}g}$ may be expanded in powers of $x_g$ as
\begin{equation}
M_{ q\overline{q}g}=\frac{1}{x_g}\sum^\infty_{n=0} C_n x_g^n\; ,
\end{equation}
where crucially both the first and second terms, $C_0$ and $C_1$, are given in terms of the the Born--level amplitude $M_{q\overline{q}}$. Thus the first non--vanishing term in the case of $J_z=0$ incoming gluons occurs for $n=2$, giving a cross section which behaves like
\begin{equation}
\frac{{\rm d}\sigma(J_z=0)}{{\rm d}E_g}\sim E_g^3\;,
\end{equation}
in the massless quark limit\footnote{Re--introducing a non--zero quark mass will give the usual infrared behaviour $\sim \frac{m_q^2}{M_X^2} \frac{1}{E_g}$ for very low $x_g$.}. This is to be contrasted with the inclusive, unpolarised case, for which we have the usual singular behaviour
\begin{equation}
\frac{{\rm d}\sigma}{{\rm d}E_g}\sim \frac{1}{E_g}\;,
\end{equation}
which will also occur in the case of $|J_z|=2$ incoming gluons, as well as in the case of $ggg$ jets, for which the corresponding Born--level ($gg$) amplitudes do not vanish. Thus we expect a quite distinct behaviour in the gluon energy distribution for $x_g \ll 1$, and we may in particular expect an enhancement of `Mercedes--like' configurations for the $q\overline{q}g$ case, where all three partons carry roughly equal energies and are well separated. More generally, it would be of much interest to investigate the difference in the predicted event shape variables (thrust, sphericity etc), which may be quite different between the experimentally distinguishable $b\overline{b}g$ and $ggg$ cases, as well as to the corresponding inclusive cases.

\subsection{Numerical results and comparison with data}

\begin{figure}[t]
\begin{center}
\includegraphics[scale=0.4]{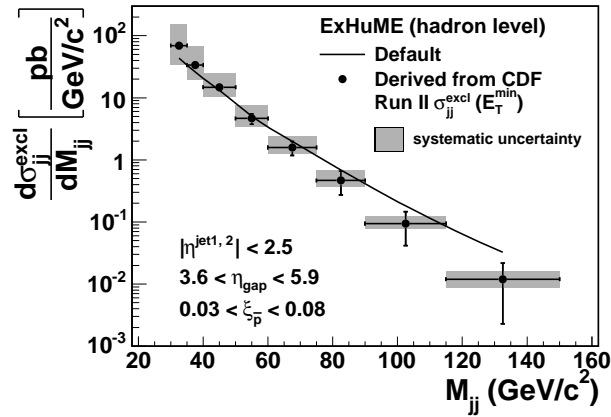}
\caption{Comparison of the \texttt{ExHuME} MC implementation~\cite{Monk:2005ji} of the Durham model and CDF data~\cite{Aaltonen:2007hs}, for exclusive dijet invariant mass distribution at the hadron--level, for $R_{jj}>0.8$. Plot taken from CDF~\cite{Aaltonen:2007hs} publication.}
\label{cdfjet}
\end{center}
\end{figure}

As discussed above, in 2008, the CDF collaboration reported~\cite{Aaltonen:2007hs} the observation and cross section measurement of exclusive dijet production using a data sample of $310$ ${\rm pb}^{-1}$, at $\sqrt{s}=1.96$ TeV and for $E_\perp^{\rm jet}>10$ GeV. It was found that both the jet $E_\perp$ and $M_{jj}$ distributions, over a range of values, were quite well described by the Durham model, with in particular the characteristic fall--off with scale $\mu \sim M_{jj}$ produced by the Sudakov factor (\ref{tsp}), being essential to describe these distributions. In Fig.~\ref{cdfjet} we show the measurement of the dijet invariant mass distribution, taken from the CDF publication~\cite{Aaltonen:2007hs}, compared to the \texttt{ExHuME} MC implementation~\cite{Monk:2005ji} of the Durham model. We can see that the agreement is good, although potentially with some discrepancy at higher $M_{jj}$: however, as discussed elsewhere~\cite{Coughlin:2009tr},  a more careful treatment of the limits on the Sudakov factor (\ref{tsp}) than was included in this MC and in earlier Durham papers, is expected to improve the agreement. 

We note that a full treatment of both exclusive dijet and trijet production, and a new MC implementation of these processes is currently the subject of ongoing work~\cite{HLfut}, and so will not present further predictions here. However, we have seen that exclusive jet production is a very interesting and rich topic for further theoretical work and experimental investigation, in particular at the LHC. Indeed, already a sample of `exclusive--like' dijet and trijet events has been collected in a combined CMS+TOTEM run~\cite{Oljemark:2013wsa,CMSeds} at 8 TeV, with results expected to be released soon. The first public event displays from these event are remarkably clean for such a hadronic environment, and are reminiscent of LEP jet events~\cite{Albrow:2013nia}.

\section{The Higgs boson}\label{cha:higgs}

\noindent Over the last decade there has been a steady interest in the CEP of new physics objects at the LHC~\cite{Khoze:2001xm,FP420,Maciula:2010tv,Albrow:2010yb,Heinemeyer:2011jb,AFP,Royon:2013en,Tasevsky:2009zza,Albrow:2012oha,Albrow:2013yka}, with the possibility of a simultaneous detection of the forward protons, using dedicated very forward detectors, and the central system opening up a window to a rich physics program, covering a variety of QCD, Electroweak and BSM processes~\cite{Khoze:2010ba,deFavereaudeJeneret:2009db,Tasevsky:2013iea}. One particularly interesting example that has received a great deal of attention  is the CEP of the Higgs boson(s).  This has played a central role in the physics targets of the FP420 LHC  project~\cite{FP420}, which proposes to complement the ATLAS and CMS experiments by additional near--beam proton detectors, located 420m away from the interaction region. This subject still remains topical even after the discovery by the CMS and ATLAS experiments of a new boson~\cite{Chatrchyan:2012ufa,Chatrchyan:2013lba,Aad:2012tfa} with a mass near 125 GeV and with production rates, decay rates, and spin-parity assignement compatible with those expected for the standard model (SM) Higgs boson. Indeed, the forward proton technique is exceptionally well suited to the investigation of crucial identification issues such as the  $CP$--parity and the $b\overline{b}$ coupling of the recently discovered object~\cite{Khoze:2001xm, Kaidalov:2003ys,Khoze:2004rc}. This approach is complementary to the mainstream strategies at the LHC, and could be useful in the study of other Higgs--like particles expected in some BSM theories~\cite{Heinemeyer:2007tu,Chaichian:2009ts,Heinemeyer:2010gs}. It is also worth recalling that the observation of even a few events corresponding to the CEP of a Higgs--like particle would confirm its $0^{++}$ nature, with the $0^{-+}$, $2^{-+}$ and $2^{++}$ assignments  (in the latter case for minimal coupling to gluons)  being strongly disfavoured~\cite{Kaidalov03,Khoze:2001xm,Khoze:2002nf}.

A further interesting possibility is the study of correlations between the outgoing proton momenta in the CEP mode, which would provide a unique opportunity to hunt for $CP$--violation effects in the Higgs sector~\cite{Khoze:2004rc}, which it should be emphasised would constitute an indisputable sign of physics beyond the SM. The contribution caused by the $CP$-odd term in the $gg \rightarrow H$ vertex is proportional to the triple-product correlation between the beam direction and the momenta of outgoing detected protons~\cite{Khoze:2004rc,kmrcp}, and in some $CP$-violating BSM scenarios~\cite{Ellis:2006eh,Carena:2000ks} the integrated counting asymmetry (based on counting events with $\phi > \pi$ and with $\phi < \pi$) can be sizeable.

An important advantage of the forward proton approach is the fact that it allows the largest decay modes,  $b\overline{b}$, $WW$ and $\tau\tau$ to be detected via a single production channel. For example, the normally challenging dominant $b\overline{b}$ decay mode for a light Higgs boson would become much more easily accessible in the CEP case, as the direct QCD background is strongly dynamically suppressed, see Section~\ref{cha:jets}. The observation of Higgs production via this decay mode is of much importance, for example because a precise knowledge of the bottom Yukawa coupling would also be crucial as an input in the determination of the Higgs couplings to other particles~\cite{Duhrssen:2004cv,LHCHiggsX1,LHCHiggsX2}.

\subsection{The Standard Model Higgs: predictions}

\begin{figure}
\begin{center}
\includegraphics[scale=0.7]{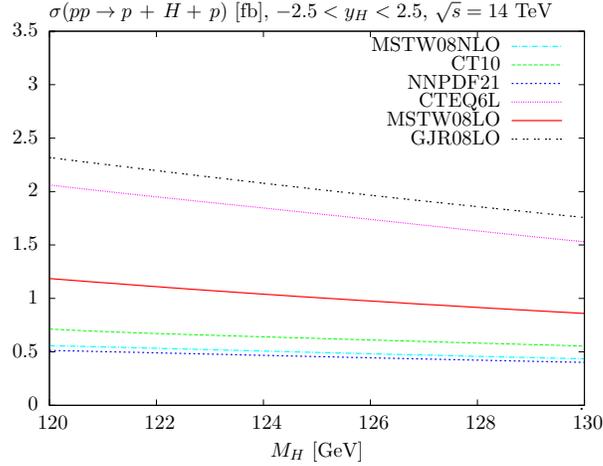}
\caption{Cross section for SM Higgs CEP as a function of the Higgs mass, $M_H$, integrated over the rapidity interval $-2.5<y_H<2.5$, for a range of PDFs (GJR08LO~\cite{Gluck:2007ck}, MSTW08LO and NLO~\cite{Martin:2009iq}, CTEQ6L~\cite{Pumplin:2002vw}, CT10~\cite{Lai:2010vv} and NNPDF2.1~\cite{Ball:2010de}). NLO K--factor included.}\label{fig:Hl}
\end{center}
\end{figure}

\begin{figure}
\begin{center}
\includegraphics[scale=0.7]{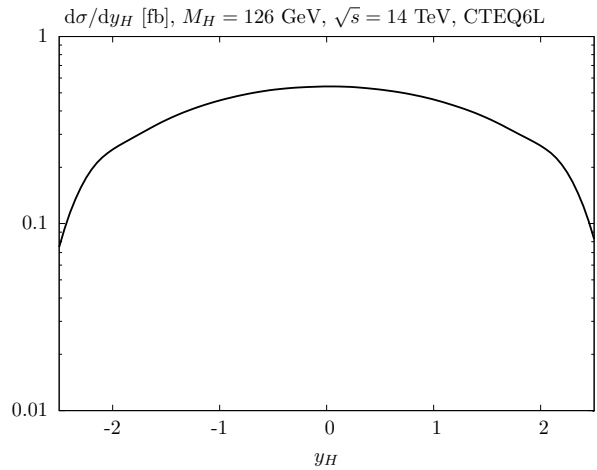}
\caption{Rapidity distribution ${\rm d}\sigma/{\rm d}y_H$ for a $M_H=126$ GeV SM Higgs boson, using CTEQ6L PDFs.}\label{fig:Hr}
\end{center}
\end{figure}

\begin{figure}
\begin{center}
\includegraphics[scale=0.7]{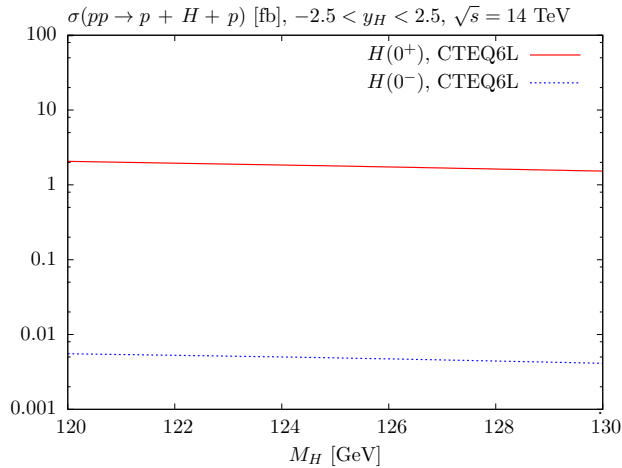}
\caption{Cross sections for the CEP of scalar $J^P=0^+$ and pseudoscalar $J^P=0^-$ particles of the Higgs sector as a function of the Higgs mass, $M_H$, integrated over the rapidity interval $-2.5<y_H<2.5$.}\label{fig:H1}
\end{center}
\end{figure}

The expectations~\cite{HarlandLang:2013jf} for the CEP of the SM Higgs boson at 14 TeV are illustrated in Figs.~\ref{fig:Hl},~\ref{fig:Hr} and~\ref{fig:H1}. For the combined enhanced\footnote{In this mass and $\sqrt{s}$ region, the suppression due to $S^2_{\rm enh}$ is expected to be weak~\cite{Ryskin:2011qe}.} and eikonal soft survival factor we take $\left \langle S^2\right\rangle=0.01$, although there is some important uncertainty in this value, and it may in particular be somewhat smaller. On the other hand~\cite{HarlandLang:2012qz} we may also expect higher--order corrections to increase the cross section by a factor of $\sim 2$ or so. We have also seen that for the LO PDFs, which give the larger cross sections in Fig.~\ref{fig:Hl}, there is good agreement with the CDF $\gamma\gamma$ data~\cite{Aaltonen:2011hi}, with the CTEQ6L~\cite{Pumplin:2002vw} set giving the closest value, see Section~\ref{gamnumer}. In Fig.~\ref{fig:Hr} we show the corresponding Higgs rapidity distribution for the CTEQ6L PDF set, for $M_H=126$ GeV. In Fig.~\ref{fig:H1} we show the cross section for the case of a scalar $J^P=0^+$ and pseudoscalar $J^P=0^-$ particle of the Higgs sector, using CTEQ6L PDFs. As expected from the $J^P_z=0^+$ selection rule~\cite{Khoze:2001xm,Khoze:2000jm}, the cross section in the case of the scalar state is much ($\sim 2$ orders of magnitude) larger. While the predicted scalar Higgs cross sections are quite small ($\sim$ fb), we recall that the CEP process provides an exceptionally clean and complementary handle on the properties of a Higgs or Higgs--like particle. 

\subsection{The MSSM Higgs}

The MSSM is one of the most widely studied BSM scenarios, and the CEP of the MSSM Higgs bosons at the LHC has in the past been the subject of detailed studies~\cite{Kaidalov:2003ys,Cox:2007sw, Heinemeyer:2007tu,Heinemeyer:2010gs}. Previously, in particular prior to the very successful Run I data taking  at the LHC, there were quite encouraging prospects for probing the MSSM Higgs sector in the forward proton mode~\cite{Heinemeyer:2007tu,Heinemeyer:2010gs,Heinemeyer:2012hr,FP420}, with the expected CEP Higgs rate strongly exceeding that for the SM light Higgs, in some particularly  promising regions in the MSSM $M_A-\tan\beta$ parameter space. However, following LHC run I, these regions are now excluded by a combination of the existing experimental bounds. A recent detailed analysis \cite{Tasevsky:2013iea} based on seven new low--energy MSSM benchmark scenarios\cite{Carena:2013qia}, has accounted for the current compilation of the LHC MSSM Higgs boson searches, and we refer the  interested reader to this study for various results on the signal cross sections, ratios of signal to background (S/B) and statistical significances for the $h/H \to b \bar b$  decays, as well as an account of  the experimental procedures and cut selection for the case that the proposed forward proton detectors are installed at ATLAS and/or CMS. In particular, one immediate observation which follows from this previous work\cite{Tasevsky:2013iea,HarlandLang:2013jf} is that the $h$ CEP yield is only weakly dependent on $M_A$, giving a cross section level of around 1 fb  (up to a factor of $\sim$ 2 theoretical uncertainty). Thus,  (contrary to earlier expectations~\cite{Kaidalov:2003ys}) the event rate for the MSSM $h$--boson cannot be sizeably higher than that for the SM Higgs. 

Considering now the heavy MSSM $H$--boson, the situation is not optimistic. Accounting for the recent LHC data and low--energy observables, and assuming that the newly observed state is a light MSSM $h$--boson, the preferred values~\cite{Bechtle:2012jw} of the heavy neutral Higgs masses are comparatively large (exceeding 250 GeV or so), which is within the acceptance of the 220--240m forward proton detectors~\cite{AFP,Royon:2013en,Tasevsky:2009zza,Albrow:2012oha}. However, the effective Pomeron--Pomeron luminosity $L^{\rm eff}$ for Higgs boson CEP decreases rapidly with the Higgs mass $M$, being given approximately by~\cite{Khoze:2001xm,Kaidalov:2003ys}
\begin{equation}
L^{\rm eff} \quad \propto \quad 1{\Big /}(M + 16~{\rm GeV})^{3.3}\;.
\label{eqH}
\end{equation}
Including the other mass dependent factors~\cite{Heinemeyer:2007tu}, we find that for a $H$--boson mass of 300 (400) GeV in the new benchmark scenarios,  the expected CEP cross section, after accounting for the experimental acceptances and efficiencies~\cite{Heinemeyer:2007tu,Tasevsky:2013iea}, are too small to produce a detectable signal within a reasonable time scale for making use of forward proton detectors\footnote{In addition, we have to keep in mind that at higher LHC luminosities the pile--up background could cause a severe problem for the Higgs CEP measurements, even if/when the fast timing detectors with precision vertex resolution~\cite{AFP,Royon:2013en,Tasevsky:2009zza,Albrow:2012oha} are installed.}. This conclusion may of course not be true for all other BSM Higgs scenarios, some of which might be more favourable. Moreover~\cite{Tasevsky:2013iea} there is still some room for improvement of the experimental techniques, e.g. the expected improvement of the gluon-b misidentification probability $P_{g/b}$ compared to the 1.3\% that was assumed previously, a sub-10~ps resolution in the timing detectors, or the use of multivariate techniques.

\section{Conclusion and Outlook}\label{conc}

In this article we have presented a review of recent studies, performed by the authors, of central exclusive production (CEP) within the Durham model, which combines a perturbative QCD based approach to model the hard production process of an object $X$ accompanied by no perturbative emission, with a model of soft physics to describe the so--called survival factor, which gives the probability that additional particles are not produced due to soft rescatterings. CEP is a quite generic mechanism, which can in principle produce any object that couples to gluons; consequently, a wide range of processes have been considered in the literature, from SM mesons to BSM Higgs bosons. In this review we have concentrated on the so--called `Standard Candle' SM processes, in particular the CEP of jets, diphotons $\gamma\gamma$, heavy $(c,b)$ quarkonia, new charmonium--like states, and meson pairs. These have sufficiently large production cross sections that they can readily be measured experimentally, thus providing a test of the Durham framework and validating predictions for new physics objects. Moreover, these processes are also of interest in their own right, and we have emphasised in this review the range of interesting theoretical features that these possess, as well as the possibilities for the clean exclusive environment to shed further light on the properties of these SM states. In addition, we have discussed the CEP of SM and BSM Higgs bosons, emphasising the implications of recent LHC data for future measurements.

A wealth of experimental measurements of high--energy CEP have been made, both at the Tevatron~\cite{Mike}, in Run I of the LHC\cite{Ronan,daSilveira:2013aaa,Zamora:2013dna,Reidt:2013kz}, as well as at RHIC, where forward proton taggers are already installed\cite{Leszek}. We have seen that these are in reasonably good agreement with the Durham expectations, but that there certainly remain some important theoretical uncertainties and unresolved issues, which future LHC measurements and analysis can clarify.

 In the future, we can expect to see an interesting and very promising program of experimental CEP studies. As discussed in the Introduction, this includes the possibility of CEP measurements with tagged protons, using the installed and proposed forward proton spectrometers. Of particular interest are the proposed PPS\cite{Albrow:2013yka} and the AFP\cite{Royon:2013en,AFP} upgrade projects, which would allow an investigation of the region of centrally produced masses around 200--800 GeV, in particular during high luminosity running. One potential measurement would be exclusive jet production, which represents an interesting and novel QCD observable sensitive to the basic ingredients of the perturbative formalism\cite{Khoze:2008cx}. A second stage with proton detectors at $\sim$420m would then allow the observation of exclusive Higgs boson production. During low--pile up runs, the addition of FSCs to the CMS detector will greatly increase the efficiency for selecting exclusive events; such detectors are also being installed at LHCb, in the so-called HERSHEL Project \cite{Ronan,Paula2}.

The study of CEP in high--energy hadronic collisions is highly topical and of great experimental and theoretical interest. The Durham program is ongoing, with further studies and developments of the available MC tools under way. We also can expect many more measurements to come, in particular from the LHC. There will therefore be many more interesting exclusive results to come in the future, and we look forward to new exciting adventures in Exclusiveland.

\section*{Acknowledgements}

The authors thank Mike Albrow, Marius Bj\o rnstad, Erik Brucken, Victor Chernyak, Paula Collins, Wlodek Guryn,
Jeff Forshaw, Ronan McNulty, Dermot Moran, Jim Pinfold, Risto Orava, Antoni Szczurek, Marek Tasevsky and Guy Wilkinson for useful discussions. 
 MGR and WJS thank the IPPP at the University of Durham for their kind hospitality. LHL and WJS acknowledge support from the Cavendish Laboratory, University of Cambridge, with which they were affiliated when performing much of the work reviewed in this article. This work was supported by the Federal Program of the Russian State RSGSS-4801.2012.2.

\bibliographystyle{ws-ijmpa}
\bibliography{references}

\begin{thebibliography}{100}

\bibitem{Khoze:2001xm}
V.~A. Khoze, A.~D. Martin and M.~G. Ryskin, {\em Eur.Phys.J.} {\bf C23}, 311
  (2002), \href{http://arxiv.org/abs/hep-ph/0111078}{{\ttfamily
  arXiv:hep-ph/0111078 [hep-ph]}}.

\bibitem{FP420}
 FP420 R \& D Collaboration (M.~G. Albrow {\em et~al.}), {\em JINST} {\bf 4},
  T10001  (2009), \href{http://arxiv.org/abs/0806.0302}{{\ttfamily
  arXiv:0806.0302 [hep-ex]}}.

\bibitem{Martin:2009ku}
A.~D. Martin, M.~G. Ryskin and V.~A. Khoze, {\em Acta Phys.Polon.} {\bf B40},
  1841  (2009), \href{http://arxiv.org/abs/0903.2980}{{\ttfamily
  arXiv:0903.2980 [hep-ph]}}.

\bibitem{Albrow:2010yb}
M.~G. Albrow, T.~D. Coughlin and J.~R. Forshaw, {\em Prog.Part.Nucl.Phys.} {\bf
  65}, 149  (2010), \href{http://arxiv.org/abs/1006.1289}{{\ttfamily
  arXiv:1006.1289 [hep-ph]}}.

\bibitem{Albrow:2013yka}
M.~Albrow  (2013), \href{http://arxiv.org/abs/1310.4529}{{\ttfamily
  arXiv:1310.4529 [physics.ins-det]}}.

\bibitem{Royon:2013ala}
C.~Royon, {\em J.Phys.Conf.Ser.} {\bf 455},   012055  (2013),
  \href{http://arxiv.org/abs/1305.0652}{{\ttfamily arXiv:1305.0652 [hep-ph]}}.

\bibitem{HarlandLang:2013jf}
L.~A. Harland-Lang, V.~A. Khoze, M.~G. Ryskin and W.~J. Stirling  (2013),
  \href{http://arxiv.org/abs/1301.2552}{{\ttfamily arXiv:1301.2552 [hep-ph]}}.

\bibitem{Ronan}
R.~McNulty [on behalf of the LHCb collaboration], `Diffractive and Forward
  Physics at LHCb', talk at the school on Diffractive and Electromagnetic
  Processes at High Energies, Heidelberg, 2-6 September 2013.

\bibitem{Kaidalov03}
A.~B. Kaidalov, V.~A. Khoze, A.~D. Martin and M.~G. Ryskin, {\em Eur. Phys. J.}
  {\bf C31}, 387  (2003), \href{http://arxiv.org/abs/hep-ph/0307064}{{\ttfamily
  arXiv:hep-ph/0307064}}.

\bibitem{Heinemeyer:2007tu}
S.~Heinemeyer, V.~A. Khoze, M.~G. Ryskin, W.~J. Stirling, M.~Tasevsky {\em
  et~al.}, {\em Eur.Phys.J.} {\bf C53}, 231  (2008),
  \href{http://arxiv.org/abs/0708.3052}{{\ttfamily arXiv:0708.3052 [hep-ph]}}.

\bibitem{HarlandLang:2010ep}
L.~A. Harland-Lang, V.~A. Khoze, M.~G. Ryskin and W.~J. Stirling, {\em
  Eur.Phys.J.} {\bf C69}, 179  (2010),
  \href{http://arxiv.org/abs/1005.0695}{{\ttfamily arXiv:1005.0695 [hep-ph]}}.

\bibitem{SuperCHIC}
The SuperCHIC code and documentation are available at {\tt
  http://projects.hepforge.org/superchic/}.

\bibitem{Mike}
M.~Albrow , `Central Exclusive Production at Hadron Colliders', lecture at the
  school on Diffractive and Electromagnetic Processes at High Energies,
  Heidelberg, 2-6 September 2013.

\bibitem{daSilveira:2013aaa}
 CMS Collaboration (G.~G. da~Silveira)  (2013),
  \href{http://arxiv.org/abs/1310.5327}{{\ttfamily arXiv:1310.5327 [hep-ex]}}.

\bibitem{Zamora:2013dna}
 ALICE Collaboration (P.~G. Zamora), {\em EPJ Web Conf.} {\bf 60},   20003
  (2013).

\bibitem{Reidt:2013kz}
 ALICE Collaboration (F.~Reidt), {\em AIP Conf.Proc.} {\bf 1523}, 17  (2012),
  \href{http://arxiv.org/abs/1301.3507}{{\ttfamily arXiv:1301.3507 [hep-ex]}}.

\bibitem{Leszek}
Leszek Adamczyk, talk at the Workshop 15th conference on Elastic and
  Diffractive scattering, EDS Blois 2013 Workshop, Saariselka, Lapland,
  September 9-13.

\bibitem{Staszewski:2011bg}
R.~Staszewski, P.~Lebiedowicz, M.~Trzebinski, J.~Chwastowski and A.~Szczurek,
  {\em Acta Phys.Polon.} {\bf B42},   1961  (2011),
  \href{http://arxiv.org/abs/1104.3568}{{\ttfamily arXiv:1104.3568 [hep-ex]}}.

\bibitem{Oljemarkeds}
Fredrik Oljemark, talk at the Workshop 15th conference on Elastic and
  Diffractive scattering, EDS Blois 2013 Workshop, Saariselka, Lapland,
  September 9-13.

\bibitem{Oljemark:2013wsa}
 TOTEM Collaboration (F.~Oljemark)  (2013),
  \href{http://arxiv.org/abs/1310.4305}{{\ttfamily arXiv:1310.4305 [hep-ex]}}.

\bibitem{CMSeds}
Christina Mesropian (CMS Collaboration) , talk at the Workshop 15th conference
  on Elastic and Diffractive scattering, EDS Blois 2013 Workshop, Saariselka,
  Lapland, September 9-13.

\bibitem{Antchev:2013hya}
 TOTEM Collaboration (G.~Antchev {\em et~al.}), {\em Int.J.Mod.Phys.} {\bf
  A28},   1330046  (2013), \href{http://arxiv.org/abs/1310.2908}{{\ttfamily
  arXiv:1310.2908 [physics.ins-det]}}.

\bibitem{Royon:2013en}
C.~Royon  (2013), \href{http://arxiv.org/abs/1302.0623}{{\ttfamily
  arXiv:1302.0623 [physics.ins-det]}}.

\bibitem{AFP}
The AFP project in ATLAS, Letter of Intent of the Phase-I Upgrade (ATLAS
  Collab.), {\tt http://cdsweb.cern.ch/record/1402470}.

\bibitem{Ponzo}
Aldo Ponzo , talk at `Forward Physics at the LHC', Reggio Calabria, Jul. 15-18,
  2013.

\bibitem{Paula2}
Paula Collins, talk at `Forward Physics at the LHC', Reggio Calabria, Jul.
  15-18, 2013.

\bibitem{Khoze97}
V.~A. Khoze, A.~D. Martin and M.~G. Ryskin, {\em Phys. Lett.} {\bf B401}, 330
  (1997), \href{http://arxiv.org/abs/hep-ph/9701419}{{\ttfamily
  arXiv:hep-ph/9701419}}.

\bibitem{Khoze00}
V.~A. Khoze, A.~D. Martin and M.~G. Ryskin, {\em Eur. Phys. J.} {\bf C14}, 525
  (2000), \href{http://arxiv.org/abs/hep-ph/0002072}{{\ttfamily
  arXiv:hep-ph/0002072}}.

\bibitem{Harland-Lang:2013xba}
L.~A. Harland-Lang, {\em Phys.Rev.} {\bf D88},   034029  (2013),
  \href{http://arxiv.org/abs/1306.6661}{{\ttfamily arXiv:1306.6661 [hep-ph]}}.

\bibitem{Gluck94}
M.~Gluck, E.~Reya and A.~Vogt, {\em Z.Phys.} {\bf C67}, 433  (1995).

\bibitem{Martin:2009iq}
A.~D. Martin, W.~J. Stirling, R.~S. Thorne and G.~Watt, {\em Eur.Phys.J.} {\bf
  C63}, 189  (2009), \href{http://arxiv.org/abs/0901.0002}{{\ttfamily
  arXiv:0901.0002 [hep-ph]}}.

\bibitem{Pumplin:2002vw}
J.~Pumplin, D.~R. Stump, J.~Huston, H.~L. Lai, P.~M. Nadolsky {\em et~al.},
  {\em JHEP} {\bf 0207},   012  (2002),
  \href{http://arxiv.org/abs/hep-ph/0201195}{{\ttfamily arXiv:hep-ph/0201195
  [hep-ph]}}.

\bibitem{Martin:1999ww}
A.~D. Martin, R.~Roberts, W.~J. Stirling and R.~S. Thorne, {\em Eur.Phys.J.}
  {\bf C14}, 133  (2000), \href{http://arxiv.org/abs/hep-ph/9907231}{{\ttfamily
  arXiv:hep-ph/9907231 [hep-ph]}}.

\bibitem{Ryskin:1992ui}
M.~G. Ryskin, {\em Z.Phys.} {\bf C57}, 89  (1993).

\bibitem{Belitsky:2005qn}
A.~V. Belitsky and A.~V. Radyushkin, {\em Phys.Rept.} {\bf 418}, 1  (2005),
  \href{http://arxiv.org/abs/hep-ph/0504030}{{\ttfamily arXiv:hep-ph/0504030
  [hep-ph]}}.

\bibitem{Shuvaev:1999ce}
A.~Shuvaev, K.~J. Golec-Biernat, A.~D. Martin and M.~G. Ryskin, {\em Phys.Rev.}
  {\bf D60},   014015  (1999),
  \href{http://arxiv.org/abs/hep-ph/9902410}{{\ttfamily arXiv:hep-ph/9902410
  [hep-ph]}}.

\bibitem{Coughlin:2009tr}
T.~D. Coughlin and J.~R. Forshaw, {\em JHEP} {\bf 1001},   121  (2010),
  \href{http://arxiv.org/abs/0912.3280}{{\ttfamily arXiv:0912.3280 [hep-ph]}}.

\bibitem{Bjorken:1992er}
J.~Bjorken, {\em Phys.Rev.} {\bf D47}, 101  (1993).

\bibitem{Khoze:2006uj}
V.~A. Khoze, A.~D. Martin and M.~G. Ryskin, {\em JHEP} {\bf 0605},   036
  (2006), \href{http://arxiv.org/abs/hep-ph/0602247}{{\ttfamily
  arXiv:hep-ph/0602247 [hep-ph]}}.

\bibitem{Ryskin:2009tk}
M.~G. Ryskin, A.~D. Martin and V.~A. Khoze, {\em Eur.Phys.J.} {\bf C60}, 265
  (2009), \href{http://arxiv.org/abs/0812.2413}{{\ttfamily arXiv:0812.2413
  [hep-ph]}}.

\bibitem{Ostapchenko:2010gt}
S.~Ostapchenko, {\em Phys.Rev.} {\bf D81},   114028  (2010),
  \href{http://arxiv.org/abs/1003.0196}{{\ttfamily arXiv:1003.0196 [hep-ph]}}.

\bibitem{Gotsman:2012rq}
E.~Gotsman, E.~Levin and U.~Maor, {\em Phys.Rev.} {\bf D85},   094007  (2012),
  \href{http://arxiv.org/abs/1203.2419}{{\ttfamily arXiv:1203.2419 [hep-ph]}}.

\bibitem{Khoze:2013dha}
V.~A. Khoze, A.~D. Martin and M.~G. Ryskin, {\em Eur.Phys.J.} {\bf C73},   2503
   (2013), \href{http://arxiv.org/abs/1306.2149}{{\ttfamily arXiv:1306.2149
  [hep-ph]}}.

\bibitem{Khoze:2014aca}
V.~A. Khoze, A.~D. Martin and M.~G. Ryskin  (2014),
  \href{http://arxiv.org/abs/1402.2778}{{\ttfamily arXiv:1402.2778 [hep-ph]}}.

\bibitem{Gotsman:2014pwa}
E.~Gotsman, E.~Levin and U.~Maor  (2014),
  \href{http://arxiv.org/abs/1403.4531}{{\ttfamily arXiv:1403.4531 [hep-ph]}}.

\bibitem{Khoze:2013jsa}
V.~A. Khoze, A.~D. Martin and M.~G. Ryskin, {\em Eur.Phys.J.} {\bf C74},   2756
   (2014), \href{http://arxiv.org/abs/1312.3851}{{\ttfamily arXiv:1312.3851
  [hep-ph]}}.

\bibitem{Ryskin:2011qe}
M.~G. Ryskin, A.~D. Martin and V.~A. Khoze, {\em Eur.Phys.J.} {\bf C71},   1617
   (2011), \href{http://arxiv.org/abs/1102.2844}{{\ttfamily arXiv:1102.2844
  [hep-ph]}}.

\bibitem{Ryskin:2009tj}
M.~G. Ryskin, A.~D. Martin and V.~A. Khoze, {\em Eur.Phys.J.} {\bf C60}, 249
  (2009), \href{http://arxiv.org/abs/0812.2407}{{\ttfamily arXiv:0812.2407
  [hep-ph]}}.

\bibitem{Ryskin:2012az}
M.~G. Ryskin, A.~D. Martin and V.~A. Khoze, {\em Eur.Phys.J.} {\bf C72},   1937
   (2012), \href{http://arxiv.org/abs/1201.6298}{{\ttfamily arXiv:1201.6298
  [hep-ph]}}.

\bibitem{Khoze:2004rc}
V.~A. Khoze, A.~D. Martin and M.~G. Ryskin, {\em Eur.Phys.J.} {\bf C34}, 327
  (2004), \href{http://arxiv.org/abs/hep-ph/0401078}{{\ttfamily
  arXiv:hep-ph/0401078 [hep-ph]}}.

\bibitem{Khoze:2000mw}
V.~A. Khoze, A.~D. Martin and M.~G. Ryskin, 592  (2000),
  \href{http://arxiv.org/abs/hep-ph/0006005}{{\ttfamily arXiv:hep-ph/0006005
  [hep-ph]}}.

\bibitem{HarlandLang:2009qe}
L.~A. Harland-Lang, V.~A. Khoze, M.~G. Ryskin and W.~J. Stirling, {\em
  Eur.Phys.J.} {\bf C65}, 433  (2010),
  \href{http://arxiv.org/abs/0909.4748}{{\ttfamily arXiv:0909.4748 [hep-ph]}}.

\bibitem{HarlandLang:2010ys}
L.~A. Harland-Lang, V.~A. Khoze, M.~G. Ryskin and W.~J. Stirling, {\em
  Eur.Phys.J.} {\bf C71},   1545  (2011),
  \href{http://arxiv.org/abs/1011.0680}{{\ttfamily arXiv:1011.0680 [hep-ph]}}.

\bibitem{Khoze:2004yb}
V.~A. Khoze, A.~D. Martin, M.~G. Ryskin and W.~J. Stirling, {\em Eur.Phys.J.}
  {\bf C35}, 211  (2004), \href{http://arxiv.org/abs/hep-ph/0403218}{{\ttfamily
  arXiv:hep-ph/0403218 [hep-ph]}}.

\bibitem{Pasechnik:2009bq}
R.~S. Pasechnik, A.~Szczurek and O.~V. Teryaev, {\em Phys.Lett.} {\bf B680}, 62
   (2009), \href{http://arxiv.org/abs/0901.4187}{{\ttfamily arXiv:0901.4187
  [hep-ph]}}.

\bibitem{Pasechnik:2009qc}
R.~S. Pasechnik, A.~Szczurek and O.~V. Teryaev, {\em Phys.Rev.} {\bf D81},
  034024  (2010), \href{http://arxiv.org/abs/0912.4251}{{\ttfamily
  arXiv:0912.4251 [hep-ph]}}.

\bibitem{Pumplin:1993xk}
J.~Pumplin, {\em Phys.Rev.} {\bf D47}, 4820  (1993),
  \href{http://arxiv.org/abs/hep-ph/9301216}{{\ttfamily arXiv:hep-ph/9301216
  [hep-ph]}}.

\bibitem{Yuan01}
F.~Yuan, {\em Phys. Lett.} {\bf B510}, 155  (2001),
  \href{http://arxiv.org/abs/hep-ph/0103213}{{\ttfamily arXiv:hep-ph/0103213}}.

\bibitem{Petrov:2004nx}
V.~A. Petrov and R.~A. Ryutin, {\em JHEP} {\bf 0408},   013  (2004),
  \href{http://arxiv.org/abs/hep-ph/0403189}{{\ttfamily arXiv:hep-ph/0403189
  [hep-ph]}}.

\bibitem{Petrov:2004hh}
V.~A. Petrov, R.~A. Ryutin, A.~E. Sobol and J.~P. Guillaud, {\em JHEP} {\bf
  0506},   007  (2005), \href{http://arxiv.org/abs/hep-ph/0409118}{{\ttfamily
  arXiv:hep-ph/0409118 [hep-ph]}}.

\bibitem{Bzdak:2005rp}
A.~Bzdak, {\em Phys.Lett.} {\bf B619}, 288  (2005),
  \href{http://arxiv.org/abs/hep-ph/0506101}{{\ttfamily arXiv:hep-ph/0506101
  [hep-ph]}}.

\bibitem{Rangel:2006mm}
M.~Rangel, C.~Royon, G.~Alves, J.~Barreto and R.~B. Peschanski, {\em
  Nucl.Phys.} {\bf B774}, 53  (2007),
  \href{http://arxiv.org/abs/hep-ph/0612297}{{\ttfamily arXiv:hep-ph/0612297
  [hep-ph]}}.

\bibitem{Bodwin:1994jh}
G.~T. Bodwin, E.~Braaten and G.~P. Lepage, {\em Phys.Rev.} {\bf D51}, 1125
  (1995), \href{http://arxiv.org/abs/hep-ph/9407339}{{\ttfamily
  arXiv:hep-ph/9407339 [hep-ph]}}.

\bibitem{Brambilla:2004jw}
N.~Brambilla, A.~Pineda, J.~Soto and A.~Vairo, {\em Rev.Mod.Phys.} {\bf 77},
  1423  (2005), \href{http://arxiv.org/abs/hep-ph/0410047}{{\ttfamily
  arXiv:hep-ph/0410047 [hep-ph]}}.

\bibitem{Vairo:2009tn}
A.~Vairo  (2009), \href{http://arxiv.org/abs/0912.4422}{{\ttfamily
  arXiv:0912.4422 [hep-ph]}}.

\bibitem{Eichten:2007qx}
E.~Eichten, S.~Godfrey, H.~Mahlke and J.~L. Rosner, {\em Rev.Mod.Phys.} {\bf
  80}, 1161  (2008), \href{http://arxiv.org/abs/hep-ph/0701208}{{\ttfamily
  arXiv:hep-ph/0701208 [hep-ph]}}.

\bibitem{Danilkin:2009hr}
I.~V. Danilkin and Y.~A. Simonov, {\em Phys.Rev.} {\bf D81},   074027  (2010),
  \href{http://arxiv.org/abs/0907.1088}{{\ttfamily arXiv:0907.1088 [hep-ph]}}.

\bibitem{Braaten:2014ata}
E.~Braaten and J.~Russ  (2014),
  \href{http://arxiv.org/abs/1401.7352}{{\ttfamily arXiv:1401.7352 [hep-ex]}}.

\bibitem{Khoze:2002nf}
V.~A. Khoze, A.~D. Martin and M.~G. Ryskin, {\em Eur.Phys.J.} {\bf C24}, 581
  (2002), \href{http://arxiv.org/abs/hep-ph/0203122}{{\ttfamily
  arXiv:hep-ph/0203122 [hep-ph]}}.

\bibitem{Aaltonen:2009kg}
 CDF Collaboration (T.~Aaltonen {\em et~al.}), {\em Phys.Rev.Lett.} {\bf 102},
   242001  (2009), \href{http://arxiv.org/abs/0902.1271}{{\ttfamily
  arXiv:0902.1271 [hep-ex]}}.

\bibitem{LY1}
L.~D. Landau, {\em Dokl. Akad. Nauk SSSR} {\bf 60},   213  (1948).

\bibitem{Yang50}
C.-N. Yang, {\em Phys. Rev.} {\bf 77}, 242  (1950).

\bibitem{LHCb}
LHCb Collaboration, CERN-LHCb-CONF-2011-022.

\bibitem{moranthesis}
D.~Moran, CERN-THESIS-2011-209.

\bibitem{Herb:1977ek}
S.~W. Herb, D.~C. Hom, L.~M. Lederman, J.~C. Sens, H.~D. Snyder {\em et~al.},
  {\em Phys.Rev.Lett.} {\bf 39}, 252  (1977).

\bibitem{Aubert:2008ba}
 BABAR Collaboration (B.~Aubert {\em et~al.}), {\em Phys.Rev.Lett.} {\bf 101},
   071801  (2008), \href{http://arxiv.org/abs/0807.1086}{{\ttfamily
  arXiv:0807.1086 [hep-ex]}}.

\bibitem{Beringer:1900zz}
 Particle Data Group Collaboration (J.~Beringer {\em et~al.}), {\em Phys.Rev.}
  {\bf D86},   010001  (2012).

\bibitem{Guryn:2008ky}
 STAR Collaboration (W.~Guryn)  (2008),
  \href{http://arxiv.org/abs/0808.3961}{{\ttfamily arXiv:0808.3961 [nucl-ex]}}.

\bibitem{meson}
W. Guryn, `Present and Future of Central Production With STAR Detector at
  RHIC', talk at 11th International Workshop on Meson Production, Properties
  and Interaction, Cracow, Poland, 10 - 15 June 2010;\\ `Glueball Searches with
  the STAR Detector at RHIC', talk at the 50th Cracow School of Theoretical
  Physics, Zakopane, Poland, 9-19 June, 2010.

\bibitem{Lee:2010zzp}
 STAR Collaboration (J.~H. Lee), {\em PoS} {\bf DIS2010},   076  (2010).

\bibitem{Brambilla:2010cs}
N.~Brambilla, S.~Eidelman, B.~K. Heltsley, R.~Vogt, G.~Bodwin {\em et~al.},
  {\em Eur.Phys.J.} {\bf C71},   1534  (2011),
  \href{http://arxiv.org/abs/1010.5827}{{\ttfamily arXiv:1010.5827 [hep-ph]}}.

\bibitem{Braaten:2013oba}
E.~Braaten  (2013), \href{http://arxiv.org/abs/1310.1636}{{\ttfamily
  arXiv:1310.1636 [hep-ph]}}.

\bibitem{Chiochia:2014qva}
 CDF Collaboration, CMS Collaboration, D0 Collaboration, LHCb Collaboration
  (V.~Chiochia)  (2014), \href{http://arxiv.org/abs/1403.0823}{{\ttfamily
  arXiv:1403.0823 [hep-ex]}}.

\bibitem{Chen:2008xia}
 Belle Collaboration (K.-F. Chen {\em et~al.}), {\em Phys.Rev.} {\bf D82},
  091106  (2010), \href{http://arxiv.org/abs/0810.3829}{{\ttfamily
  arXiv:0810.3829 [hep-ex]}}.

\bibitem{Bondar:2011ev}
A.~Bondar, A.~Garmash, A.~Milstein, R.~Mizuk and M.~Voloshin, {\em Phys.Rev.}
  {\bf D84},   054010  (2011), \href{http://arxiv.org/abs/1105.4473}{{\ttfamily
  arXiv:1105.4473 [hep-ph]}}.

\bibitem{Navarra:2011xa}
F.~S. Navarra, M.~Nielsen and J.-M. Richard, {\em J.Phys.Conf.Ser.} {\bf 348},
   012007  (2012), \href{http://arxiv.org/abs/1108.1230}{{\ttfamily
  arXiv:1108.1230 [hep-ph]}}.

\bibitem{Aad:2011ih}
 ATLAS Collaboration (G.~Aad {\em et~al.}), {\em Phys.Rev.Lett.} {\bf 108},
  152001  (2012), \href{http://arxiv.org/abs/1112.5154}{{\ttfamily
  arXiv:1112.5154 [hep-ex]}}.

\bibitem{Abazov:2012gh}
 D0 Collaboration (V.~M. Abazov {\em et~al.}), {\em Phys.Rev.} {\bf D86},
  031103  (2012), \href{http://arxiv.org/abs/1203.6034}{{\ttfamily
  arXiv:1203.6034 [hep-ex]}}.

\bibitem{Ferretti:2014xqa}
J.~Ferretti, G.~Galat\`{a} and E.~Santopinto  (2014),
  \href{http://arxiv.org/abs/1401.4431}{{\ttfamily arXiv:1401.4431 [nucl-th]}}.

\bibitem{Kuhn79}
J.~H. Kuhn, J.~Kaplan and E.~G.~O. Safiani, {\em Nucl. Phys.} {\bf B157},   125
   (1979).

\bibitem{alekseev}
A.I. Alekseev, Sov. Phys. JETP 34 (1958) 826.

\bibitem{Pasechnik:2007hm}
R.~S. Pasechnik, A.~Szczurek and O.~V. Teryaev, {\em Phys.Rev.} {\bf D78},
  014007  (2008), \href{http://arxiv.org/abs/0709.0857}{{\ttfamily
  arXiv:0709.0857 [hep-ph]}}.

\bibitem{Pasechnik:2009if}
R.~S. Pasechnik, A.~Szczurek and O.~V. Teryaev, {\em PoS} {\bf EPS-HEP2009},
  335  (2009), \href{http://arxiv.org/abs/0909.4498}{{\ttfamily arXiv:0909.4498
  [hep-ph]}}.

\bibitem{Stein93}
E.~Stein and A.~Schafer, {\em Phys.Lett.} {\bf B300}, 400  (1993).

\bibitem{Peng95}
H.~A. Peng, Z.~M. He and C.~S. Ju, {\em Phys.Lett.} {\bf B351}, 349  (1995).

\bibitem{HarlandLang:2011qd}
L.~A. Harland-Lang, V.~A. Khoze, M.~G. Ryskin and W.~J. Stirling, {\em
  Eur.Phys.J.} {\bf C71},   1714  (2011),
  \href{http://arxiv.org/abs/1105.1626}{{\ttfamily arXiv:1105.1626 [hep-ph]}}.

\bibitem{KMRsoft}
V.~A. Khoze, A.~D. Martin and M.~G. Ryskin, {\em Eur.Phys.J.} {\bf C18}, 167
  (2000), \href{http://arxiv.org/abs/hep-ph/0007359}{{\ttfamily
  arXiv:hep-ph/0007359 [hep-ph]}}.

\bibitem{Ablikim:2012xi}
 BESIII Collaboration (M.~Ablikim {\em et~al.}), {\em Phys.Rev.} {\bf D85},
  112008  (2012), \href{http://arxiv.org/abs/1205.4284}{{\ttfamily
  arXiv:1205.4284 [hep-ex]}}.

\bibitem{HarlandLang:2012qz}
L.~A. Harland-Lang, V.~A. Khoze, M.~G. Ryskin and W.~J. Stirling  (2012),
  \href{http://arxiv.org/abs/1204.4803}{{\ttfamily arXiv:1204.4803 [hep-ph]}}.

\bibitem{Albrow:2013mva}
M.~Albrow, A.~Swiech and M.~Zurek  (2013),
  \href{http://arxiv.org/abs/1310.3839}{{\ttfamily arXiv:1310.3839 [hep-ex]}}.

\bibitem{Mikeeds}
Mike Albrow, talk at EDS Blois 2013 Workshop, Saariselka, Lapland, September
  9-13.

\bibitem{Harland-Lang:2013dia}
L.~A. Harland-Lang, M.~G. Khoze and M.~G. Ryskin  (2013),
  \href{http://arxiv.org/abs/1312.4553}{{\ttfamily arXiv:1312.4553 [hep-ph]}}.

\bibitem{Khoze:2000jm}
V.~A. Khoze, A.~D. Martin and M.~G. Ryskin, {\em Eur.Phys.J.} {\bf C19}, 477
  (2001), \href{http://arxiv.org/abs/hep-ph/0011393}{{\ttfamily
  arXiv:hep-ph/0011393 [hep-ph]}}.

\bibitem{Uehara:2009tx}
 Belle Collaboration (S.~Uehara {\em et~al.}), {\em Phys.Rev.Lett.} {\bf 104},
   092001  (2010), \href{http://arxiv.org/abs/0912.4451}{{\ttfamily
  arXiv:0912.4451 [hep-ex]}}.

\bibitem{Abe:2004zs}
 Belle Collaboration (K.~Abe {\em et~al.}), {\em Phys.Rev.Lett.} {\bf 94},
  182002  (2005), \href{http://arxiv.org/abs/hep-ex/0408126}{{\ttfamily
  arXiv:hep-ex/0408126 [hep-ex]}}.

\bibitem{Aubert:2007vj}
 BaBar Collaboration (B.~Aubert {\em et~al.}), {\em Phys.Rev.Lett.} {\bf 101},
   082001  (2008), \href{http://arxiv.org/abs/0711.2047}{{\ttfamily
  arXiv:0711.2047 [hep-ex]}}.

\bibitem{Albuquerque:2013owa}
R.~Albuquerque, J.~Dias, M.~Nielsen and C.~Zanetti  (2013),
  \href{http://arxiv.org/abs/1311.6411}{{\ttfamily arXiv:1311.6411 [hep-ph]}}.

\bibitem{Sreethawong:2013qua}
W.~Sreethawong, K.~Xu and Y.~Yan  (2013),
  \href{http://arxiv.org/abs/1306.2780}{{\ttfamily arXiv:1306.2780 [hep-ph]}}.

\bibitem{Choi:2003ue}
 Belle Collaboration (S.~K. Choi {\em et~al.}), {\em Phys.Rev.Lett.} {\bf 91},
   262001  (2003), \href{http://arxiv.org/abs/hep-ex/0309032}{{\ttfamily
  arXiv:hep-ex/0309032 [hep-ex]}}.

\bibitem{Aaij:2013zoa}
 LHCb Collaboration (R.~Aaij {\em et~al.}), {\em Phys. Rev. Lett. 110,} {\bf
  222001}  (2013), \href{http://arxiv.org/abs/1302.6269}{{\ttfamily
  arXiv:1302.6269 [hep-ex]}}.

\bibitem{Voloshin:2007dx}
M.~Voloshin, {\em Prog.Part.Nucl.Phys.} {\bf 61}, 455  (2008),
  \href{http://arxiv.org/abs/0711.4556}{{\ttfamily arXiv:0711.4556 [hep-ph]}}.

\bibitem{Chatrchyan:2013cld}
 CMS Collaboration (S.~Chatrchyan {\em et~al.}), {\em JHEP} {\bf 1304},   154
  (2013), \href{http://arxiv.org/abs/1302.3968}{{\ttfamily arXiv:1302.3968
  [hep-ex]}}.

\bibitem{Artoisenet:2009wk}
P.~Artoisenet and E.~Braaten, {\em Phys.Rev.} {\bf D81},   114018  (2010),
  \href{http://arxiv.org/abs/0911.2016}{{\ttfamily arXiv:0911.2016 [hep-ph]}}.

\bibitem{Bignamini:2009sk}
C.~Bignamini, B.~Grinstein, F.~Piccinini, A.~D. Polosa and C.~Sabelli, {\em
  Phys.Rev.Lett.} {\bf 103},   162001  (2009),
  \href{http://arxiv.org/abs/0906.0882}{{\ttfamily arXiv:0906.0882 [hep-ph]}}.

\bibitem{Khoze:2004ak}
V.~A. Khoze, A.~D. Martin, M.~G. Ryskin and W.~J. Stirling, {\em Eur.Phys.J.}
  {\bf C38}, 475  (2005), \href{http://arxiv.org/abs/hep-ph/0409037}{{\ttfamily
  arXiv:hep-ph/0409037 [hep-ph]}}.

\bibitem{Khoze:2007td}
V.~A. Khoze, A.~D. Martin and M.~G. Ryskin, {\em Frascati Phys.Ser.} {\bf 44},
  147  (2007), \href{http://arxiv.org/abs/0705.2314}{{\ttfamily arXiv:0705.2314
  [hep-ph]}}.

\bibitem{Aaltonen:2007am}
 CDF Collaboration (T.~Aaltonen {\em et~al.}), {\em Phys.Rev.Lett.} {\bf 99},
  242002  (2007), \href{http://arxiv.org/abs/0707.2374}{{\ttfamily
  arXiv:0707.2374 [hep-ex]}}.

\bibitem{Aaltonen:2011hi}
 CDF Collaboration Collaboration (T.~Aaltonen {\em et~al.}), {\em
  Phys.Rev.Lett.} {\bf 108},   081801  (2012),
  \href{http://arxiv.org/abs/1112.0858}{{\ttfamily arXiv:1112.0858 [hep-ex]}}.

\bibitem{CMSgam}
CMS Collaboration (2012), CMS-PAS-FWD-11-004.

\bibitem{Bern:2001dg}
Z.~Bern, A.~De~Freitas, L.~J. Dixon, A.~Ghinculov and H.~L. Wong, {\em JHEP}
  {\bf 0111},   031  (2001),
  \href{http://arxiv.org/abs/hep-ph/0109079}{{\ttfamily arXiv:hep-ph/0109079
  [hep-ph]}}.

\bibitem{Jikia:1993tc}
G.~Jikia and A.~Tkabladze, {\em Phys.Lett.} {\bf B323}, 453  (1994),
  \href{http://arxiv.org/abs/hep-ph/9312228}{{\ttfamily arXiv:hep-ph/9312228
  [hep-ph]}}.

\bibitem{Lai:2010vv}
H.-L. Lai, M.~Guzzi, J.~Huston, Z.~Li, P.~M. Nadolsky {\em et~al.}, {\em
  Phys.Rev.} {\bf D82},   074024  (2010),
  \href{http://arxiv.org/abs/1007.2241}{{\ttfamily arXiv:1007.2241 [hep-ph]}}.

\bibitem{Ball:2010de}
R.~D. Ball, L.~Del~Debbio, S.~Forte, A.~Guffanti, J.~I. Latorre {\em et~al.},
  {\em Nucl.Phys.} {\bf B838}, 136  (2010),
  \href{http://arxiv.org/abs/1002.4407}{{\ttfamily arXiv:1002.4407 [hep-ph]}}.

\bibitem{Spira:1995rr}
M.~Spira, A.~Djouadi, D.~Graudenz and P.~M. Zerwas, {\em Nucl.Phys.} {\bf
  B453}, 17  (1995), \href{http://arxiv.org/abs/hep-ph/9504378}{{\ttfamily
  arXiv:hep-ph/9504378 [hep-ph]}}.

\bibitem{Kunszt:1996yp}
Z.~Kunszt, S.~Moretti and W.~J. Stirling, {\em Z.Phys.} {\bf C74}, 479  (1997),
  \href{http://arxiv.org/abs/hep-ph/9611397}{{\ttfamily arXiv:hep-ph/9611397
  [hep-ph]}}.

\bibitem{Barbieri:1981xz}
R.~Barbieri, M.~Caffo, R.~Gatto and E.~Remiddi, {\em Nucl.Phys.} {\bf B192},
  ~61  (1981).

\bibitem{Brodsky:1981rp}
S.~J. Brodsky and G.~P. Lepage, {\em Phys.Rev.} {\bf D24},   1808  (1981).

\bibitem{Benayoun:1989ng}
M.~Benayoun and V.~L. Chernyak, {\em Nucl.Phys.} {\bf B329},   285  (1990).

\bibitem{Harland-Lang:2013ncy}
L.~A. Harland-Lang, V.~A. Khoze, M.~G. Ryskin and W.~J. Stirling, {\em
  Eur.Phys.J.} {\bf C73},   2429  (2013),
  \href{http://arxiv.org/abs/1302.2004}{{\ttfamily arXiv:1302.2004 [hep-ph]}}.

\bibitem{Harland-Lang:2013qia}
L.~A. Harland-Lang, V.~A. Khoze, M.~G. Ryskin and W.~J. Stirling, {\em
  Phys.Lett.} {\bf B725}, 316  (2013),
  \href{http://arxiv.org/abs/1304.4262}{{\ttfamily arXiv:1304.4262 [hep-ph]}}.

\bibitem{Thomas:2007uy}
C.~E. Thomas, {\em JHEP} {\bf 0710},   026  (2007),
  \href{http://arxiv.org/abs/0705.1500}{{\ttfamily arXiv:0705.1500 [hep-ph]}}.

\bibitem{DiDonato:2011kr}
C.~Di~Donato, G.~Ricciardi and I.~Bigi, {\em Phys.Rev.} {\bf D85},   013016
  (2012), \href{http://arxiv.org/abs/1105.3557}{{\ttfamily arXiv:1105.3557
  [hep-ph]}}.

\bibitem{Lepage:1980fj}
G.~P. Lepage and S.~J. Brodsky, {\em Phys.Rev.} {\bf D22},   2157  (1980).

\bibitem{Aubert:2009mc}
 BABAR Collaboration (B.~Aubert {\em et~al.}), {\em Phys.Rev.} {\bf D80},
  052002  (2009), \href{http://arxiv.org/abs/0905.4778}{{\ttfamily
  arXiv:0905.4778 [hep-ex]}}.

\bibitem{Druzhinin:2009gq}
V.~P. Druzhinin, {\em PoS} {\bf EPS-HEP2009},   051  (2009),
  \href{http://arxiv.org/abs/0909.3148}{{\ttfamily arXiv:0909.3148 [hep-ex]}}.

\bibitem{Uehara:2012ag}
 Belle Collaboration (S.~Uehara {\em et~al.})  (2012),
  \href{http://arxiv.org/abs/1205.3249}{{\ttfamily arXiv:1205.3249 [hep-ex]}}.

\bibitem{Chernyak:1981zz}
V.~L. Chernyak and A.~R. Zhitnitsky, {\em Nucl.Phys.} {\bf B201},   492
  (1982).

\bibitem{Ohrndorf:1981uz}
T.~Ohrndorf, {\em Nucl.Phys.} {\bf B186},   153  (1981).

\bibitem{Atkinson:1983yh}
G.~W. Atkinson, J.~Sucher and K.~Tsokos, {\em Phys.Lett.} {\bf B137},   407
  (1984).

\bibitem{Baier:1985wv}
V.~N. Baier and A.~G. Grozin, {\em Z.Phys.} {\bf C29}, 161  (1985).

\bibitem{Baier:1981pm}
V.~N. Baier and A.~G. Grozin, {\em Nucl.Phys.} {\bf B192}, 476  (1981).

\bibitem{Feldmann:1997vc}
T.~Feldmann and P.~Kroll, {\em Eur.Phys.J.} {\bf C5}, 327  (1998),
  \href{http://arxiv.org/abs/hep-ph/9711231}{{\ttfamily arXiv:hep-ph/9711231
  [hep-ph]}}.

\bibitem{Kiselev:1992ms}
A.~V. Kiselev and V.~A. Petrov, {\em Z.Phys.} {\bf C58}, 595  (1993).

\bibitem{Leutwyler:1997yr}
H.~Leutwyler, {\em Nucl.Phys.Proc.Suppl.} {\bf 64}, 223  (1998),
  \href{http://arxiv.org/abs/hep-ph/9709408}{{\ttfamily arXiv:hep-ph/9709408
  [hep-ph]}}.

\bibitem{Feldmann:1998vh}
T.~Feldmann, P.~Kroll and B.~Stech, {\em Phys.Rev.} {\bf D58},   114006
  (1998), \href{http://arxiv.org/abs/hep-ph/9802409}{{\ttfamily
  arXiv:hep-ph/9802409 [hep-ph]}}.

\bibitem{Heyssler:1997ng}
M.~Heyssler and W.~J. Stirling, {\em Eur.Phys.J.} {\bf C5}, 475  (1998),
  \href{http://arxiv.org/abs/hep-ph/9712314}{{\ttfamily arXiv:hep-ph/9712314
  [hep-ph]}}.

\bibitem{Brown:1982xx}
R.~W. Brown, K.~L. Kowalski and S.~J. Brodsky, {\em Phys.Rev.} {\bf D28},   624
   (1983).

\bibitem{Brodsky:1981kj}
S.~J. Brodsky and G.~P. Lepage, {\em Phys.Rev.} {\bf D24},   2848  (1981).

\bibitem{Mangano:1990by}
M.~L. Mangano and S.~J. Parke, {\em Phys.Rept.} {\bf 200}, 301  (1991),
  \href{http://arxiv.org/abs/hep-th/0509223}{{\ttfamily arXiv:hep-th/0509223
  [hep-th]}}.

\bibitem{Parke:1986gb}
S.~J. Parke and T.~R. Taylor, {\em Phys.Rev.Lett.} {\bf 56},   2459  (1986).

\bibitem{Berends:1987me}
F.~A. Berends and W.~T. Giele, {\em Nucl.Phys.} {\bf B306},   759  (1988).

\bibitem{Britto:2004ap}
R.~Britto, F.~Cachazo and B.~Feng, {\em Nucl.Phys.} {\bf B715}, 499  (2005),
  \href{http://arxiv.org/abs/hep-th/0412308}{{\ttfamily arXiv:hep-th/0412308
  [hep-th]}}.

\bibitem{Georgiou:2004by}
G.~Georgiou, E.~N. Glover and V.~V. Khoze, {\em JHEP} {\bf 0407},   048
  (2004), \href{http://arxiv.org/abs/hep-th/0407027}{{\ttfamily
  arXiv:hep-th/0407027 [hep-th]}}.

\bibitem{Chernyak:2009dj}
V.~L. Chernyak  (2009), \href{http://arxiv.org/abs/0912.0623}{{\ttfamily
  arXiv:0912.0623 [hep-ph]}}.

\bibitem{Nakazawa:2004gu}
 BELLE Collaboration (H.~Nakazawa {\em et~al.}), {\em Phys.Lett.} {\bf B615},
  39  (2005), \href{http://arxiv.org/abs/hep-ex/0412058}{{\ttfamily
  arXiv:hep-ex/0412058 [hep-ex]}}.

\bibitem{Kroll:2013iwa}
P.~Kroll and K.~Passek-Kumericki, {\em J.Phys.} {\bf G40},   075005  (2013),
  \href{http://arxiv.org/abs/1206.4870}{{\ttfamily arXiv:1206.4870 [hep-ph]}}.

\bibitem{Ochs:2013gi}
W.~Ochs, {\em J.Phys.} {\bf G40},   043001  (2013),
  \href{http://arxiv.org/abs/1301.5183}{{\ttfamily arXiv:1301.5183 [hep-ph]}}.

\bibitem{Kaidalov:1974qi}
A.~Kaidalov and K.~Ter-Martirosyan, {\em Nucl.Phys.} {\bf B75}, 471  (1974).

\bibitem{Azimov:1974fa}
Y.~I. Azimov, V.~A. Khoze, E.~M. Levin and M.~G. Ryskin, {\em Sov.J.Nucl.Phys.}
  {\bf 21},   215  (1975).

\bibitem{Pumplin:1976dm}
J.~Pumplin and F.~Henyey, {\em Nucl.Phys.} {\bf B117},   377  (1976).

\bibitem{Desai:1978rh}
B.~R. Desai, B.~C. Shen and M.~Jacob, {\em Nucl.Phys.} {\bf B142},   258
  (1978).

\bibitem{Lebiedowicz:2009pj}
P.~Lebiedowicz and A.~Szczurek, {\em Phys.Rev.} {\bf D81},   036003  (2010),
  \href{http://arxiv.org/abs/0912.0190}{{\ttfamily arXiv:0912.0190 [hep-ph]}}.

\bibitem{Lebiedowicz:2012nk}
P.~Lebiedowicz and A.~Szczurek  (2012),
  \href{http://arxiv.org/abs/1212.0166}{{\ttfamily arXiv:1212.0166 [hep-ph]}}.

\bibitem{Collins:1977jyp}
P. D. B. Collins, \emph{An Introduction to Regge Theory and High-Energy
  Physics} (1977).

\bibitem{Breakstone:1989ty}
 Ames-Bologna-CERN-Dortmund-Heidelberg-Warsaw Collaboration (A.~Breakstone {\em
  et~al.}), {\em Z.Phys.} {\bf C42},   387  (1989).

\bibitem{Breakstone:1990at}
 Ames-Bologna-CERN-Dortmund-Heidelberg-Warsaw Collaboration (A.~Breakstone {\em
  et~al.}), {\em Z.Phys.} {\bf C48}, 569  (1990).

\bibitem{enterria}
David d'Enterria, private communication.

\bibitem{Sykoraeds}
Tomas Sykora, talk at EDS Blois 2013 Workshop, Saariselka, Lapland, September
  9-13.

\bibitem{Paula}
Paula Collins, talk at `Results and Prospects of Forward Physics at the LHC',
  CERN, Feb. 11-13, 2013.

\bibitem{fortheCOMPASS:2013vda}
 COMPASS Collaboration (A.~Austregesilo)  (2013),
  \href{http://arxiv.org/abs/1310.3190}{{\ttfamily arXiv:1310.3190 [hep-ex]}}.

\bibitem{Albrowpriv1}
Mike Albrow and Erik Brucken, private communication.

\bibitem{Martin:1997kv}
A.~D. Martin, M.~G. Ryskin and V.~A. Khoze, {\em Phys.Rev.} {\bf D56}, 5867
  (1997), \href{http://arxiv.org/abs/hep-ph/9705258}{{\ttfamily
  arXiv:hep-ph/9705258 [hep-ph]}}.

\bibitem{Khoze:2006iw}
V.~A. Khoze, A.~D. Martin and M.~G. Ryskin, {\em Eur.Phys.J.} {\bf C48}, 467
  (2006), \href{http://arxiv.org/abs/hep-ph/0605113}{{\ttfamily
  arXiv:hep-ph/0605113 [hep-ph]}}.

\bibitem{Aaltonen:2007hs}
 CDF Collaboration (T.~Aaltonen {\em et~al.}), {\em Phys.Rev.} {\bf D77},
  052004  (2008), \href{http://arxiv.org/abs/0712.0604}{{\ttfamily
  arXiv:0712.0604 [hep-ex]}}.

\bibitem{Abazov:2010bk}
 D0 Collaboration Collaboration (V.~M. Abazov {\em et~al.}), {\em Phys.Lett.}
  {\bf B705}, 193  (2011), \href{http://arxiv.org/abs/1009.2444}{{\ttfamily
  arXiv:1009.2444 [hep-ex]}}.

\bibitem{Khoze:2006um}
V.~A. Khoze, M.~G. Ryskin and W.~J. Stirling, {\em Eur.Phys.J.} {\bf C48}, 477
  (2006), \href{http://arxiv.org/abs/hep-ph/0607134}{{\ttfamily
  arXiv:hep-ph/0607134 [hep-ph]}}.

\bibitem{Tasevsky:2009zza}
M.~Tasevsky, {\em Nucl.Phys.Proc.Suppl.} {\bf 179-180}, 187  (2008).

\bibitem{Albrow:2012oha}
M.~G. Albrow, {\em AIP Conf.Proc.} {\bf 1523}, 320  (2012).

\bibitem{Shuvaev:2008yn}
A.~G. Shuvaev, V.~A. Khoze, A.~D. Martin and M.~G. Ryskin, {\em Eur.Phys.J.}
  {\bf C56}, 467  (2008), \href{http://arxiv.org/abs/0806.1447}{{\ttfamily
  arXiv:0806.1447 [hep-ph]}}.

\bibitem{Khoze:2009er}
V.~A. Khoze, M.~G. Ryskin and A.~D. Martin, {\em Eur.Phys.J.} {\bf C64}, 361
  (2009), \href{http://arxiv.org/abs/0907.0966}{{\ttfamily arXiv:0907.0966
  [hep-ph]}}.

\bibitem{Low:1958sn}
F.~Low, {\em Phys.Rev.} {\bf 110}, 974  (1958).

\bibitem{Burnett:1967km}
T.~Burnett and N.~M. Kroll, {\em Phys.Rev.Lett.} {\bf 20},  ~86  (1968).

\bibitem{Monk:2005ji}
J.~Monk and A.~Pilkington, {\em Comput.Phys.Commun.} {\bf 175}, 232  (2006),
  \href{http://arxiv.org/abs/hep-ph/0502077}{{\ttfamily arXiv:hep-ph/0502077
  [hep-ph]}}.

\bibitem{HLfut}
L.A. Harland-Lang, V. A. Khoze, and M. G. Ryskin, future publication.

\bibitem{Albrow:2013nia}
M.~Albrow  (2013), \href{http://arxiv.org/abs/1310.7047}{{\ttfamily
  arXiv:1310.7047 [hep-ex]}}.

\bibitem{Maciula:2010tv}
R.~Maciula, R.~Pasechnik and A.~Szczurek, {\em Phys.Rev.} {\bf D83},   114034
  (2011), \href{http://arxiv.org/abs/1011.5842}{{\ttfamily arXiv:1011.5842
  [hep-ph]}}.

\bibitem{Heinemeyer:2011jb}
S.~Heinemeyer, V.~A. Khoze, M.~G. Ryskin, M.~Tasevsky and G.~Weiglein  (2011),
  \href{http://arxiv.org/abs/1106.3450}{{\ttfamily arXiv:1106.3450 [hep-ph]}}.

\bibitem{Khoze:2010ba}
V.~A. Khoze, A.~D. Martin, M.~G. Ryskin and A.~Shuvaev, {\em Eur.Phys.J.} {\bf
  C68}, 125  (2010), \href{http://arxiv.org/abs/1002.2857}{{\ttfamily
  arXiv:1002.2857 [hep-ph]}}.

\bibitem{deFavereaudeJeneret:2009db}
J.~de~Favereau~de Jeneret, V.~Lemaitre, Y.~Liu, S.~Ovyn, T.~Pierzchala {\em
  et~al.}  (2009), \href{http://arxiv.org/abs/0908.2020}{{\ttfamily
  arXiv:0908.2020 [hep-ph]}}.

\bibitem{Tasevsky:2013iea}
M.~Tasevsky, {\em Eur.Phys.J.} {\bf C73},   2672  (2013),
  \href{http://arxiv.org/abs/1309.7772}{{\ttfamily arXiv:1309.7772 [hep-ph]}}.

\bibitem{Chatrchyan:2012ufa}
 CMS Collaboration (S.~Chatrchyan {\em et~al.}), {\em Phys.Lett.} {\bf B716},
  30  (2012), \href{http://arxiv.org/abs/1207.7235}{{\ttfamily arXiv:1207.7235
  [hep-ex]}}.

\bibitem{Chatrchyan:2013lba}
 CMS Collaboration (S.~Chatrchyan {\em et~al.}), {\em JHEP} {\bf 1306},   081
  (2013), \href{http://arxiv.org/abs/1303.4571}{{\ttfamily arXiv:1303.4571
  [hep-ex]}}.

\bibitem{Aad:2012tfa}
 ATLAS Collaboration (G.~Aad {\em et~al.}), {\em Phys.Lett.} {\bf B716}, 1
  (2012), \href{http://arxiv.org/abs/1207.7214}{{\ttfamily arXiv:1207.7214
  [hep-ex]}}.

\bibitem{Kaidalov:2003ys}
A.~Kaidalov, V.~A. Khoze, A.~D. Martin and M.~G. Ryskin, {\em Eur.Phys.J.} {\bf
  C33}, 261  (2004), \href{http://arxiv.org/abs/hep-ph/0311023}{{\ttfamily
  arXiv:hep-ph/0311023 [hep-ph]}}.

\bibitem{Chaichian:2009ts}
M.~Chaichian, P.~Hoyer, K.~Huitu, V.~A. Khoze and A.~D. Pilkington, {\em JHEP}
  {\bf 0905},   011  (2009), \href{http://arxiv.org/abs/0901.3746}{{\ttfamily
  arXiv:0901.3746 [hep-ph]}}.

\bibitem{Heinemeyer:2010gs}
S.~Heinemeyer, V.~A. Khoze, M.~G. Ryskin, M.~Tasevsky and G.~Weiglein, {\em
  Eur.Phys.J.} {\bf C71},   1649  (2011),
  \href{http://arxiv.org/abs/1012.5007}{{\ttfamily arXiv:1012.5007 [hep-ph]}}.

\bibitem{kmrcp}
V.~A.~Khoze, A.~D.~Martin and M.~G.~Ryskin, In:``Workshop on CP Studies and
  Non-Standard Higgs Physics, hep-ph/0608079, pp 144-149.

\bibitem{Ellis:2006eh}
J.~R. Ellis, J.~S. Lee and A.~Pilaftsis, {\em Mod.Phys.Lett.} {\bf A21}, 1405
  (2006), \href{http://arxiv.org/abs/hep-ph/0605288}{{\ttfamily
  arXiv:hep-ph/0605288 [hep-ph]}}.

\bibitem{Carena:2000ks}
M.~S. Carena, J.~R. Ellis, A.~Pilaftsis and C.~Wagner, {\em Phys.Lett.} {\bf
  B495}, 155  (2000), \href{http://arxiv.org/abs/hep-ph/0009212}{{\ttfamily
  arXiv:hep-ph/0009212 [hep-ph]}}.

\bibitem{Duhrssen:2004cv}
M.~Duhrssen, S.~Heinemeyer, H.~Logan, D.~Rainwater, G.~Weiglein {\em et~al.},
  {\em Phys.Rev.} {\bf D70},   113009  (2004),
  \href{http://arxiv.org/abs/hep-ph/0406323}{{\ttfamily arXiv:hep-ph/0406323
  [hep-ph]}}.

\bibitem{LHCHiggsX1}
 LHC Higgs Cross Section Working Group Collaboration (A.~David {\em et~al.})
  (2012), \href{http://arxiv.org/abs/1209.0040}{{\ttfamily arXiv:1209.0040
  [hep-ph]}}.

\bibitem{LHCHiggsX2}
 LHC Higgs Cross Section Working Group Collaboration (S.~Heinemeyer {\em
  et~al.})  (2013), \href{http://arxiv.org/abs/1307.1347}{{\ttfamily
  arXiv:1307.1347 [hep-ph]}}.

\bibitem{Gluck:2007ck}
M.~Gluck, P.~Jimenez-Delgado and E.~Reya, {\em Eur.Phys.J.} {\bf C53}, 355
  (2008), \href{http://arxiv.org/abs/0709.0614}{{\ttfamily arXiv:0709.0614
  [hep-ph]}}.

\bibitem{Cox:2007sw}
B.~Cox, F.~Loebinger and A.~D. Pilkington, {\em JHEP} {\bf 0710},   090
  (2007), \href{http://arxiv.org/abs/0709.3035}{{\ttfamily arXiv:0709.3035
  [hep-ph]}}.

\bibitem{Heinemeyer:2012hr}
S.~Heinemeyer, V.~A. Khoze, M.~Tasevsky and G.~Weiglein, 493  (2012),
  \href{http://arxiv.org/abs/1206.0183}{{\ttfamily arXiv:1206.0183 [hep-ph]}}.

\bibitem{Carena:2013qia}
M.~Carena, S.~Heinemeyer, O.~St\r{a}Cl, C.~Wagner and G.~Weiglein, {\em Eur.
  Phys. J.} {\bf C73},   2552  (2013),
  \href{http://arxiv.org/abs/1302.7033}{{\ttfamily arXiv:1302.7033 [hep-ph]}}.

\bibitem{Bechtle:2012jw}
P.~Bechtle, S.~Heinemeyer, O.~Stal, T.~Stefaniak, G.~Weiglein {\em et~al.},
  {\em Eur.Phys.J.} {\bf C73},   2354  (2013),
  \href{http://arxiv.org/abs/1211.1955}{{\ttfamily arXiv:1211.1955 [hep-ph]}}.

\bibitem{Khoze:2008cx}
V.~A. Khoze, A.~D. Martin and M.~G. Ryskin, {\em Eur.Phys.J.} {\bf C55}, 363
  (2008), \href{http://arxiv.org/abs/0802.0177}{{\ttfamily arXiv:0802.0177
  [hep-ph]}}.

\end{thebibliography}

\end{document}